\documentclass{article}
\usepackage{graphicx}
\usepackage[english]{babel}
\usepackage{blindtext}
\usepackage{amsmath}
\usepackage{amssymb}
\usepackage{amsfonts}
\usepackage{amsthm}
\usepackage{color}
\usepackage{cases}
\usepackage{setspace}
\usepackage{fullpage}

\def\theequation{{\arabic{section}}.{\arabic{equation}}}

\DeclareMathAlphabet{\itbf}{OML}{cmm}{b}{it}

\def\bD{{{\itbf D}}}

\newcommand{\mb}[1]{\mathbf{#1}}
\usepackage{hyperref}

\usepackage{subcaption}

\usepackage{algorithm}
\usepackage{algorithmic}

\newcommand{\nc}{\newcommand}
\nc{\dsp}{\displaystyle}
\nc{\R}{\Bbb{R}}
\nc{\Z}{\Bbb{Z}}
\nc{\Pp}{\Bbb{P}}
\nc{\Ap}{\Bbb{A}}
\nc{\Wp}{\Bbb{W}}
 \nc{\brho}{\boldsymbol \rho}
\nc{\va}{\vec{\boldsymbol \mu}}
\nc{\ve}{\vec{\boldsymbol \epsilon}}
\nc{\bk}{{\bf k}}
\nc{\vrho}{\vec{\brho}} 
\nc{\vr}{\vec{\bf r}}
\nc{\vx}{\vec{\bf x}}
\nc{\om}{\omega}
\nc{\brhoi}{\brho^{\cal I}}
\nc{\bzetai}{\bzeta^{\cal I}}
\nc{\vzeta}{\vec{\bzeta}}
\nc{\vzetai}{\vec{\bzeta}^{\cal I}}
\nc{\cI}{{\cal I}}
\nc{\cJ}{{\cal J}}
\nc{\cF}{{\cal F}}
\nc{\cW}{{\cal W}}
\nc{\cA}{{\cal A}}
\nc{\cL}{{\cal L}}
\nc{\cS}{{\cal S}}
\nc{\cC}{{\cal C}}
\nc{\cN}{{\cal N}}
\nc{\vrhoi}{\vec{\brho}^{\,\cal I}}
\nc{\xii}{\xi^{\cI}}
\nc{\etai}{\eta^{\cI}}
\nc{\la}{\lambda}
\nc{\de}{\delta}
\nc{\ep}{\varepsilon}
\nc{\vu}{\vec{\bf u}}
\nc{\bu}{{\bf u}}
\nc{\vui}{\vec{\bf u}^{\cI}}
\nc{\bui}{{\bf u}^{\cI}}
\nc{\bt}{{\bf t}}
\nc{\vt}{\vec{\bt}}
\nc{\bn}{{\bf n}}
\nc{\vn}{\vec{\bn}}
\nc{\bm}{{\bf m}}
\nc{\vm}{\vec{\bm}}
\nc{\vrp}{\vec{{\bf r}'}}
\nc{\vrc}{\vec{{\bf r}^c_p}}
\nc{\ts}{\tilde s}
\nc{\os}{\overline s}
\nc{\tom}{\tilde \om}
\nc{\tO}{\tilde \Omega}
\nc{\tS}{\tilde S}
\nc{\oS}{\overline S}
\nc{\vrhos}{\vrho_{\star}}
\nc{\vrhosi}{\vrho_{\star}^{\cI}}
\nc{\brhos}{\brho_{\star}}
\nc{\brhosi}{\brhos^{\cI}}
\nc{\vms}{\vm_{\star}}
\nc{\vmi}{\vm_{_{\cI}}}
\nc{\vM}{\vec{\bf M}}
\nc{\vMi}{\vM_{_{\cI}}}
\nc{\Ppi}{\Pp_{\cI}}
\nc{\vxi}{\vec{\bxi}}
\nc{\bK}{{\bf K}}
\nc{\bmi}{\bm_{_{\cI}}}
\nc{\Pppi}{\Ppi^p}
\nc{\be}{{\bf e}}
\nc{\bep}{{\bf e}^p}
\renewcommand{\hat}{\widehat}

\begin{document}
	\title{Generalized correlation based Imaging for satellites}\author{Matan Leibovich\footnotemark[1], George Papanicolaou\footnotemark[2], and Chrysoula Tsogka\footnotemark[3]
	}\renewcommand{\thefootnote}{\fnsymbol{footnote}}
	\footnotetext[2]{Institute for Computational and Mathematical Engineering, Stanford University, Stanford, CA 94305. \\\   (matanle@stanford.edu)} 
	\footnotetext[2]{Department of
		Mathematics, Stanford University, Stanford, CA 94305.
		(papanicolalou@stanford.edu)}\footnotetext[3]{Applied Math Unit,
  University of California, Merced, 5200 North Lake Road, Merced, CA
  95343 (ctsogka@ucmerced.edu)}\date{}
	\maketitle  \date{}
	
	
	\begin{abstract}
We consider imaging of fast moving small objects in space, such as low earth orbit satellites. The imaging system consists of ground based, asynchronous sources of radiation and several passive receivers above the dense atmosphere. We use the cross correlation of the received signals to reduce distortions from ambient medium fluctuations. Imaging with correlations also has the advantage of not requiring any knowledge about the probing pulse 
and depends weakly on the emitter positions. We account for the target's orbital velocity by introducing the necessary Doppler compensation. We show that over limited imaging regions, a constant Doppler factor can be used, resulting in an efficient data structure for the correlations of the recorded signals. We then investigate and analyze different imaging methods using the cross-correlation data structure. Specifically, we show that using a generalized two point migration of the cross correlation data, the top eigenvector of the migrated data matrix provides superior image resolution compared to the usual single-point migration scheme. We carry out a theoretical analysis that illustrates the role of the two point migration methods as well as that of the inverse aperture in improving resolution. Extensive numerical simulations support the theoretical results and assess the scope of the imaging methodology. 
	\end{abstract}
\section{Introduction}
\label{sec:intro}

The imaging of satellites in low earth orbit is motivated by the need to detect and closely track small debris (1-10cm) revolving around the earth at altitudes in the range of $200$km-$2000$km, \cite{sandia}. 
The amount of debris has been growing steadily in recent years, substantially increasing the risk of satellite damage from collisions \cite{esa, amos15}. 
There are roughly $700{,}000$ debris of size larger than $1$cm in Low Earth Orbit (LEO) and there is concern that future collisions 
may lead to a chain reaction that will generate an unacceptably risky environment \cite{wormnes},\cite{kessler1978collision}. 
In this paper, we model the small fast moving debris as point-like reflectors moving with constant velocity, $\mb v_\mb T$. 

The recorded data are the scattered signals from a train of incident pulses emitted by transmitters located on the ground. 
The receivers are assumed to be located at a height of 15 km or more. They span an area of diameter $a$, 
which acts as the physical aperture of the imaging system. In synthetic aperture radar (SAR), 
a single airborne transmit/receive element is moving and its trajectory defines the synthetically created aperture of the imaging system \cite{curlander, lanari99}. 
In a similar way, the trajectory of the moving target defines an inverse synthetic aperture (iSAR) of length $S|\mb v_\mb T|$,  
with $S$ denoting the total time of data acquisition. 
Other important parameters of the imaging system are the central frequency, $f_o$, and the bandwidth, $B$ of the probing signals as well as the pulse repetition frequency.
We consider here emitters operating at high frequency with a relatively large bandwidth, such as the X-Band (8-12 GHz) with a bandwidth of up to 600 MHz.

%
Correlation-based imaging uses the cross correlations of the recorded signals between pairs of receivers. 
The signals are also Doppler compensated and synchronized. 
However, in this case the synchronization does not require knowledge of the emitter locations since only time differences matter in cross correlations. 
In correlation-based imaging we also do not need to know the pulse profile or the emission times but we need to record an extended train of scattered pulses. 
We assume that recording is done at a sufficiently high sampling rate so that signals can be recorded with no loss of information \cite{garnier17, issueieee}.
Correlation-based imaging has been shown to be more robust to ambient medium fluctuations such as atmospheric lensing \cite{lawrence} and aberrations \cite{mcmillan}. 
This is true for receivers that are not located on the ground but are located above the turbulent atmosphere, on drones for example.
Indeed, considering airborne receivers transforms the passive correlation-based problem to a virtual source array imaging problem that has been studied for stationary receiver arrays in \cite{garnier12a,garnier16,virtual2}. 
The key idea is that passive correlation-based imaging becomes equivalent to having a virtual active array at the location of the passive receivers. 
By moving the receivers above the turbulent atmosphere, the atmospheric fluctuation effects on imaging are minimized and imaging resolution is as if we were in a homogeneous, fluctuation free medium. 

Correlation-based imaging is, in addition, passive because it can be carried out using opportunistic emitters with largely unknown properties. 
In the imaging setting considered in this paper, opportunistic sources could be global navigation satellite systems (GNSS). 
This has been considered in \cite{amos16_leo} and it is shown that the main challenge in this case is the low signal-to-noise ratio 
because the scattered signals received at terrestrial receivers are very weak. 

The use of correlation based imaging for satellites is considered in \cite{fournier2017matched}, where its resolution is analyzed and compared to matched-filter based imaging. Matched-filter imaging depends linearly on the measurements, but requires knowing the position of the emitters with high accuracy, as well as the emitted pulse trains. It also lacks the robustness to medium fluctuations, contrary to correlation based imaging.  It is shown  in \cite{fournier2017matched} that under certain conditions matched-filter and correlation based imaging attain comparable resolutions, indicating that correlation based imaging is a good method for LEO satellite imaging. The imaging functions introduced in \cite{fournier2017matched} are a generalization of the well known Kirchhoff migration (KM), which forms an image by superposing the recorded signals after translating them by the travel time to a search location in the image region. 

As was noted in \cite{fournier2017matched}, the resolution analysis is limited to well separated point scatterers. While this is a suitable choice with a linear imaging method, where more complicated scenarios are considered as superpositions of the single target case, this is no longer so when dealing with correlations, especially when there is interference between neighboring targets. In correlation based imaging, the abilities to localize and separate targets need to be addressed separately. We show in this paper that the single point scatterer resolution results do not generalize to multiple closely spaced scatterers in a straightforward manner. Another key result of \cite{fournier2017matched} is that the resolution of the target depends only weakly on the inverse synthetic aperture. This seems to be counter intuitive as the correlation data still contains a significant amount of information. These observations have motivated us to consider other possible extensions of the imaging function.

\subsection{Main result}
In this paper we present a generalization of the migration method for correlations, which is motivated by the structure of the data. We first define a correlation data structure, which compensates for the Doppler effect on the measurements. We then show that out of this data structure a natural migration scheme arises, which translates the data to two points rather than one. This is natural when considering data correlations, and a similar migration scheme was recently used in \cite{borcea2019twopoint}. The result of this migration is not an image but rather a  \textbf{two-point interference pattern}. We show that there are several possible ways to derive an image from the two-point interference pattern, with one in particular that we call the \textbf{rank-1 image}.
This rank-1 image is the result of taking the first eigenvector of the interference pattern, and it provides superior resolution. We demonstrate its superiority both in terms of favorable mathematical properties, such as having an optimization interpretation, as well as by providing analytical explanations of its performance. This is the main result of the paper.

Numerical simulations confirm that the resolution of the rank-1 image is better than that of the previously used correlation based imaging function, and achieves resolution comparable to that of linear migration, while retaining the benefits of imaging with correlations rather than with the data directly. The method is stable with respect to additive noise
and can be applied to many other imaging problems.

The rest of the paper is as follows. In Section~\ref{sec:corr_imaging} we briefly review the model for our data (further expanded in Appendix~\ref{app:forward_wave}), and introduce the cross-correlation data structure. In Section~\ref{sec:generalized_migration} we recast the problem in matrix form and introduce the interference pattern, and the possible extensions of KM to cross-correlation data, particularly the rank-1 image. In Section~\ref{sec:numerical_simulations} we present numerical simulations which confirm the superior performance of the rank-1 image. We motivate the results by further simulations in Section~\ref{sec:prop_filt}, and analysis in Appendix~\ref{app:stat_phase} and Appendix~\ref{app:kernel}. We conclude with a summary and conclusions in Section~\ref{sec:summary}.

\section{Correlation based imaging of fast moving objects}
\label{sec:corr_imaging}
In this section we present the general form of the recorded signals and then review the basic algorithms of correlation based imaging with Doppler effects. We also introduce a cross-correlation data structure that is more efficient for imaging, which will be used in subsequent sections for introducing new imaging methods and comparing them with existing methods.
\subsection{Scattered signal model}
We want to image a cluster of objects, moving in Keplerian low earth orbit at a high velocity $\mb v_L$, so that without rotation for all the objects we have
\begin{equation}
\mb x_\mb T(s)=\mb x_\mb T+\mb v_\mb Ts.
\label{eq:pont_scatter}
\end{equation}

\begin{figure}[htbp]
	\centering
		\includegraphics[width=0.55\textwidth]{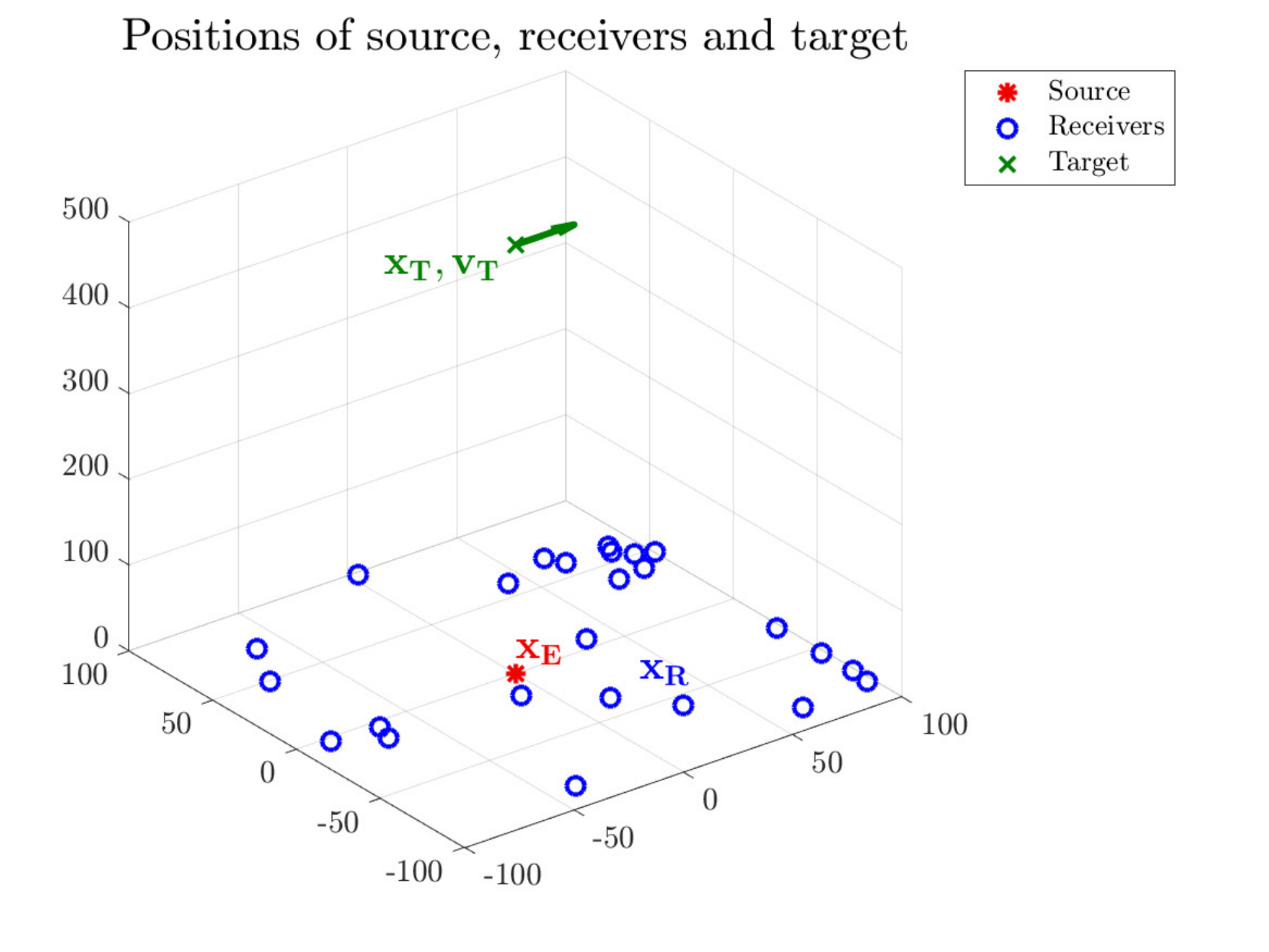}
\caption{iSAR imaging schematic. A network of receivers with positions $\mb x_\mb R$, is randomly distributed over an area of 100 km $\times$ 100 km at an altitude of 15 km.  The source $\mb x_\mb E$  is on the ground, and the target $\mb x_\mb T$ is at 500 km elevation and moving at $7$ km/s.}
\label{fig:layout}
\end{figure}

The data is the collection of signals recorded at ground based or at low elevation receivers, with positions $\mb x_\mb R$. Successive pulses are emitted at a slow time $s$ by a source located at  $\mb x_\mathbf{E}$ on the ground,  as illustrated in Figure~\ref{fig:layout}. 
The receiver emits a series of pulses $f(t)=\cos(\omega_o t)e^{-B^2t^2/2}$, at slow time intervals of $\Delta s$, with a total aperture size $S$, such that the recorded signal at the receiver location $\mb x_\mb R$ due to a pulse, $f(s+t)$, emitted at slow time $s\in[-S/2,S/2]$, is  
\begin{equation}
u_{\mathbf{R}}(s,t)=-\rho \frac{f''(s+\gamma_{\mathbf{R}}(\mb x_\mathbf{T}(s),\mb x_\mathbf{E},\mathbf{v}_\mb T) t-t_{\mathbf{R}}(\mb x_\mathbf{T}(s),\mb x_\mathbf{E},\mathbf{v}_\mb T))}{(4\pi|\mb x_{\mathbf{T}}(s)-\mb x_\mathbf{R}|)^2}.
\label{eq:scattered_field}
\end{equation}
The derivation of  \eqref{eq:scattered_field} is in Appendix~\ref{app:forward_wave}.
With $c_0$ the speed of light,  $\gamma_{\mathbf{R}}$ is the Doppler scale correction factor and $t_\mathbf{R}$ is the signal travel time, which to first order in $|\mb v_{\mb T}|/c_0$ are given by
\begin{equation}
\begin{split}
&\gamma_{\mathbf{R}}(\mb x_\mathbf{T},\mb x_\mathbf{E},\mathbf{v}_\mb T)=1-\frac{\mathbf{v}_\mb T}{c_0}\cdot\left(\frac{\mb x_\mathbf{T}-\mb x_\mathbf{E}}{|\mb x_\mathbf{T}-\mb x_\mathbf{E}|}+(\frac{\mb x_\mathbf{T}-\mb x_\mathbf{R}}{|\mb x_\mathbf{T}-\mb x_\mathbf{R}|}\right),\\
&t_{\mathbf{R}}(\mb x_\mathbf{T},\mb x_\mathbf{E},\mathbf{v}_\mb T)=\frac{|\mb x_\mathbf{T}-\mb x_\mathbf{E}|}{c_0}+\frac{|\mb x_\mathbf{T}-\mb x_\mathbf{R}|}{c_0}\gamma_{\mathbf{R}}(\mb x_\mathbf{T},\mb x_\mathbf{E},\mathbf{v}_\mb T).
\end{split}
\label{eq:gamma_t}
\end{equation}
The formula (\ref{eq:scattered_field}) for the recorded data will be used in the theoretical resolution analysis.

An important consideration in the data recording process is the sampling frequency, which in currently available microwave acquisition systems is limited to about 2 GHz. Signals with higher carrier frequency need to be downramped before sampling, to avoid information loss. This is done with a heterodyne oscillator that shifts down the carrier frequency followed by low pass filtering. When using asynchronous emitters, the central frequency and bandwidth might not be known accurately. In that case the frequency of the oscillator and the sampling frequency need to be chosen so that loss of information is minimal.

\subsection{Imaging with Doppler-scaled cross correlations}
We use the cross correlations of the signals received at different receivers to image. In \cite{fournier2017matched} the imaging function was defined via scaling and translating the signals at the receivers by different Doppler factors $\gamma _\mb R$ for every point $\mb x,\mb v$ in the imaging domain
\begin{equation}
\mathcal{I}^{CC}(\mb x,\mb v)=\sum\limits_{s_j,\mb R,\mb R'}\int dt u_{\mb R}\left(s_j, \frac{|\mb x-\mb x_\mb R|}{c_0}+\frac{t}{\gamma_{\mathbf{R}}(\mb x,\mb x_\mb E,\mb v)}\right) u_{\mb R'}\left(s_j, \frac{|\mb x-\mb x_{\mb R'}|}{c_0}+\frac{t}{\gamma_{\mathbf{R'}}(\mb x,\mb x_\mb E,\mb v)}\right). 
\label{eq:CC_singlepoint}
\end{equation}
This imaging function is a  generalization of the regular Kirchhoff migration, which back propagates the recorded signals by the travel time to a specific search point in the image domain, but now at the level of cross correlations. The idea is that when migrating and correlating the data from the different receivers, the peaks of the signals sum coherently if a reflector exists at a point $\mb x$, with velocity $\mb v$, resulting in a peak for $\mathcal{I}^{CC}(\mb x,\mb v)$ at that point. Since the rapid velocity affects the signal via $\gamma_{\mathbf{R}}$, the Doppler factor must be used when migrating the data. At low speeds and without the Doppler correction this imaging function reduces to the classical correlation based migration \cite{garnier2016passive}.

We assume here that the locations of the receivers $\mb x_\mb R,\mb x_{\mb R'}$ are accurately known. The source location $\mb x_\mb E$ is also assumed to be known, but a lesser degree of accuracy is needed, since $\mb x_\mb E$ only affects $\gamma_{\mathbf{R}}$ (see \cite{fournier2017matched}). 
The formulation \eqref{eq:CC_singlepoint} is less advantageous for two reasons: 
\begin{enumerate}
	\item It is computationally ineffective to recompute the cross correlation for any point $\mb x$ in the imaging domain. It would be much more efficient to compute the cross correlations once for any triplet $(s_j,\mb R,\mb R')$ as a function of $\tau$, the translation, and use that for imaging \cite{garnier2016passive}.
	\item Important features in the image manifest themselves in the raw cross correlation data. In order to use the data intelligently one must define a data structure, as an intermediate step, before applying the imaging function.
\end{enumerate}
We next define the Doppler-scaled cross correlation function $C_{\mb R\mb R'}(s,\tau)$, and show how it can be used in the imaging problem. 

\subsection{Relative variation of the Doppler factor}
In general, \eqref{eq:gamma_t} shows that $\gamma_{\mb R}$ is dependent on the target's position and velocity. Having an accurate value for $\gamma_{\mb R}$ is extremely important in calculating cross correlations, as small changes in the frequency support would greatly affect the outcome of the correlation integral and hence the resolution of the image.
However, we can use prior information at our disposal to limit the image region. For example, $\mb v_\mb T$  can be estimated from the fact that the objects are assumed to be in a Keplerian orbit, i.e., $|\mb v_\mb T|\approx \sqrt{\frac{2G M_{\text{Earth}}}{\mb R_\mb T}}$. Accurate range measurements based on the signal's bandwidth can be used to estimate $\mb R_\mb T$ and subsequently $\mb v_\mb T$. 
Moreover, when forming an image we are usually interested in a limited search domain in space, based on auxiliary information. Of course $\gamma_{\mb R}$ is a continuous function of $\mb x_\mb T$, $\mb v_\mb T$, such that if $|\mb x_\mb T-\mb x_0|$, $|\mb v_\mb T-\mb v_0|$, are small enough then
\begin{equation}
\label{eq:gamma_var}
\frac{\gamma_{\mathbf{R}}(\mb x_\mathbf{T},\mb x_\mathbf{E},\mathbf{v_T})}{\gamma_{\mathbf{R}}(\mb x_\mathbf{0},\mb x_\mathbf{E},\mathbf{v}_0)}\approx 1.
\end{equation}

In our numerical simulations over an image window of size $10$m$\times 10$m$\times 10$m, $\gamma_\mb R$ varies at most by $0.00001$\%. For signals with our frequency support and time duration, $\gamma_\mb R$ can be very well approximated by a constant in the imaging domain.

\subsection{The cross-correlation data structure in iSAR}

Relying upon the weak variation of $\gamma_\mb R$,  we use a rescaled signal to image over the image window, rather than rescaling the signal for every point in the image window as was done in \cite{fournier2017matched}(eq. 3.6). 
Specifically, we construct $C_{\mathbf{RR}'}(s,\tau)$ by scaling the signal received by a reference Doppler factor,

\begin{equation}
\tilde{u}_{\mb R,\mathbf{x}_0,\mb v_0}(s,t)=u_\mathbf{R}\left(s,\frac{t}{\gamma_\mb R(\mb x_0,\mb x_\mb E,\mathbf{v_0})}\right).
\end{equation}
Note that for a reflector with a trajectory $\mb x_{\mb T}, \mb v_\mb T$, the received signal has the form
\begin{equation}
\tilde{u}_{\mb R,\mathbf{x}_0,\mb v_0}(s,t)=-\rho \frac{f''(s+\frac{\gamma_{\mathbf{R}}(\mb x_\mathbf{T},\mb x_\mathbf{E},\mathbf{v_T})}{\gamma_{\mathbf{R}}(\mb x_0,\mb x_\mathbf{E},\mathbf{v_0})} t-t_{\mathbf{R}})}{(4\pi|\mb x_{\mathbf{T}}-\mb x_\mathbf{R}|)^2}\approx -\rho \frac{f''(s+t-t_{\mathbf{R}})}{(4\pi|\mb x_{\mathbf{T}}-\mb x_\mathbf{R}|)^2}.
\label{eq:field_scaled}
\end{equation}

We define the cross correlation function as well,
\begin{equation}
\begin{split}
C_{\mathbf{RR}'}(s,\tau)=\int dt \tilde u_{\mb R,\mb x_0,\mb v_0}(s,t+t_\mb R(\mb x_0+s \mb v_0, \mb x_ \mb E, \mb v)) \tilde u_{\mb R',\mb x_0,\mb v_0}(s,t+t_{\mb R'}(\mb x_0+s \mb v_0, \mb x_ \mb E, \mb v)+\tau).
\end{split}
\label{eq:scaled_CC}
\end{equation}

Using \eqref{eq:field_scaled}, the imaging function \eqref{eq:CC_singlepoint} reduces to
\begin{equation}
\mathcal{I}^{CC}(\mb x, \mb v)=\sum\limits_{s,\mathbf{R},\mathbf{R}'}C_{\mathbf{RR}'}(s,\tau^s(\mb x,\mb v)),
\label{eq:single_point_CC}
\end{equation}
where 
\begin{equation}
\begin{split}
&\tau^s(\mb x,\mb v)=t_\mathbf{R'}^{\mb x(s)}-t_\mathbf{R}^{\mb x(s)}-\Delta\tau^s_{\mathbf{RR}'},\quad  \mb x (s)=\mb x+s\mb v,\\
&t_\mathbf{R}^{\mb x}(\mb x,\mb x_\mathbf{E},\mathbf{v})=\frac{|\mb x -\mb x_\mathbf{E}|}{c_0}+\frac{|\mb x -\mb x_\mathbf{R}|}{c_0}\gamma_{\mathbf{R}}(\mb x,\mb x_\mathbf{E},\mathbf{v}),\\
&\Delta \tau^s_\mathbf{RR'}=t_{\mb R'}(\mb x_0+s\mb v_0, \mb x_ \mb E, \mb v)-t_\mb R(\mb x_0+s \mb v_0, \mb x_ \mb E, \mb v).
\end{split}
\end{equation}
Notice that $t^\mb x_\mb R$ is still dependent on $\gamma_{\mathbf{R}}(\mb x,\mb x_\mb E,\mb v)$. 
The imaging function in \eqref{eq:single_point_CC} migrates the cross correlation data to a search point $\mb x$, based on the expected travel time difference between the signals recorded at both receivers $\mb R,\mb R'$. This is possible because we have adequately scaled the fields with respect to a specific search region, such that $C_{\mb R\mb R'}(s,\tau)$ is only dependent on travel time differences. Note also that the recorded signals are time shifted by the sum of the travel times from the source to the center of the image window plus the travel time from the center of the image window to the receiver. This ensures that  $C_{\mathbf{RR}'}(s,\tau)$ is supported around $\tau=0$.

The imaging function (\ref{eq:single_point_CC}) is an extension of the classical Kirchhoff Migration (KM) imaging function in the linear case, where the travel times are used to migrate the field
\begin{equation}
\mathcal{I}^{\text{KM}}(\mb x,\mb v)=\sum\limits_{s,\mathbf{R}}\tilde{u}_{\mb R,\mb x_0,\mb v_0}(s,t_\mathbf{R}^{\mb x}(\mb x,\mb x_\mathbf{E},\mathbf{v})).
\label{eq:KM_space}
\end{equation}
Thus, we can break down the imaging problem into two separate steps:
\begin{enumerate}
	\item In the first step, the trajectory and fast linear motion of the object is estimated and thereafter assumed known. The relevant imaging domain can then be taken to be small enough so that the approximation of the ratio (\ref{eq:gamma_var}) holds. We express the estimated motion of the center of the image window as  $\mb x_ \mb L (s),(\mb x_\mb L,\mb v_\mb L)$, where 
	\begin{equation}
	\label{eq:x_L}
	\mb x_\mb L(s)=\mb x_\mb L+s \mb v_\mb L.
	\end{equation} 
	Alternatively, we can use prior information about the trajectory to center the image window. 
	\item We then substitute in (\ref{eq:scaled_CC}) $\mb x_\mb L(s)$ for $\mb x_0$ and $\mb v_\mb L$ for $\mb v_0$ to compute the cross-correlation $C_{\mb R\mb R'}(s,\tau)$.
\end{enumerate}

\section{Generalized migration schemes for correlation based imaging}
\label{sec:generalized_migration}
The imaging function defined in \eqref{eq:CC_singlepoint} has been analyzed in \cite{fournier2017matched}, where its resolution was also compared to that of the matched filter imaging function. While Kirchhoff migration arises naturally in regular SAR/iSAR as the action of the adjoint forward operator applied to the recorded data, this is not necessarily a natural choice when the image is to be formed using correlations of the data.
Moreover, as mentioned Section~\ref{sec:intro}, unlike in classic KM the resolution only weakly depends on the synthetic aperture.
Numerical simulations with correlation based imaging suggest deterioration of the resolution when multiple reflectors are not well separated, or when there is a contrast in the reflectivity, with strong and weak reflectors in the same image window. The two-point migration introduced below in Section~\ref{sec:twopt_scheme} addresses these issues.

As a first step, we want to express the scaled cross correlation data structure $C_{\mb{RR'}}(s,\tau)$ in (\ref{eq:scaled_CC}) in terms of the medium reflectivity. We shall use that to introduce extensions of the migration imaging function to cross correlation data.

\subsection{The model for the cross correlation data}

Assume we have discretized the medium in a small image window relative to its moving center $\mb x _\mb L(s)$, as in (\ref{eq:x_L}), with grid points $\mb y_k$ and $\mb y_k=0$ corresponds to the center of the window. The unknown reflectivity is discretized by its values on this grid $$\rho_k=\rho(\mb y_k),\hspace{.1em}k=1,\dots,K.$$
The unknown reflectivity vector has dimension $K$, which is the number of pixels in the image window. Most of these reflectivities are zero because there are usually few relatively strong reflectors that can be imaged, that is, we can estimate their location in the image window and their strength.
We assume the image window is small enough so that the Doppler term $\gamma_{\mathbf{R}}$ is constant over it, as explained in Section~\ref{sec:corr_imaging}. 

We saw in \eqref{eq:field_scaled} that the signal recorded at receiver location $\mb x_\mb R$ can be written, after an appropriate scaling, as
\begin{equation}
\tilde{u}_{\mb R,\mathbf{x}_\mb L(s),\mb v_\mb L}(s,t)\approx-\sum\limits_{k=1}^K\rho_k \frac{f''(s+ t-(t^k_{\mathbf{R}}(s)-t_\mb R(s)))}{(4\pi|\mb x_{\mathbf{L}(s)}-\mb x_\mathbf{R}|)^2},
\end{equation}
where 
\begin{equation}
\begin{split}
t_\mb R^k(s) &=t_{\mathbf{R}}(\mb x_\mb L(s)+\mb y_k,\mb x_\mathbf{E},\mathbf{v})=\frac{|\mb x_\mathbf{L}(s)+\mb y_k-\mb x_\mathbf{E}|}{c_0}+\frac{|\mb x_\mathbf{L}(s)+\mb y_k-\mb x_\mathbf{R}|}{c_0}\gamma_{\mathbf{R}}(\mb x_\mathbf{L}(s)+\mb y_k,\mb x_\mathbf{E},\mathbf{v}),\\
t_\mb R(s) &=t_{\mathbf{R}}(\mb x_\mb L(s),\mb x_\mathbf{E},\mathbf{v}).
\end{split}
\end{equation}
In the frequency domain the recorded signal is
\begin{equation}
\label{eq:u_freq}
\hat{u}_{\mb R,\mathbf{x}_L(s),\mb v_L}(s,\omega)\approx\sum\limits_{k=1}^K\rho_k \omega^2 \frac{\hat f( \omega)}{(4\pi|\mb x_{\mathbf{L}}(s)-\mb x_\mathbf{R}|)^2}e^{i\omega (t_\mb R^k(s)-t_\mb R(s))}\equiv \sum\limits_{k=1}^K \frac{\omega^2 \hat{f} (\omega)}{(4\pi|\mb x_{\mathbf{L}}(s)-\mb x_\mathbf{R}|)^2}A_{\mb R,k}(s,\omega)\rho _k,
\end{equation}
with 
\begin{equation}
A_{\mb R,k}(s,\omega)=e^{i\omega (t_\mb R^k(s)-t_\mb R(s))}.
\end{equation}
The phase $A_{\mb R,k}(s,\omega)$ comes from the reduced travel time from the target to the receiver, relative to that of the image window center. 

We assume the distance from the reflectors to the different receivers doesn't vary greatly, hence we can approximate
\begin{equation}
\frac{\omega^2 \hat{f} (\omega)}{(4\pi|\mb x_{\mathbf{L}}(s)-\mb x_\mathbf{R}|)^2}\approx \xi(\omega,s),
\end{equation}
i.e, we neglect the dependence of the amplitude factor on a specific receiver. By using this approximation we compromise the accuracy in which the amplitude of the reflector is retrieved. However, for most applications, the support and the relative reflectivities of the scatterers are of the greatest importance.
Using this notation, and the fact that correlation in time is equivalent to multiplication in frequency, we get that
\begin{equation}
\begin{split}
\hat C_{\mb R\mb R'}(s,\omega)=&\hat u_{\mb R,\mb x_L(s),\mb v_L}(s,\omega)\overline{\hat u_{\mb R',\mb x_L(s),\mb v_L}(s,\omega)}=|\xi(\omega,s)|^2\sum\limits_{k,k'=1}^K A_{\mb R,k}(s,\omega)\overline{A_{\mb R',k'}(s,\omega)}\rho_k \rho_{k'}\\
&=|\xi(\omega,s)|^2\sum\limits_{k,k'=1}^K A_{\mb R,k}(s,\omega)\overline{A_{\mb R',k'}(s,\omega) }\rho_k \rho_{k'}.
\end{split}
\label{eq:C_form_element}
\end{equation}

\subsection{Matrix formulation of the forward model}
	
	We introduce matrix notation that relates our model for reflectivities to the cross correlation data. 
Denote by $\pmb \rho$ the unknown reflectivities in vector form $$\pmb \rho=[\rho_1,\dots,\rho _K]^T\in\mathbb{R}^K.$$
Denote $\mb A (s,\omega)$, our model for the migration matrix, has dimensions $N_R \times K$ and entries $A_{\mb R,k}(s,\omega)$. This matrix $\mb A(s,\omega)$ acts on the reflectivities and returns data.  
 Denote the recorded signal data as an $N_R$ vector vector $\hat{\mb u}_R(s,\omega)$ whose entries are given by (\ref{eq:u_freq}).
The cross correlation data is also a matrix, of dimension $N_R \times N_R$, $\hat {\mb C}(s,\omega)$ with entries 
$\hat C_{\mb R\mb R'}(s,\omega).$

Combining these,  we have in matrix form the following model for the recorded signal data vector $\hat{\mb u }_{\mb R}(s,\omega)$ and cross correlation data matrix $\hat{\mb C}(s,\omega)$
\begin{equation}
\hat{\mb u }_{\mb R}(s,\omega)=\mb A(s,\omega)\pmb \rho,
\label{eq:forward_field}
\end{equation}
\begin{equation}
\hat{\mb C}(s,\omega)=\hat{\mb u}_\mb R(s,\omega)\overline{\hat{\mb u}_\mb R(s,\omega)}^T=|\xi(\omega,s)|^2(\mb A (s,\omega)\pmb \rho)\overline{(\mb A (s,\omega)\pmb \rho)}^T=|\xi(\omega,s)|^2\mb A (s,\omega)\pmb \rho \pmb \rho ^T \overline{\mb A (s,\omega)}^T.
\label{eq:C_mat_exp}
\end{equation}
We denote by $\mb X=\pmb \rho \pmb \rho ^T, X_{kk'}=\rho _k \rho_{k'}$, the outer product of reflectivities, then our model in matrix form is
\begin{equation}
\hat{\mb C}(s,\omega)=|\xi(\omega,s)|^2\mb A (s,\omega) \mb X \overline{\mb A(s,\omega)}^T.
\label{eq:C_form}
\end{equation}
The cross correlations depend on the reflectivities $\pmb \rho$ through their outer product $\mb X=\pmb \rho \pmb \rho^T$. As a result there can be several different extensions of Kirchhoff migration to cross correlations, which we investigate in the next section. 

\subsection{Imaging functions for cross correlation data}
\label{sec:twopt_scheme}
Given the data $\hat{\mb u}_{\mb R}(s,\omega)$ and the model $\hat{\mb u}_{\mb R}(s,\omega)=\mb A(s,\omega )\pmb \rho$, Kirchhoff migration of the data $\hat{\mb u}_\mb R(s,\omega) $ is given by
\begin{equation}
\tilde{\pmb{\rho}}=\sum\limits_{s,\omega}\overline{\mb A(s,\omega)}^T\hat{\mb u}_{\mb R}(s,\omega).
\label{eq:KM_freq}
\end{equation}
Identifying $\tilde{\pmb\rho}_k=\tilde{\rho}(\mb y_k)$, \eqref{eq:KM_freq} is a discretized version of the imaging function in \eqref{eq:KM_space}, $\mathcal{I}^{KM}(\mb y_k,\mb v_\mb T)$.

Kirchhoff migration is usually derived by considering the solution of the ordinary least square (OLS) optimization problem. While the solution of OLS involves the pseudo-inverse of the forward matrix, it is common to approximate it as diagonal for imaging problems, which is well justified for separated targets. In that case, OLS and Kirchhoff migration yield the same support. Kirchhoff migration can also be interpreted as the solution of a different constrained optimization problem, related to the forward models of \eqref{eq:forward_field}. Neglecting the scales $\xi(\omega,s)$, we can interpret migration as looking for the set of reflectivities that yield the set of signals most correlated with our measurements, under a norm constraint.
\begin{equation}
	\begin{split}
		\tilde{\pmb \rho}=\arg\max\limits_{\|\pmb \varrho\|_2=c}\sum\limits_{s,\omega}\langle \mb A(s,\omega) \pmb \varrho, \hat{\mb u}_\mb R \rangle =\arg\max\limits_{\|\pmb \varrho\|_2=c}\langle  \pmb \varrho, \sum\limits_{s,\omega} \overline{\mb A(s,\omega)} ^T\hat{\mb u}_\mb R \rangle \Rightarrow  \hat{\pmb \rho}\propto\sum\limits_{s,\omega}\overline{\mb A(s,\omega)}^T\hat{\mb u}_\mb R(s,\omega).
	\end{split}
	\label{eq:km_opt}
\end{equation}

While the solution of OLS for the cross correlation model \eqref{eq:C_form} involves writing $\hat{\mb C}(s,\omega)$ as a vector and the matrix multiplication operations as matrix-vector multiplication, \eqref{eq:km_opt} has a direct extension for the model of\eqref{eq:C_form}. With abuse of notation for matrices $A,B$ $\langle A,B\rangle =\text{tr}(B^TA)$. Neglecting the scales $\xi(\omega,s)$ again, we have
\begin{equation}
\begin{split}
\tilde{\mb X}&=\arg\max\limits_{\|\mb X\|_2=c}\sum\limits_{s,\omega}\langle \mb A(s,\omega) \mb X\overline{\mb A(s,\omega)}^T, \hat{\mb C}_{\mb R\mb R'}(s,\omega) \rangle =\sum\limits_{s,\omega}\langle  \mb X, \overline{\mb A(s,\omega)}^T\hat{\mb C}_{\mb R\mb R'}(s,\omega) \mb A(s,\omega)\rangle \\&
=\langle  \mb X, \sum\limits_{s,\omega}\overline{\mb A(s,\omega)}^T\hat{\mb C}_{\mb R\mb R'}(s,\omega) \mb A(s,\omega)\rangle \Rightarrow  \tilde{\mb X}\propto\sum\limits_{s,\omega}\overline{\mb A(s,\omega)}^T\hat{\mb C}(s,\omega)\mb A (s,\omega). 
\end{split}
\label{eq:X_opt}
\end{equation}

As a result, a natural extension of the migration of \eqref{eq:KM_freq}  for migrating cross correlations is the matrix $\tilde{\mb X}$
\begin{equation}
\tilde{\mb X}=\sum\limits_{s,\omega}\overline{\mb A(s,\omega)}^T\hat{\mb C}(s,\omega)\mb A (s,\omega) ,
\end{equation}
with elements
\begin{equation}
\tilde{X}_{kk'}=\sum\limits_{s,\omega,\mb R,\mb R'}\overline{ A}_{\mb R,k}(s,\omega)\hat{C}_{\mb R,\mb R'}(s,\omega)A_{\mb R',k'} (s,\omega).
\label{eq:M_tilde}
\end{equation}

As expected, the result of the migration is an estimation of $\mb {X}$ rather than $\pmb \rho$, as our model is quadratic with respect to the reflectivities.
$\tilde{\mb X}$ is a square matrix with dimensions $K\times K$, where $K$ is the number of search points in the imaging domain. If points $\mb y_k,\mb y_{k '}$ are associated with reflectivities $\rho_k, \rho_{k'}$, we can think about $\tilde{\mb X}$ as a two-variable generalized cross correlation imaging function
\begin{equation}
\mathcal{I}^{GCC}(\mb y_k,\mb y_{k'})=\tilde{X}_{kk'}.
\label{eq:2_point_mat_imag}
\end{equation}
Note that $\tilde{\mb X}\in \mathbb{C}^{K\times K}$  is Hermitian positive definite by definition. We next see how can an image be extracted from $\mathcal{I}^{GCC}$.
\subsection{Imaging with migrated cross correlation data}
The imaging function \eqref{eq:2_point_mat_imag} lacks a direct physical interpretation. It evaluates the outer product of reflectivities rather than the reflecitivities themselves. We shall see that several ways to extract $\pmb \rho$ exist, not all equal in general. 

If $\tilde{\mb X}$ was exactly rank-1, i.e. $\tilde{\mb X}=\pmb \rho \pmb \rho^T$, there would be several ways to extract $\pmb \rho$. 

\begin{figure}[htbp]
	\centering
	\begin{subfigure}[t]{0.29\textwidth}
		\includegraphics[width=\textwidth]{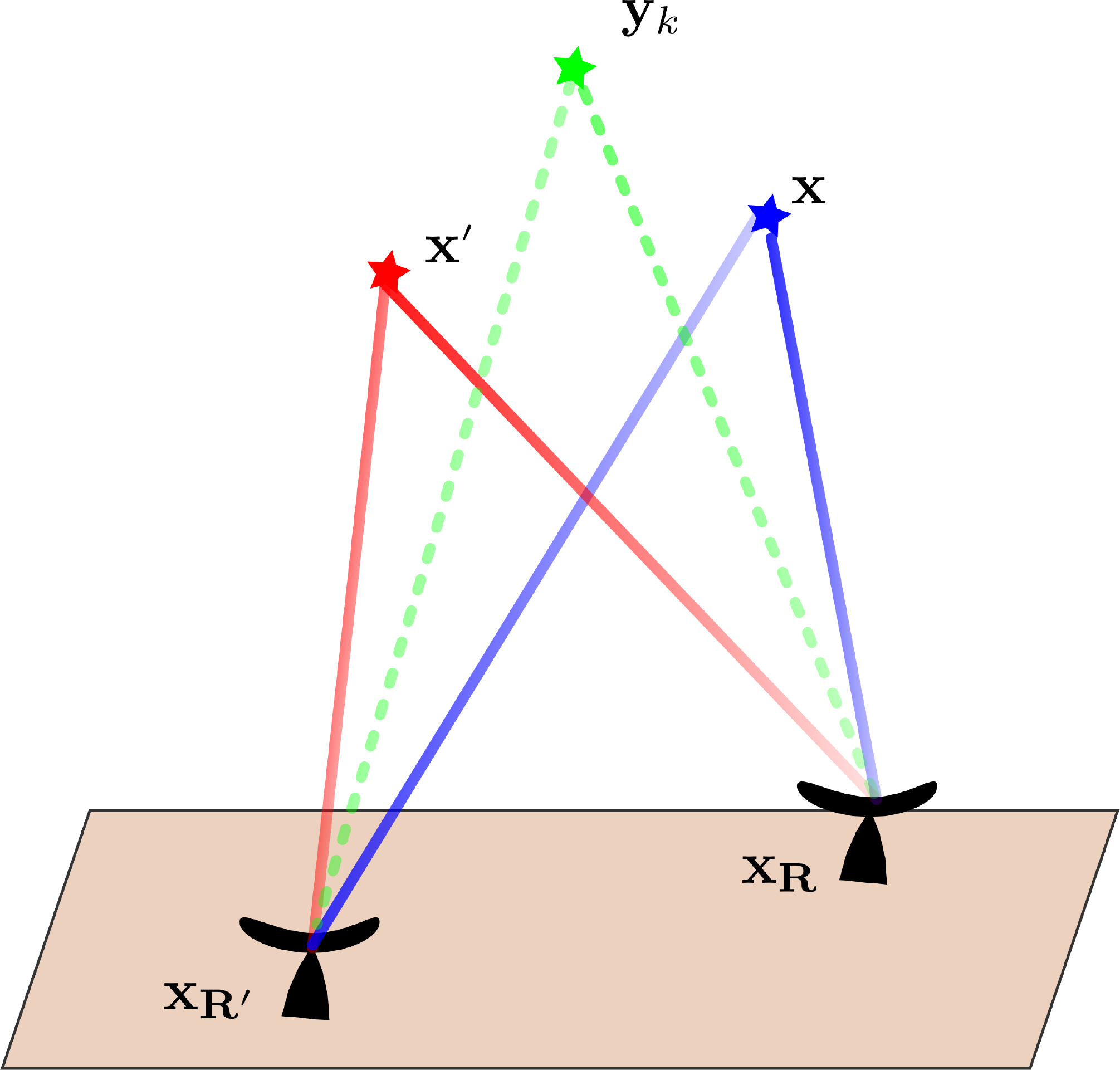}
		\caption{ }
		\label{fig:single_point_migration}	\end{subfigure}
	\begin{subfigure}[t]{0.29\textwidth}
		\includegraphics[width=1\textwidth]{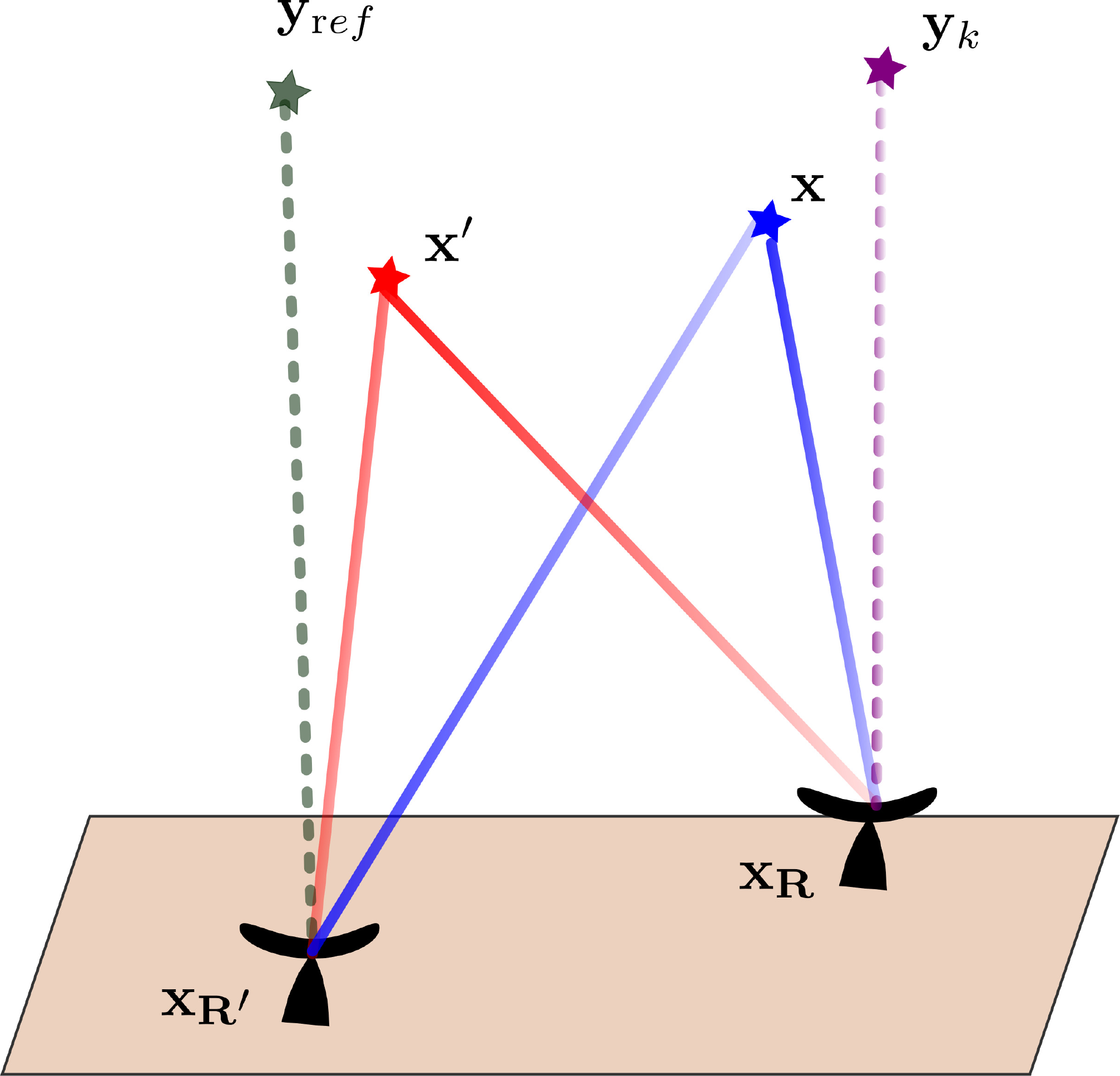}
		\caption{ }
		\label{fig:reference_point_migration}
	\end{subfigure}
	\begin{subfigure}[t]{0.29\textwidth}
		\includegraphics[width=1\textwidth]{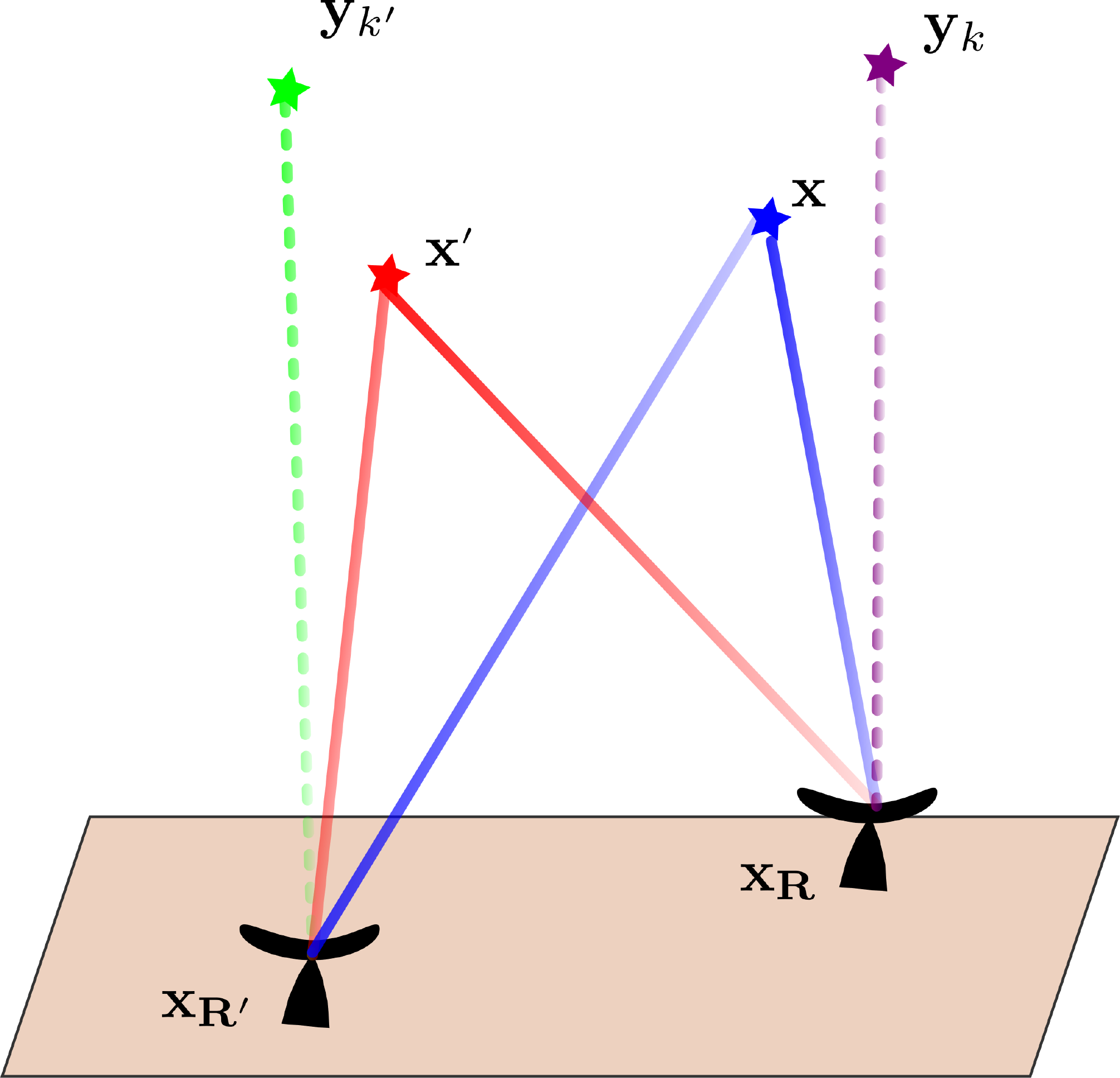}
		\caption{ }
		\label{fig:two_point_migration}
	\end{subfigure}
	\caption{Illustration of different migration schemes. The cross correlated data of receiver pairs contain both signals generated by the same reflector and signals that were generate by different reflectors. The migration scheme determines how the data is to be treated.  $(a)$ Single point migration: Only the signals that arrive at both receivers from the same reflector would be summed coherently. The result of the migration is $\mathcal{I}^{GCC}(\mb y_k,\mb y_k)=\mathcal{I}^{CC}(\mb y_k)$; $(b)$ Reference point migration: One of the receiver pairs is always migrated to the same point- if a strong reflector exists there, one could expect strong signals when migrating with respect to the other receiver. The result of the migration is $\mathcal{I}^{GCC}(\mb y_k,\mb y_{\text ref})=\mathcal{I}^{\text{ref}CC}(\mb y_k)$; $(c)$ Two point migration: Signals from all possible reflector pairs would  summed coherently. The result of the migration is $\mathcal{I}^{GCC}(\mb y_k,\mb y_{k'} )$. The rank-1 image $\mathcal{I}^{R1CC}(\mb y_k)$ is extracted through eigendecomposition.}
	\label{fig:migration_schemes}
\end{figure}

\begin{enumerate}
	\item $|\rho_k|^2=\tilde{X}_{kk}$. This is equivalent to \eqref{eq:single_point_CC}, since the diagonal terms recreate the image generate by migrating the data to the same point $\mb y_{k}=\mb y_{k'}$
	\begin{equation}
	\begin{split}
	&\sum\limits_{s,\omega,\mb R,\mb R'}\overline{ A}_{\mb R,k}(s,\omega)\hat{C}_{\mb R,\mb R'}(s,\omega)A_{\mb R',k'} (s,\omega)=\sum\limits_{s,\mb R,\mb R',\omega}\hat{\mb C}_{\mb R\mb R'}(s,\omega)e^{i\omega (t_\mb R^k(s)-t_\mb R(s))}e^{-i\omega (t_{\mb R'}^k(s)- t_{\mb R'}(s))}\\&\propto  \sum\limits_{s,\mb R,\mb R'}\mb C_{\mb R\mb R'}(s, t_\mb R^k(s)-t_\mb R(s)-(t_{\mb R'}^k(s)- t_{\mb R'}(s))).
	\end{split}
	\end{equation}
	In terms of \eqref{eq:2_point_mat_imag} the image is evaluated by plugging in the same search point in both variables
	$$\mathcal{I}^{CC}(\mb y_k)=\mathcal{I}^{GCC}(\mb y_k,\mb y_k).$$
	\item $|\rho_k|^2\propto \tilde{X}_{kk_{\text ref}}, k_{\text ref}\in[1,\dots,K]$. This is equivalent to migrating the data to a reference point with respect to one of the receiver pairs and forming an image by migrating the other receiver to a search point. In terms of \eqref{eq:2_point_mat_imag} the image is evaluated by plugging a reference term in the first variable  and a search point in the other
	$$\mathcal{I}^{{\text ref}CC}(\mb y_k)=\mathcal{I}^{GCC}(\mb y_k,\mb y_{\text{ref}}).$$
	\item $|\rho_k|^2=|\mb v_1(\tilde{\mb X})|^2_k$, i.e., calculate the top eigenvector of $\tilde{\mb X}$.  In terms of \eqref{eq:2_point_mat_imag} the image is evaluated by taking, $\mathcal{V}(\mb y_k)$, the first eigenvector of $\mathcal{I}^{GCC}(\mb y_k,\mb y_{k'})$, thought of as a matrix
	$$\mathcal{I}^{R1CC}(\mb y_k)=\mathcal V(\mb y_k).$$
	We call this the \textbf{rank-1 image}.
\end{enumerate}
The different methods are illustrated in Figure~\ref{fig:migration_schemes}.

All of these methods reconstruct the reflectivity up to a global phase as expected. The first two methods have the advantage that they are only dependent on a small subdomain of the elements of $\tilde{\mb X}$. However, as mentioned in Section~\ref{sec:corr_imaging}, they suffer from limitations in the achieved image resolution. The eigenvector on the other hand is a \textbf{global} method, relying on the entire cross correlation data to reconstruct the image. 

A physical interpretation would be the following- a cluster of scatterers moving at the same speed would exhibit coherent correlations not only when correlating reflections from the same scatterer at different receivers, but also when  correlating reflections from different scatterers at different receivers. Thus,  the entire data can be used to estimate the most probable locations for the scatterers, given that the fields measured at both receivers were generated by the same set of scatterers. Note that single point migration can also not resolve any of the relative phases of the reflectors, while the eigenvector is only invariant with respect to a global phase.

The single point migration and two point migration are related by
\begin{equation}
\tilde{\mb X}=\mb V \mb \Lambda \overline{\mb V}^T \Rightarrow \tilde X_{kk}=\sum\limits_{i=1}^r\lambda_i |\mb v_i(\tilde{\mb X})\hspace{0.01em}_k|^2,
\label{eq:eig_to_diag}
\end{equation}
with $r$ being the rank of $\tilde{\mb X}$. i.e., the single point migration is a weighted sum of all the eigenvectors, squared, by their respective eigenvalues. When there is a rapid decay in the singular values, we expect similar results for both migration methods. However, when this is not the case, we expect a different result, as we show through analysis and numerical simulations in Section~\ref{sec:prop_filt}. 

\subsection{The rank-1 imaging function}
\label{subsec:rank_1_image}
Based on the observations of the previous section we suggest the following imaging function
\begin{enumerate}
	\item Backpropagate the cross correlation $\hat{\mb C}(s,\omega)$ to two different points using the operator $\mb A (s,\omega)$ to form the two-point image $\tilde{\mb X}$
	$$  \tilde{\mb X}=\sum\limits_{s,\omega}\overline{\mb A(s,\omega)}^T \hat{\mb C}(s,\omega) \mb A(s,\omega).$$
	\item Take the rank-1 image to be the top eigenvector of $\tilde{\mb X}$
	$$\mathcal I^{R1CC} (\mb x)=\mb v_1(\tilde{\mb X})$$
\end{enumerate}
We can reduce the cost of the algorithm by considering a rectangular $\tilde{\mb X}$ rather than a square matrix. This means that when migrating we don't use the same grid for both receivers. We replace our candidate image with the top left singular vector (for a tall matrix). As we show in Section~\ref{sec:numerical_simulations}, numerical simulations suggest that we can downsample one dimension by a factor of 10 and still retain the same resolution.
There are several motivations to consider the rank-1 image:
\begin{enumerate}
	\item \textbf{Robustness to noise}. Since the eigendecomposition uses the entire cross correlation data, it is less sensitive to additive noise or even incomplete data. One could use low-rank matrix completion algorithms to reconstruct the image.  
\item \textbf{Optimization interpretation}. The top singular vector has an optimization interpretation:
plug in the expression for $\hat{\mb C}$ in \eqref{eq:C_mat_exp}, then
\begin{equation}
\tilde{\mb X}= \sum\limits_{s,\omega}\overline{\mb A(s,\omega)}^T\mb A (s,\omega) \pmb \rho \bar{\pmb \rho} ^T \overline{\mb A(s,\omega)}^T\mb A (s,\omega) .
\end{equation}
The top eigenvector  can be expressed as the solution to an optimization problem. 
The top eigenvector solves:
\begin{equation}
\begin{split}
\mb v _1(\tilde{\mb {X}}) =&\arg\max\limits_{\|\pmb \varrho\|_2=1}\bar{\pmb \varrho}^T\tilde{\mb X}\pmb \varrho= \arg\max\limits_{\|\pmb \varrho\|_2=1}\sum\limits_{s,\omega}\bar{\pmb \varrho}^T\overline{\mb A(s,\omega)}^T\mb A (s,\omega) \pmb \rho \bar{\pmb \rho} ^T \overline{\mb A(s,\omega)}^T\mb A (s,\omega) \pmb \varrho\\
=&\arg\max\limits_{\|\pmb \varrho\|_2=1}\sum\limits_{s,\omega}|\langle \mb A(s,\omega) \pmb \varrho, \mb A(s,\omega) \pmb \rho \rangle |^2
=\arg\max\limits_{\|\pmb \varrho\|_2=1}\sum\limits_{s,\omega}|\langle \mb A(s,\omega) \pmb \varrho, \hat{\mb u}_\mb R (s,\omega)\rangle |^2
\end{split}
\label{eq:svd_opt}
\end{equation}
i.e, we are trying to find the $\pmb \varrho$ which is the most correlated with the observed field measurement, given our forward model. Notice that while similar to \eqref{eq:km_opt}, this is a nonlinear optimization problem, and the absolute value in \eqref{eq:svd_opt} removes the sensitivity to a global phase.
\end{enumerate}

\subsection{Performance of the rank-1 image} \label{sec:perf}
In Sections~\ref{sec:numerical_simulations} and \ref{sec:prop_filt} we compare through simulations the performance of the two-point and single-point migration schemes for plane images, that is, images with a fixed $z$ coordinate (height), so that the image coordinate $\mb y_k\in \mathbb{R}^2$. We show that as the synthetic aperture increases, the resolution of the rank-1 image tends to improve compared to the usual single point migration image.
We explain this by analyzing the stationary points of the interference pattern $\tilde{\mb X}$, in Appendix~\ref{app:stat_phase}. 

The main result in this appendix is that the synthetic aperture induces an anisotropy in the resolution of $\tilde{\mb X}$ in the space $\mb y_k\times \mb y_{k'}\in \mathbb{R}^4$. For a small synthetic aperture, the width of the stationary point is approximately $\frac{\lambda H_\mb T}{a}$ in all directions, where $\lambda=c_0/f_0$ is the wavelength of the carrier frequency, $H_\mb T$ is the height of the target (the average distance from the receivers), and $a$ is the diameter of the imaging array spanned by the receivers. This is consistent with the resolution analysis in \cite{fournier2017matched}, where it was evaluated for a relatively short synthetic aperture, so that the resolution is dominated by the size of the array of receivers. 

\begin{figure}
	\centering
	\begin{subfigure}[t]{0.3\textwidth}
	\includegraphics[width=1\textwidth]{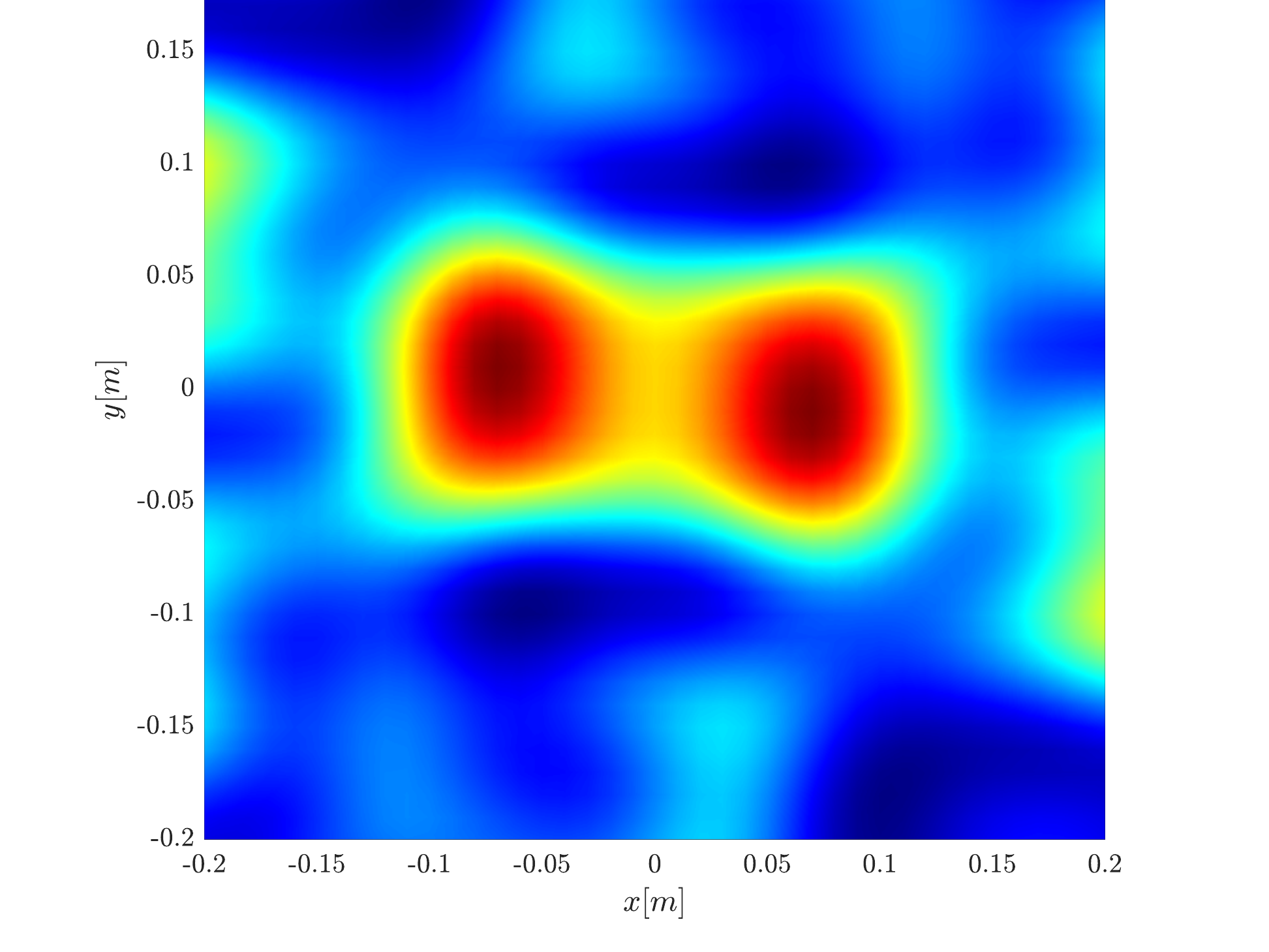}
	\caption{}
	\end{subfigure}
	\begin{subfigure}[t]{0.3\textwidth}
	\includegraphics[width=1\textwidth]{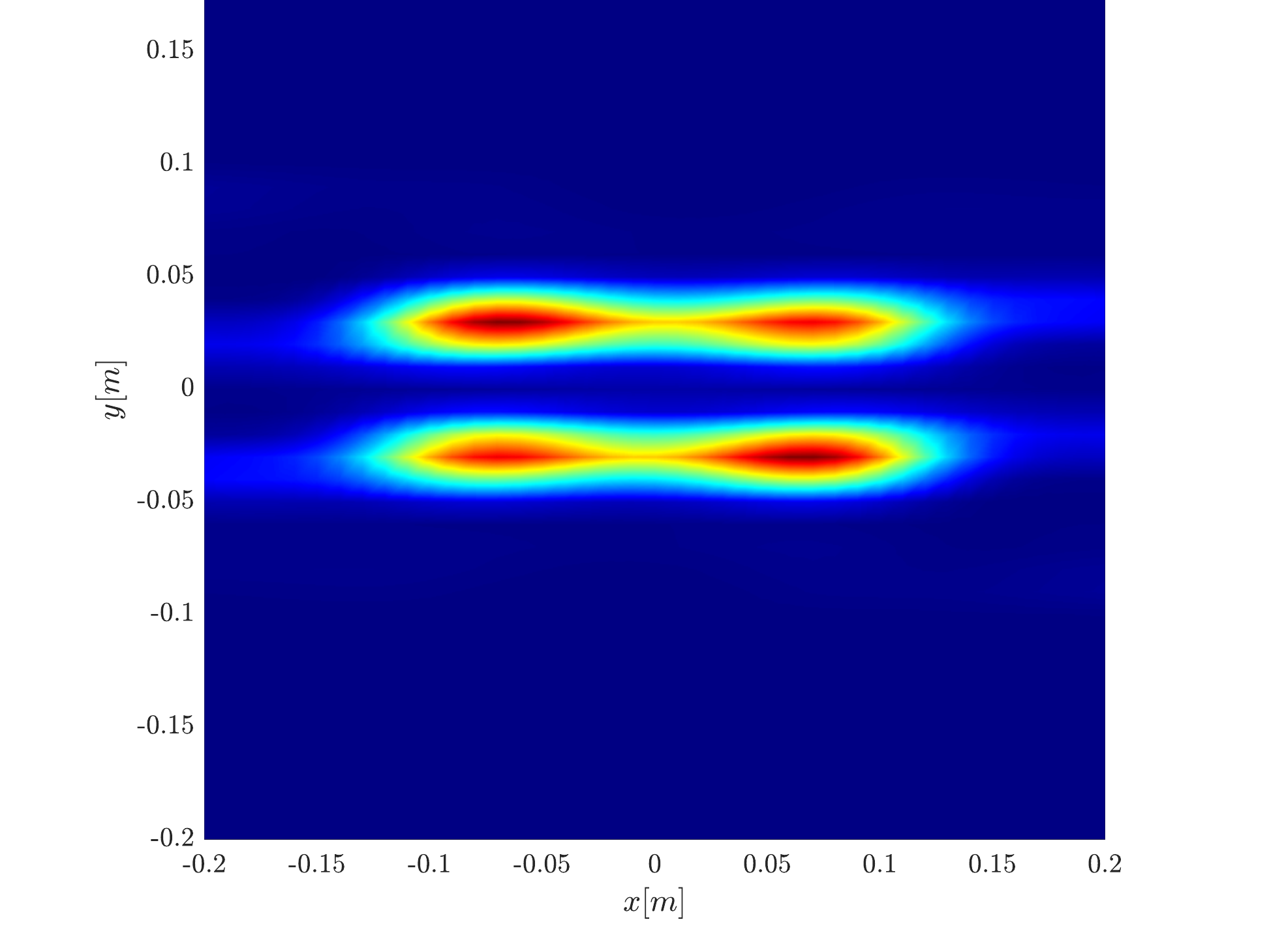}
	\caption{}
\end{subfigure}
	\begin{subfigure}[t]{0.3\textwidth}
	\includegraphics[width=1\textwidth]{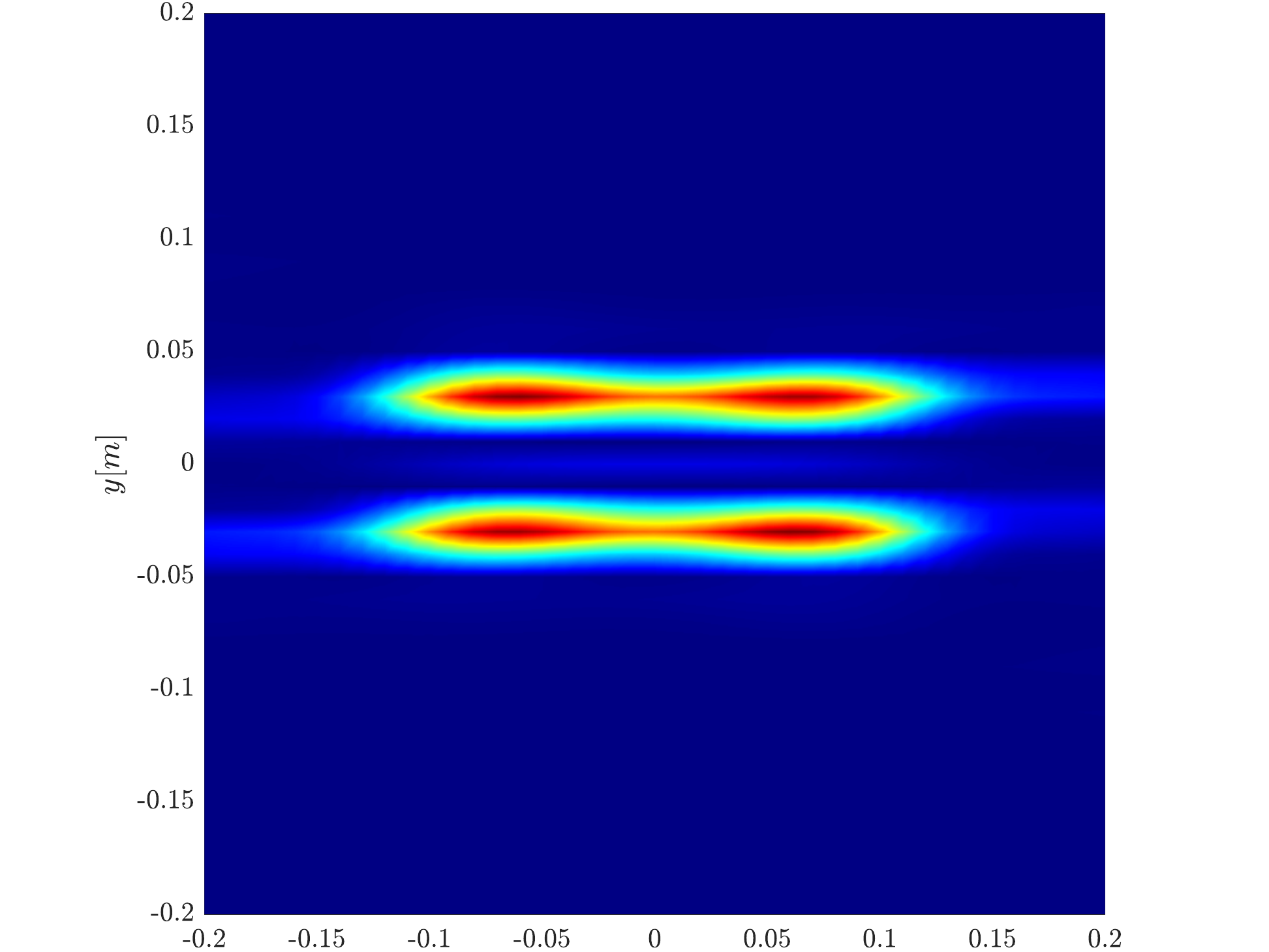}
	\caption{}
\end{subfigure}

	\begin{subfigure}[t]{0.65\textwidth}
	\centering
	\includegraphics[width=\textwidth]{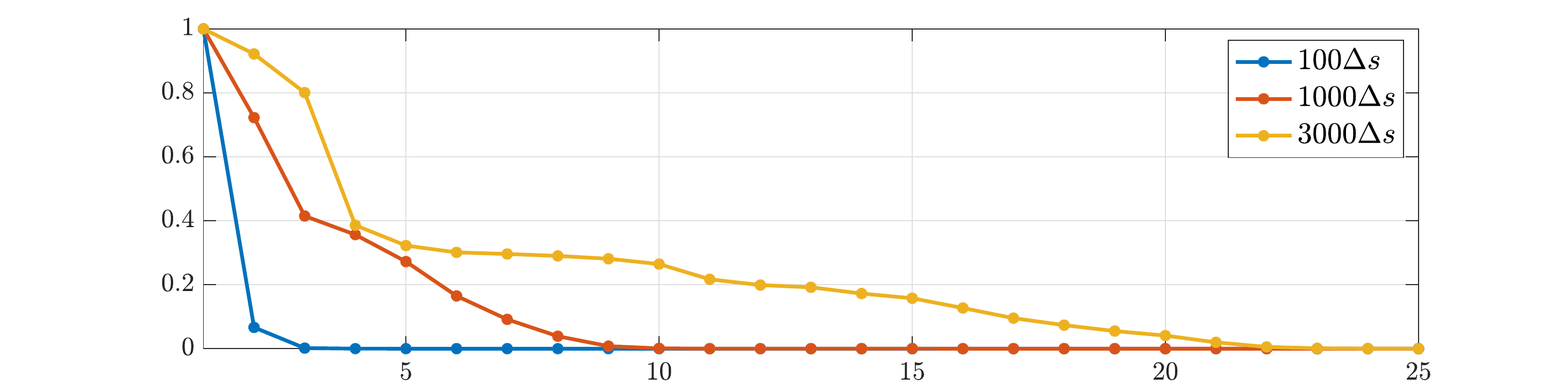}
	\caption{}
	\end{subfigure}
\caption{Imaging a cluster composed of four scatterers. $(a)$ Single-point migration, $(b)$ rank-1 image, $(c)$ Linear KM. The synthetic aperture is $3000\Delta s$. We can see that there is a dramatic improvement in the resolution attained by the rank-1 image, which is comparable to the one obtained by KM  $(d)$ Top 25 eigenvalues of $\tilde{\mb X}$ for different synthetic aperture sizes, normalized such that the top eigenvalue is always 1. We can see that, as the synthetic aperture size increases, so does the number of significant eigenvalues.}
\label{fig:num_exam}
\end{figure}

However, as the synthetic aperture grows larger and becomes comparable in size to the physical aperture, the anisotropy becomes significant. As a result, the stationary points have different widths in different directions. They exhibit the narrowest spot size in the direction $\hat{\mb y}_k\cdot \mb v_\mb T=-\hat{\mb y}_{k'}\cdot \mb v_\mb T$, which is approximately $\frac{\lambda H_\mb T}{\mb v_\mb T S}$, with $\mb v_\mb TS$ the size of the synthetic aperture. The width is unchanged in $\hat{\mb y}_k=\hat{\mb y}_{k'}$, which is the direction corresponding to the diagonal/single point migration \eqref{eq:single_point_CC}. This is again consistent with the resolution analysis of \cite{fournier2017matched}, which suggests that the resolution of planar images is independent of the synthetic aperture size. Anisotropic localized kernels have eigenvectors whose width is smaller then the maximum width, as we show in Appendix~\ref{app:kernel} for several one dimensional cases. In the case of the point interference patters the width tends to the harmonic mean, which greaty improves on the maximal width. Thus, the top eigenvector $\mathcal{I}^{R1CC}(\mb y_k)$ can provide an image with better resolution. This is indeed the case, as illustrated in Figure~\ref{fig:num_exam}. In numerical simulations, shown in Section~\ref{sec:numerical_simulations}, the resolution is comparable to the linear KM image.

In Section~\ref{sec:numerical_simulations} we present extensive numerical simulations that confirm the superior performance of the rank-1 image with an increasing synthetic aperture size. Then, in Section~\ref{sec:prop_filt} we further investigate the performance by considering the case of a single target. The numerical results are in accordance with the analysis of Appendix~\ref{app:stat_phase} and Appendix~\ref{app:kernel}. 

\section{Numerical simulations}
\label{sec:numerical_simulations}

In this section we compare with numerical simulations the performance of the single point migration, the rank-1 image and the linear Kirchhoff migration.  The object to be imaged is a cluster composed of either two or four point scatterers. The scatterers are at $z=500$km height and they all move with the same linear speed. In the inertial frame of references the scatterers are in the $x-y$ plane. The distance between the scatterers is $11$cm in the $x$ direction, and $6$cm in the $y$ direction. We show reconstructions in the $x-y$ plane at a fixed height $z$ equal to the true height of the scatterers. 



We compute imaging results for different synthetic aperture sizes corresponding to 100 pulses (1.5 seconds), 1000 pulses (15 seconds) and 3000 pulses (45 seconds). The calculations were performed in the frequency domain using as sampling frequencies $\omega_i=\omega_0 + i B/30$, $i=-90,89,\dots,0,\ldots 89,90$ with $\omega_0=2\pi f_0$, $f_0=9.6$GHz and $B=300$MHz corresponding to half the bandwidth of the pulse $f(t)$.

In the following figures we demonstrate the improvement in resolution and target separation when using the rank-1 image, compared to the classical single-point migration. The observed performance is further investigated in Section~\ref{sec:prop_filt}, and given an analytical explanation in Appendix~\ref{app:stat_phase}.

\begin{figure}[htbp]
	\centering
	\begin{subfigure}[t]{0.18\textwidth}
		\includegraphics[width=\textwidth]{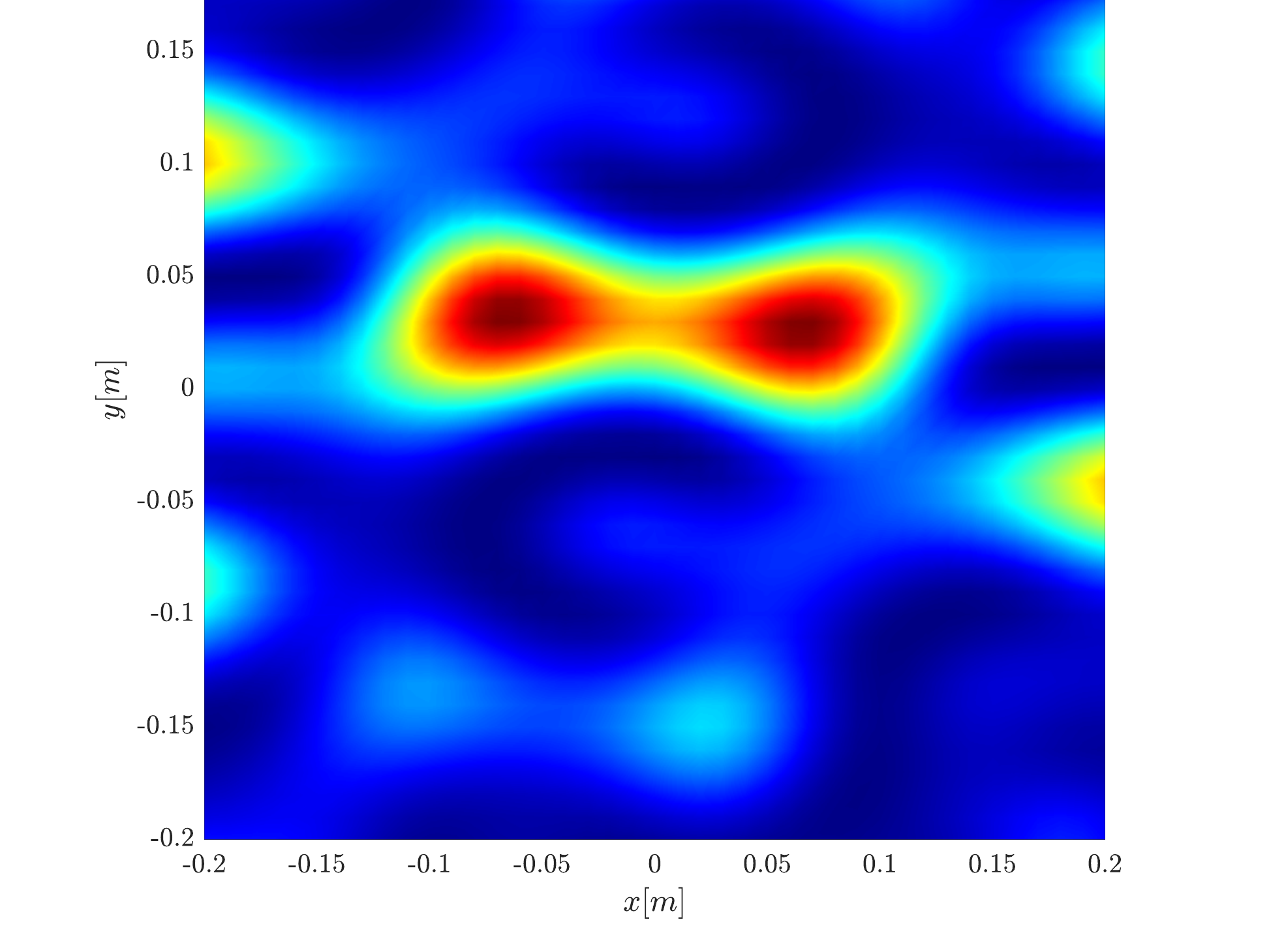}
	\end{subfigure}
	\begin{subfigure}[t]{0.18\textwidth}
		\includegraphics[width=\textwidth]{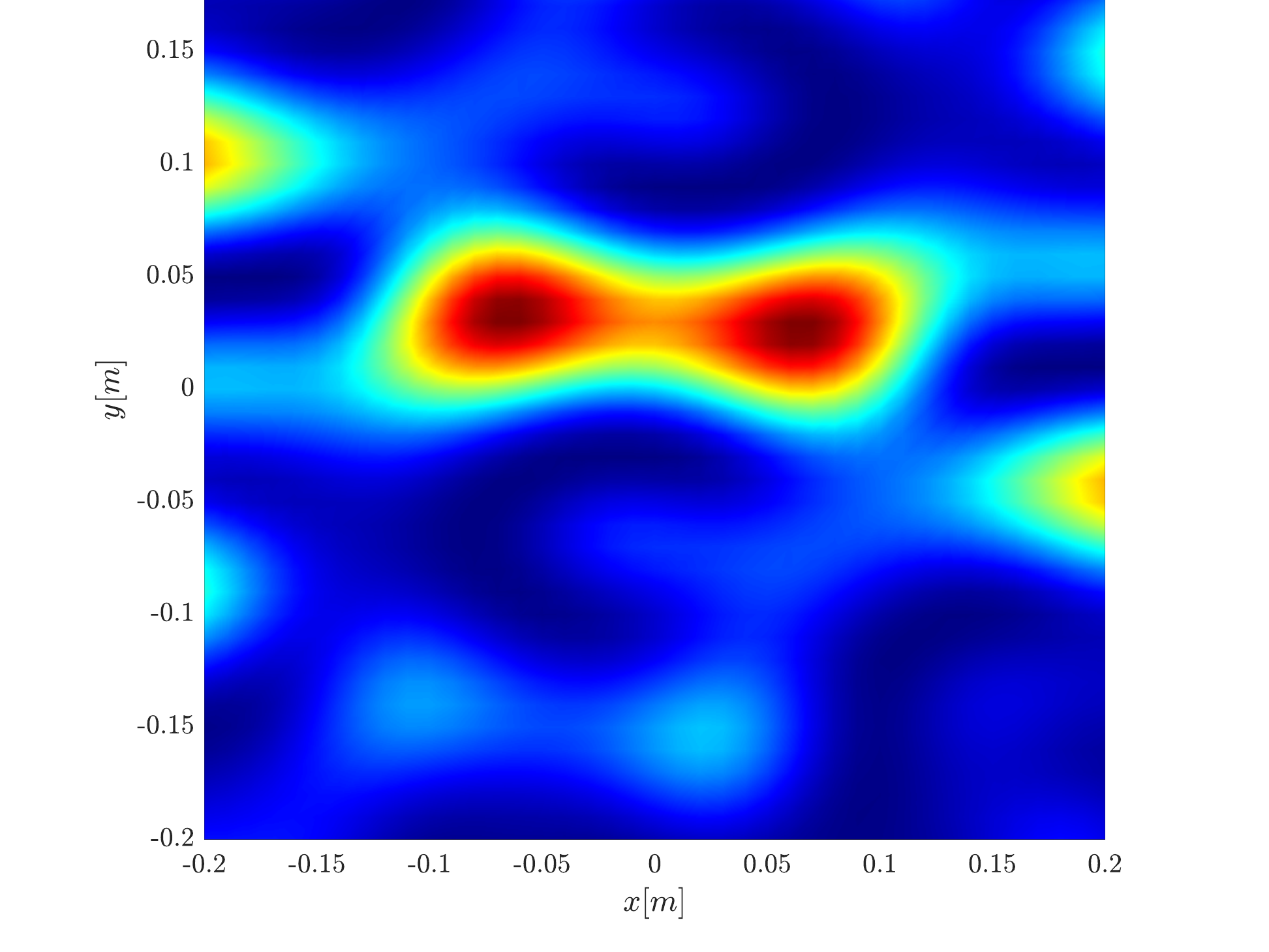}
	\end{subfigure}
	\begin{subfigure}[t]{0.18\textwidth}
		\includegraphics[width=\textwidth]{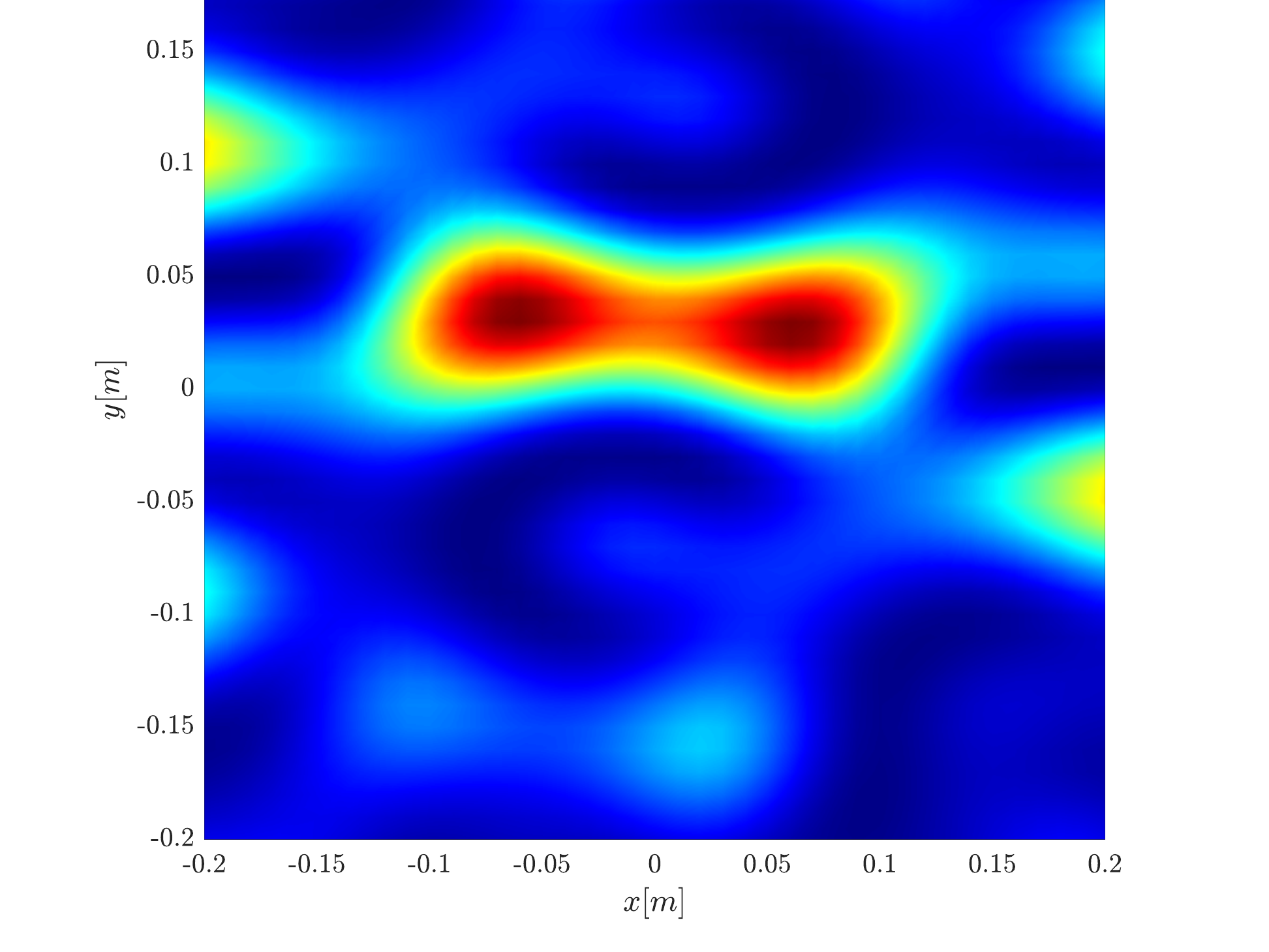}
	\end{subfigure}
	
	\begin{subfigure}[t]{0.18\textwidth}
		\includegraphics[width=\textwidth]{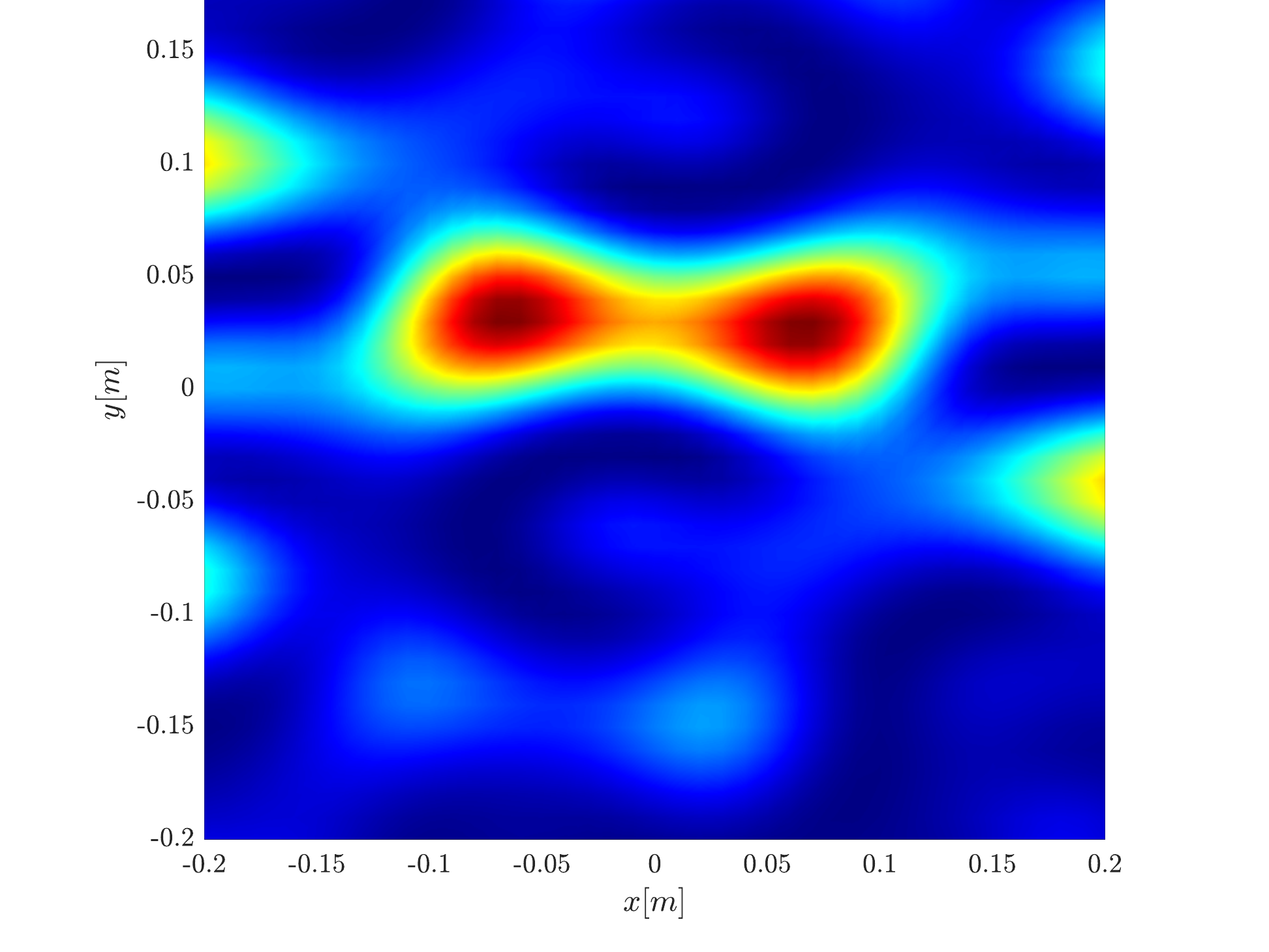}
	\end{subfigure}
	\begin{subfigure}[t]{0.18\textwidth}
		\includegraphics[width=\textwidth]{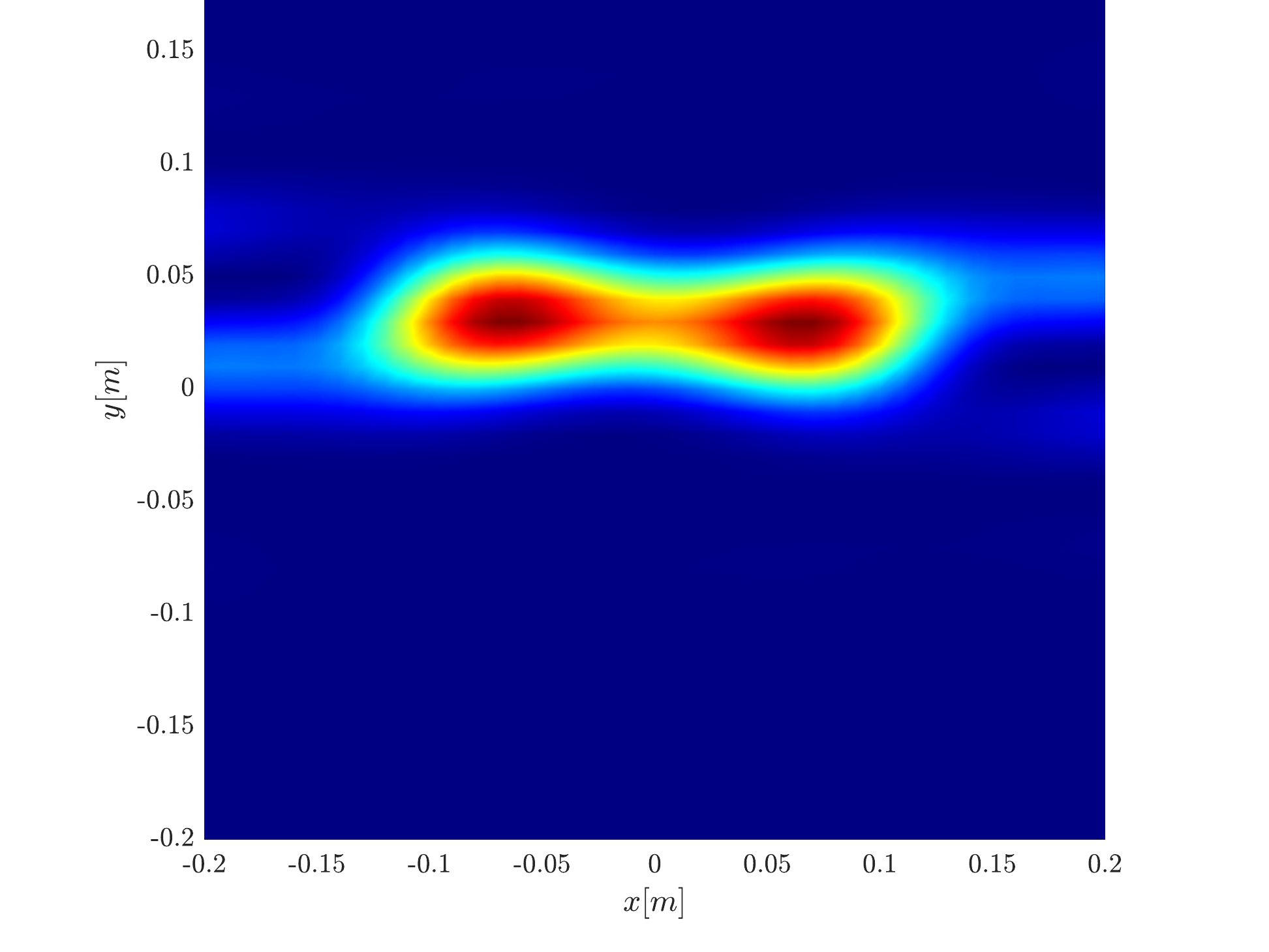}
	\end{subfigure}
	\begin{subfigure}[t]{0.18\textwidth}
		\includegraphics[width=\textwidth]{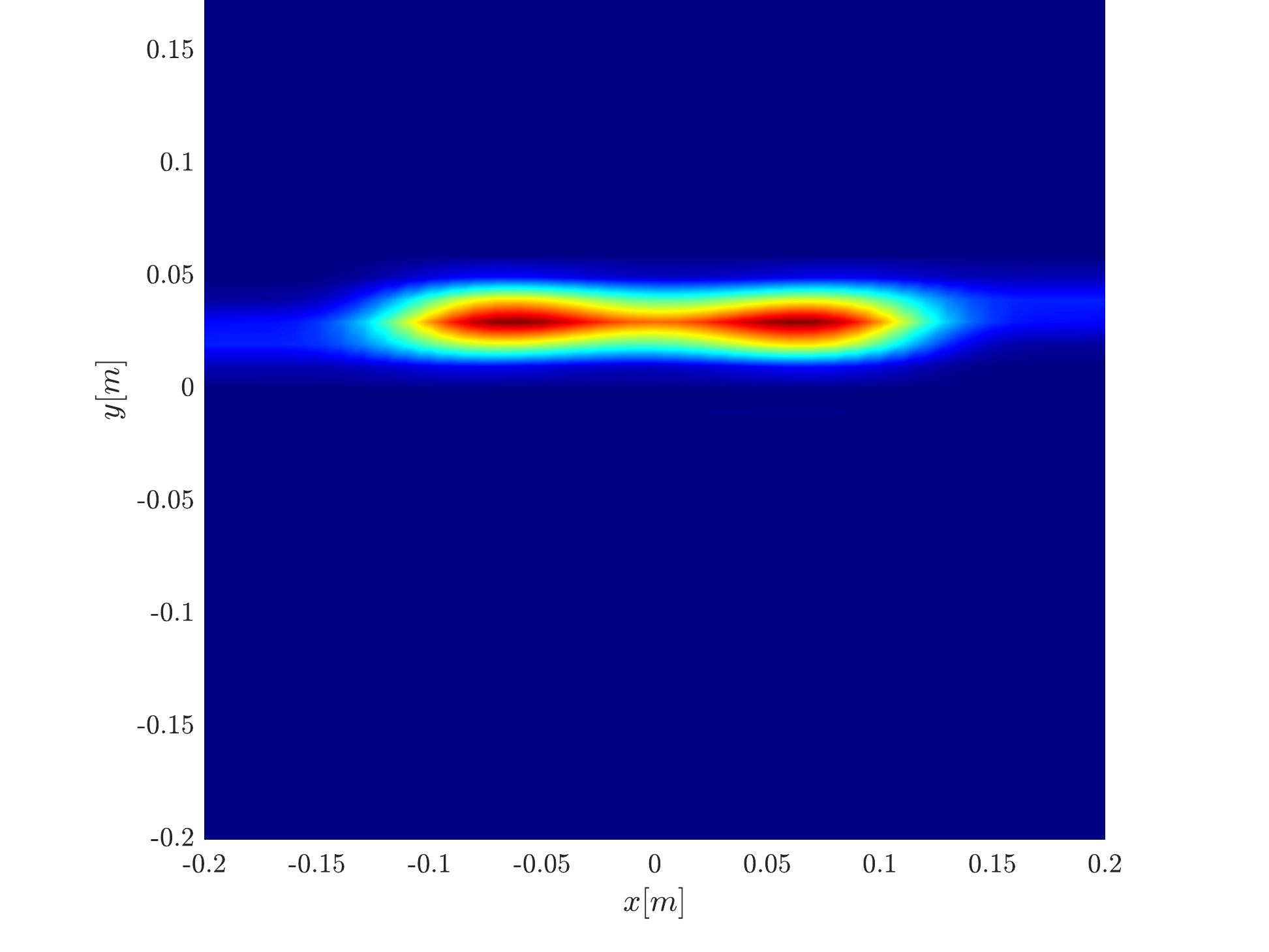}
	\end{subfigure}

\begin{subfigure}[t]{0.18\textwidth}
	\includegraphics[width=\textwidth]{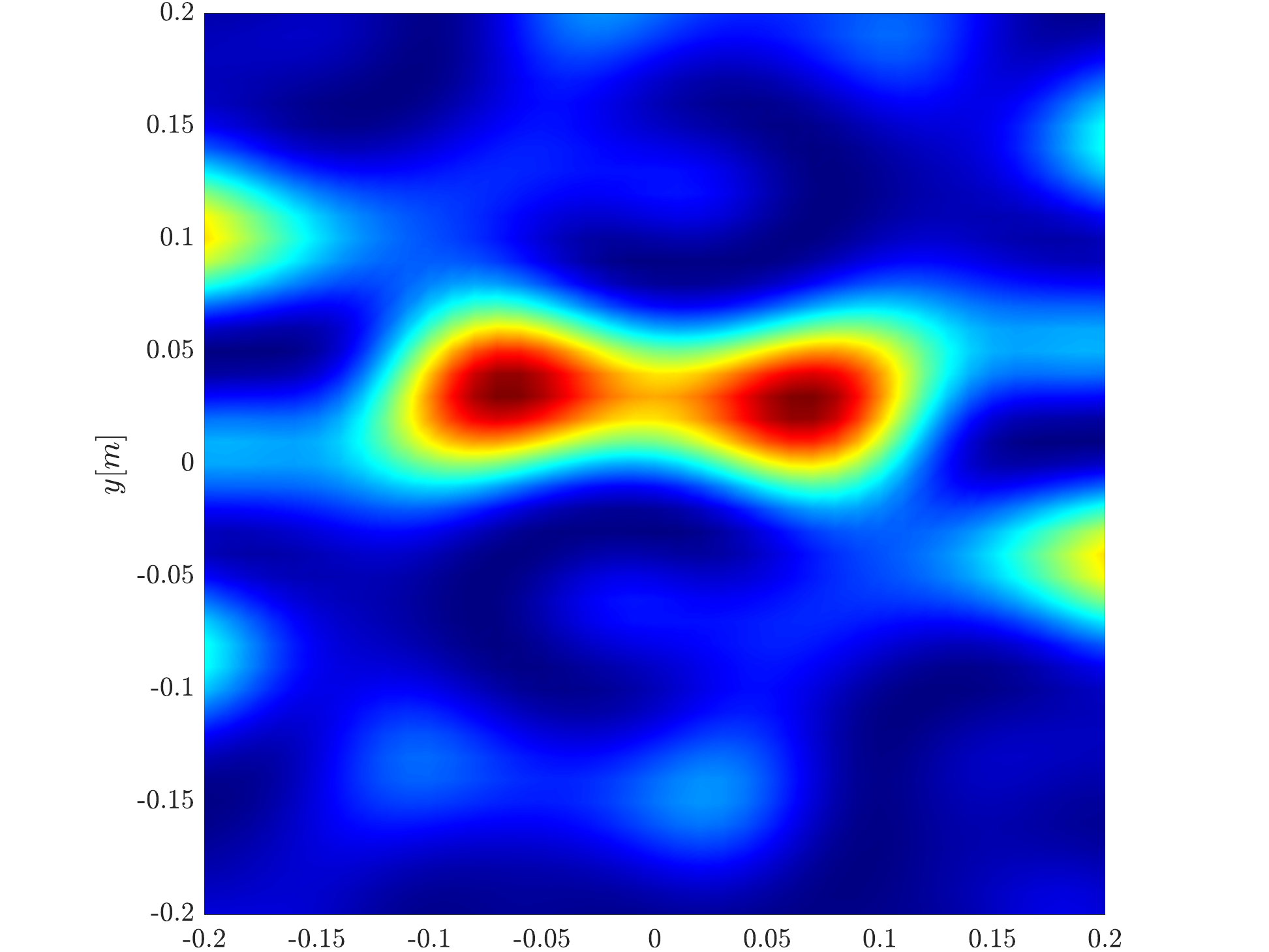}
	\caption{}
\end{subfigure}
\begin{subfigure}[t]{0.18\textwidth}
	\includegraphics[width=\textwidth]{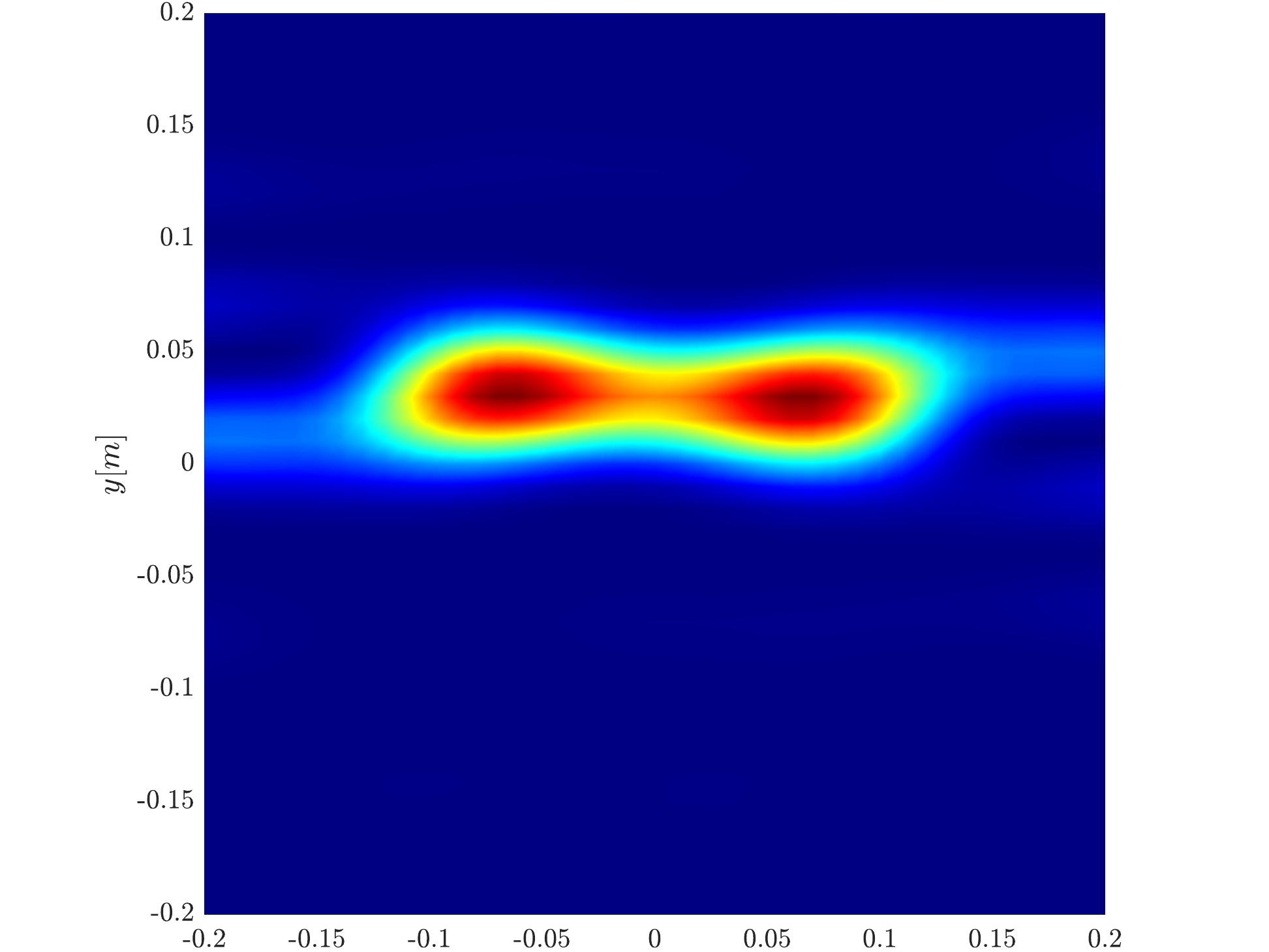}
	\caption{}
\end{subfigure}
\begin{subfigure}[t]{0.18\textwidth}
	\includegraphics[width=\textwidth]{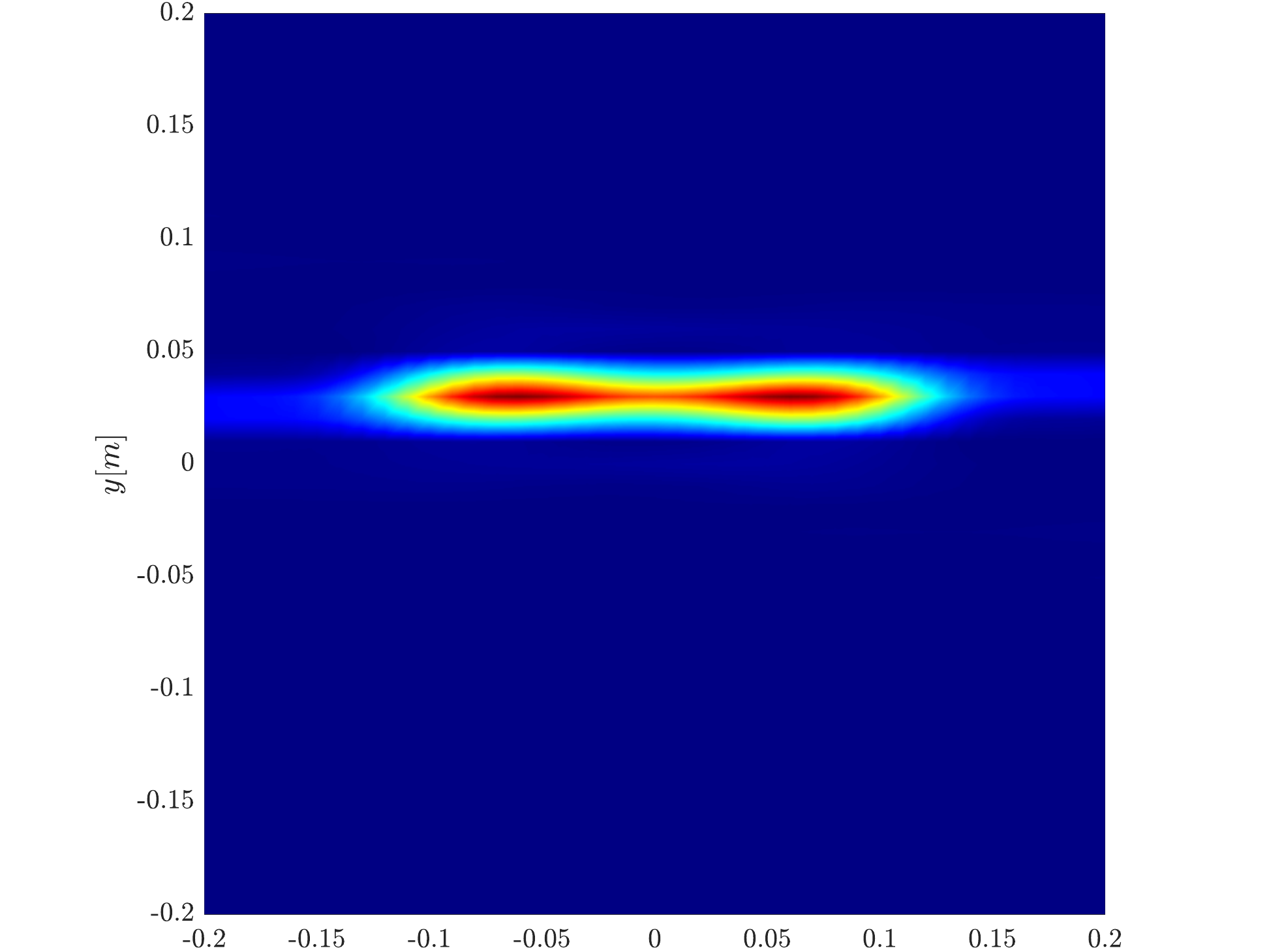}
	\caption{}
\end{subfigure}
		\caption{Comparison of single point migration (top row), rank-1 image (middle row), and linear Kirchhoff migration (bottom row), for increasing synthetic aperture size of $(a)$ $100\Delta s$,$(b)$ $1000\Delta s$ and $(c)$ $3000\Delta s$. We can see that as the synthetic aperture increases there is a dramatic improvement in the resolution attained by the rank-1 image compared to the single point one, whose resolution does not significantly change as the synthetic aperture increases. The resolution of the rank-1 image is comparable to the linear Kirchhoff migration.}
	\label{fig:eig_v_diag_double}
\end{figure}

We first consider the two scatterer cluster illustrated in Figure~\ref{fig:eig_v_diag_double}. The scatterers are located at $(\pm 0.05,0.03)$ with respect to the center of the image window. We observe that as the synthetic aperture increases there is a dramatic improvement in the resolution attained by the rank-1 image compared to the single point one, whose resolution does not significantly change as the synthetic aperture increases. Note that the resolution of the rank-1 image is comparable to that of the linear Kirchhoff migration.

\begin{figure}[htbp]
	\centering
	\begin{subfigure}[t]{0.18\textwidth}
		\includegraphics[width=\textwidth]{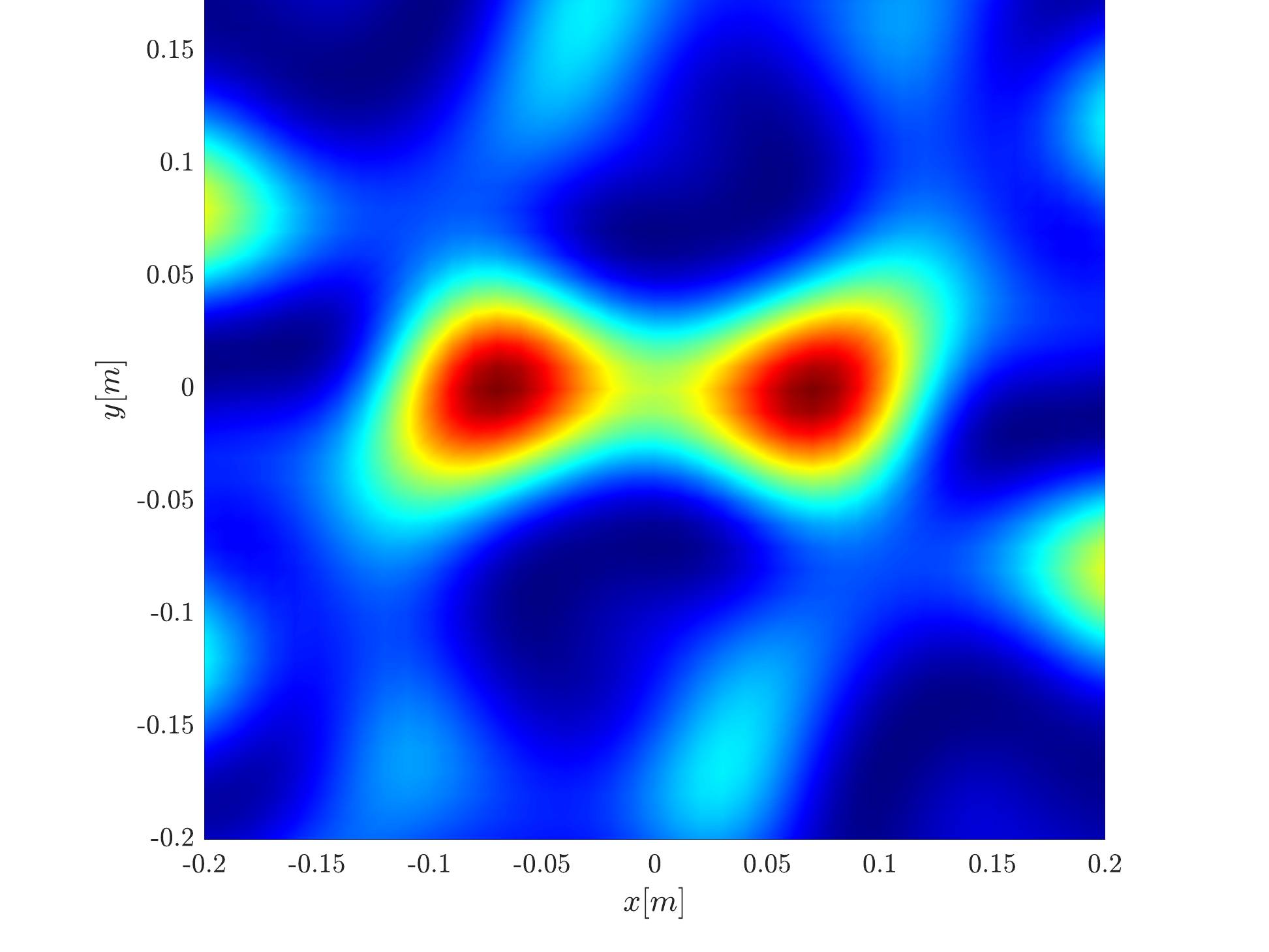}
	\end{subfigure}
	\begin{subfigure}[t]{0.18\textwidth}
		\includegraphics[width=\textwidth]{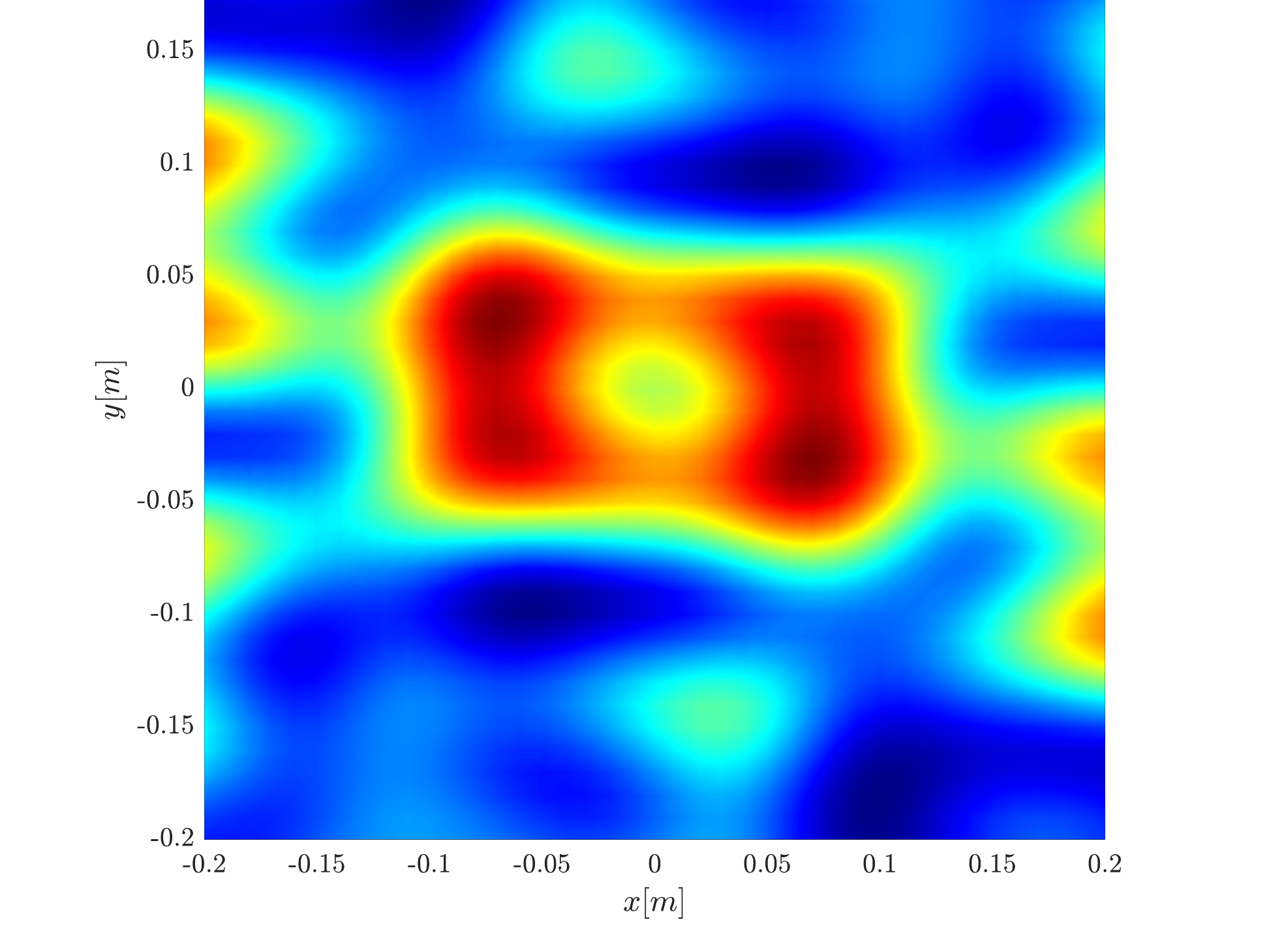}
	\end{subfigure}
	\begin{subfigure}[t]{0.18\textwidth}
		\includegraphics[width=\textwidth]{Figures_new/quad/diag_3000-eps-converted-to.pdf}
	\end{subfigure}
	
	\begin{subfigure}[t]{0.18\textwidth}
		\includegraphics[width=\textwidth]{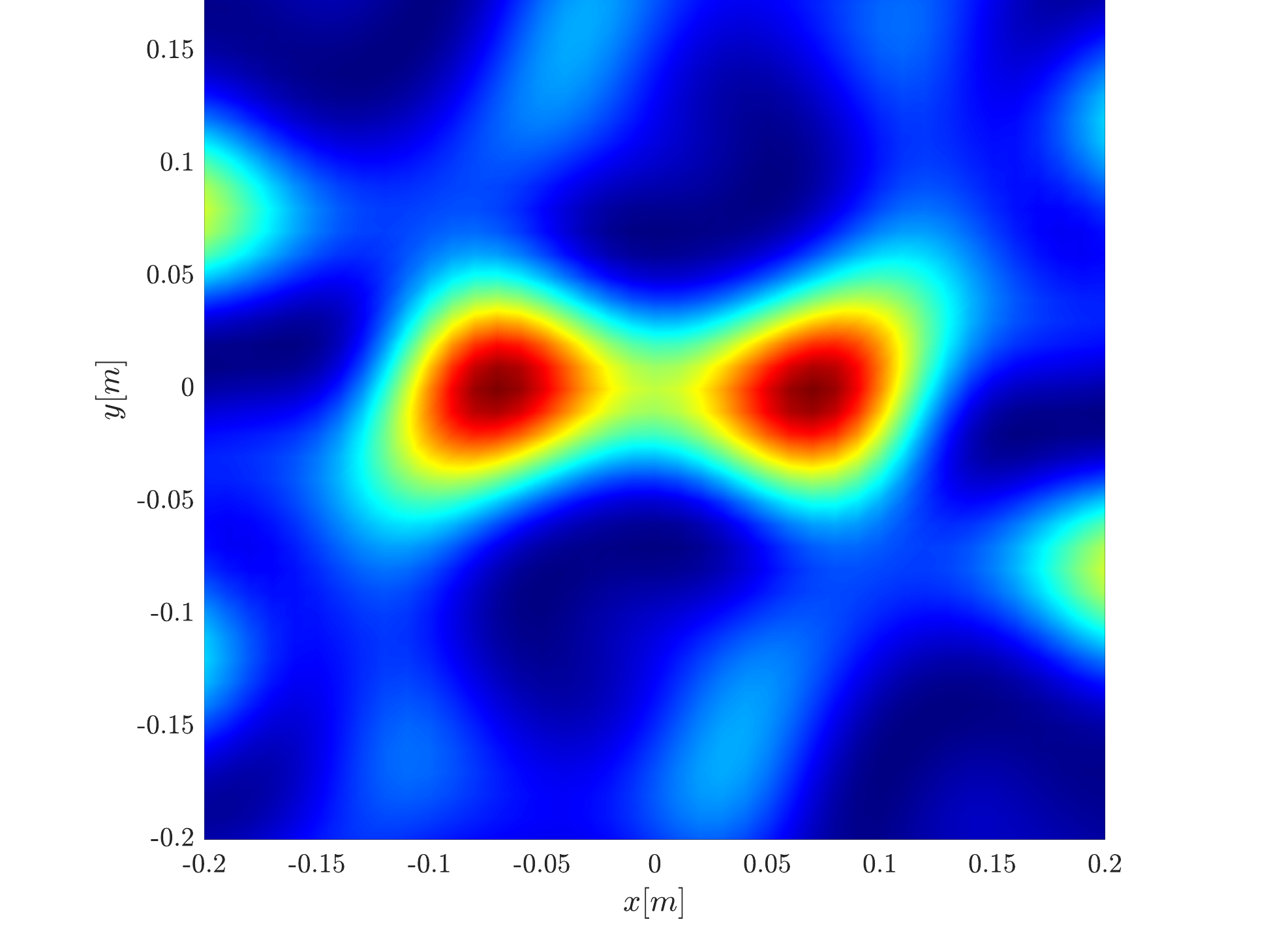}
	\end{subfigure}
	\begin{subfigure}[t]{0.18\textwidth}
		\includegraphics[width=\textwidth]{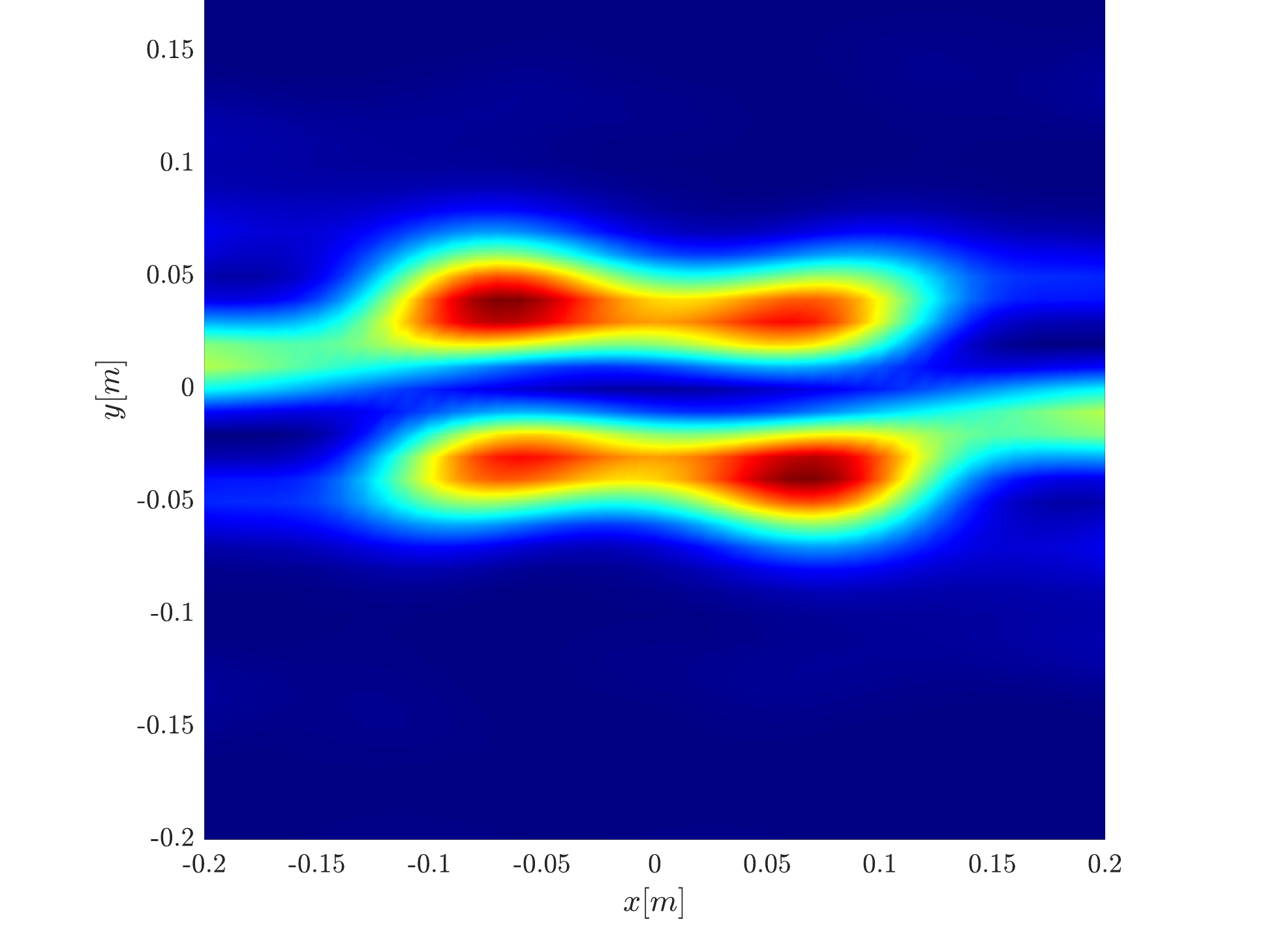}
	\end{subfigure}
	\begin{subfigure}[t]{0.18\textwidth}
		\includegraphics[width=\textwidth]{Figures_new/quad/1stmode_3000-eps-converted-to.pdf}
	\end{subfigure}
	
	\begin{subfigure}[t]{0.18\textwidth}
		\includegraphics[width=\textwidth]{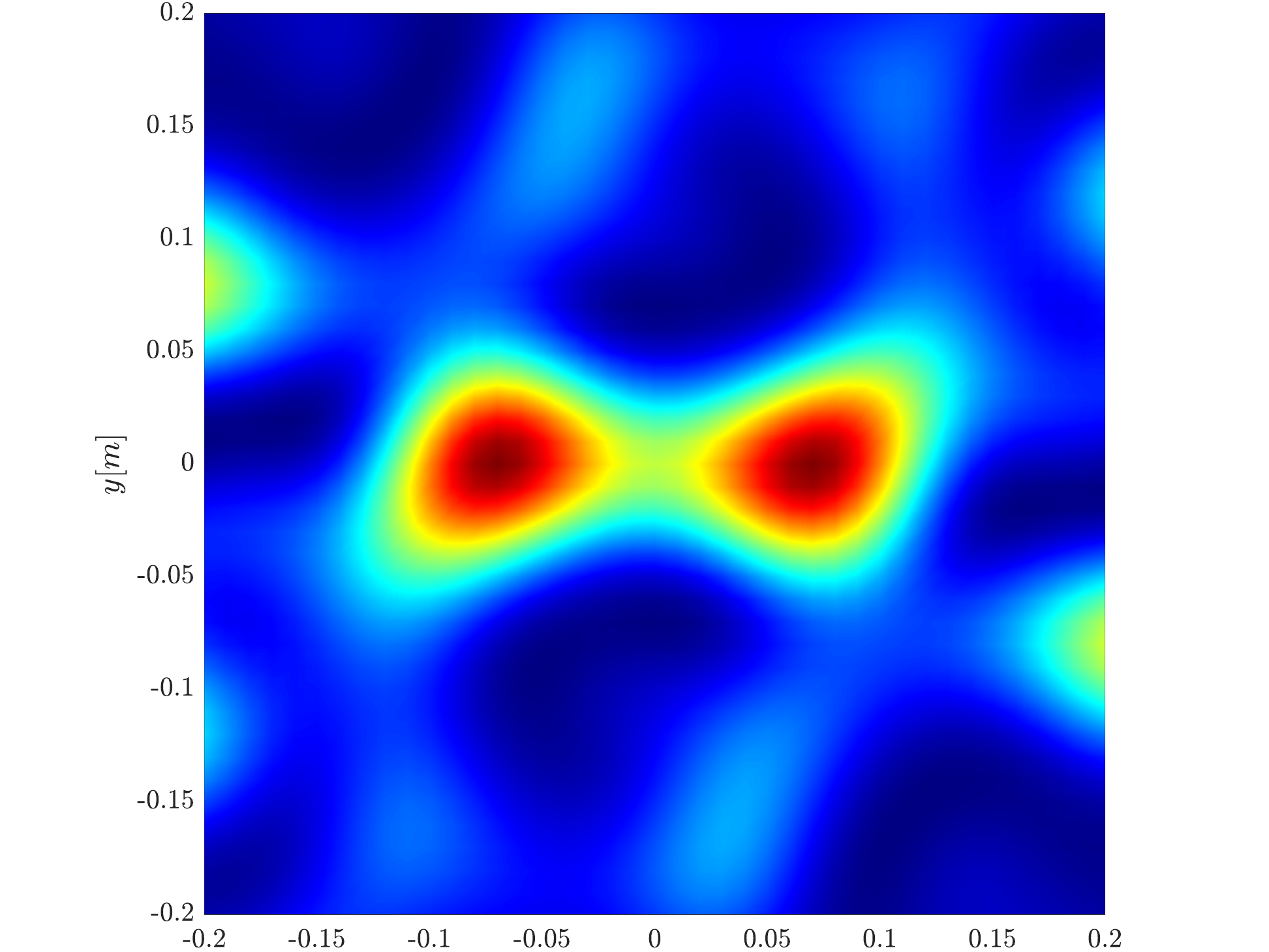}
		\caption{}
	\end{subfigure}
	\begin{subfigure}[t]{0.18\textwidth}
		\includegraphics[width=\textwidth]{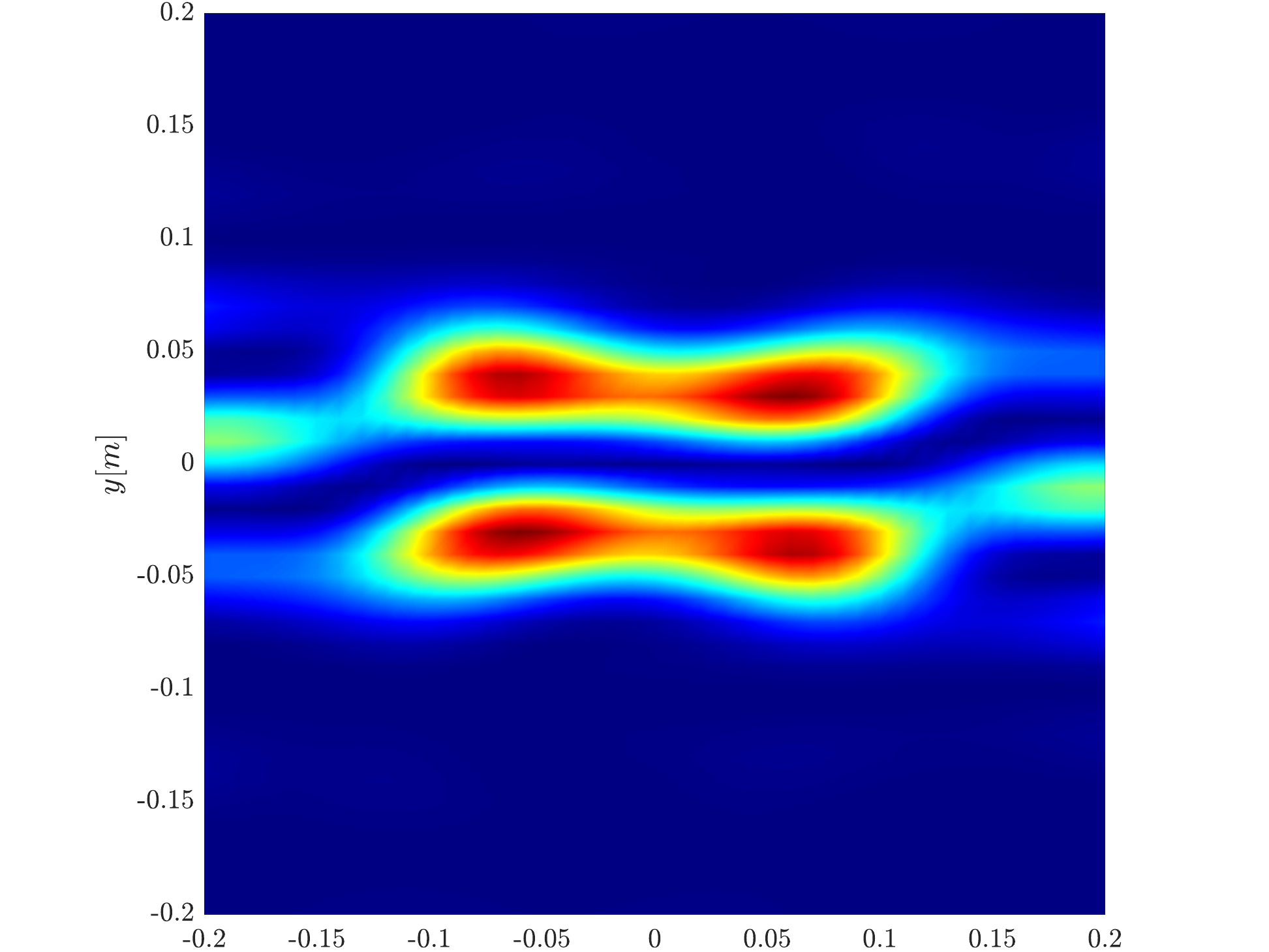}
		\caption{}
	\end{subfigure}
	\begin{subfigure}[t]{0.18\textwidth}
		\includegraphics[width=\textwidth]{Figures_new/quad/linear_3000-eps-converted-to.pdf}
		\caption{}
	\end{subfigure}
	\caption{Comparison of single point migration (top row), rank-1 image (middle row), and linear Kirchhoff migration (bottom row), for increasing synthetic aperture size of $(a)$ $100\Delta s$,$(b)$ $1000\Delta s$ and $(c)$ $3000\Delta s$. As for Figure \ref{fig:eig_v_diag_double} we observe a dramatic improvement in the resolution of the rank-1 image as the synthetic aperture increases. }
	\label{fig:eig_v_diag_quad}
\end{figure}

Next we consider the four scatterer cluster illustrated in Figure~\ref{fig:eig_v_diag_quad}. The scatterers are located at $(\pm0.05,\pm0.03)$ with respect to the center of the image window. Again, the resolution of the rank-1 image improves significantly as the synthetic aperture increases while this is not the case for the single point migration. For all the synthetic apertures considered, the single point migration cannot resolve the scatterers in the $y$-direction and provides a blurry image of the cluster of scatterers. As for Figure~\ref{fig:eig_v_diag_double}  the resolution of the rank-1 image is comparable to that of the linear Kirchhoff migration.  The top 25 eigenvalues of $\tilde{\mb X}$ for the different synthetic aperture sizes are shown in Figure \ref{fig:num_exam}-(d).  We observe that, as the synthetic aperture size increases, so does the number of significant eigenvalues. That is due to the anisotropy in the resolution of the interference pattern $\tilde{\mb X}$. As mentioned in Section \ref{sec:perf}, in the direction $\hat{\mb y}_k=\hat{\mb y}_{k'}$ the resolution is $\frac{\lambda H_T}{a}$ while it is $\frac{\lambda H_\mb T}{\mb v_\mb T S}$ in the direction $\hat{\mb y}_k\cdot \mb v_\mb T=-\hat{\mb y}_{k'}\cdot \mb v_\mb T$. Therefore as the size of the synthetic aperture increases ($\mb v_\mb TS$) the anisotropy becomes more significant. As a consequence the rank of $\tilde{\mb X}$ increases and we expect a differentiation between the single-point migration and the rank-1 image. As we explain in Appendix~\ref{app:kernel} the rank-1 image provides a resolution which is in-between  $\frac{\lambda H_T}{a}$  and $\frac{\lambda H_\mb T}{\mb v_\mb T S}$. 

\begin{figure}[htbp]
	\centering
	\begin{subfigure}[t]{0.18\textwidth}
		\includegraphics[width=\textwidth]{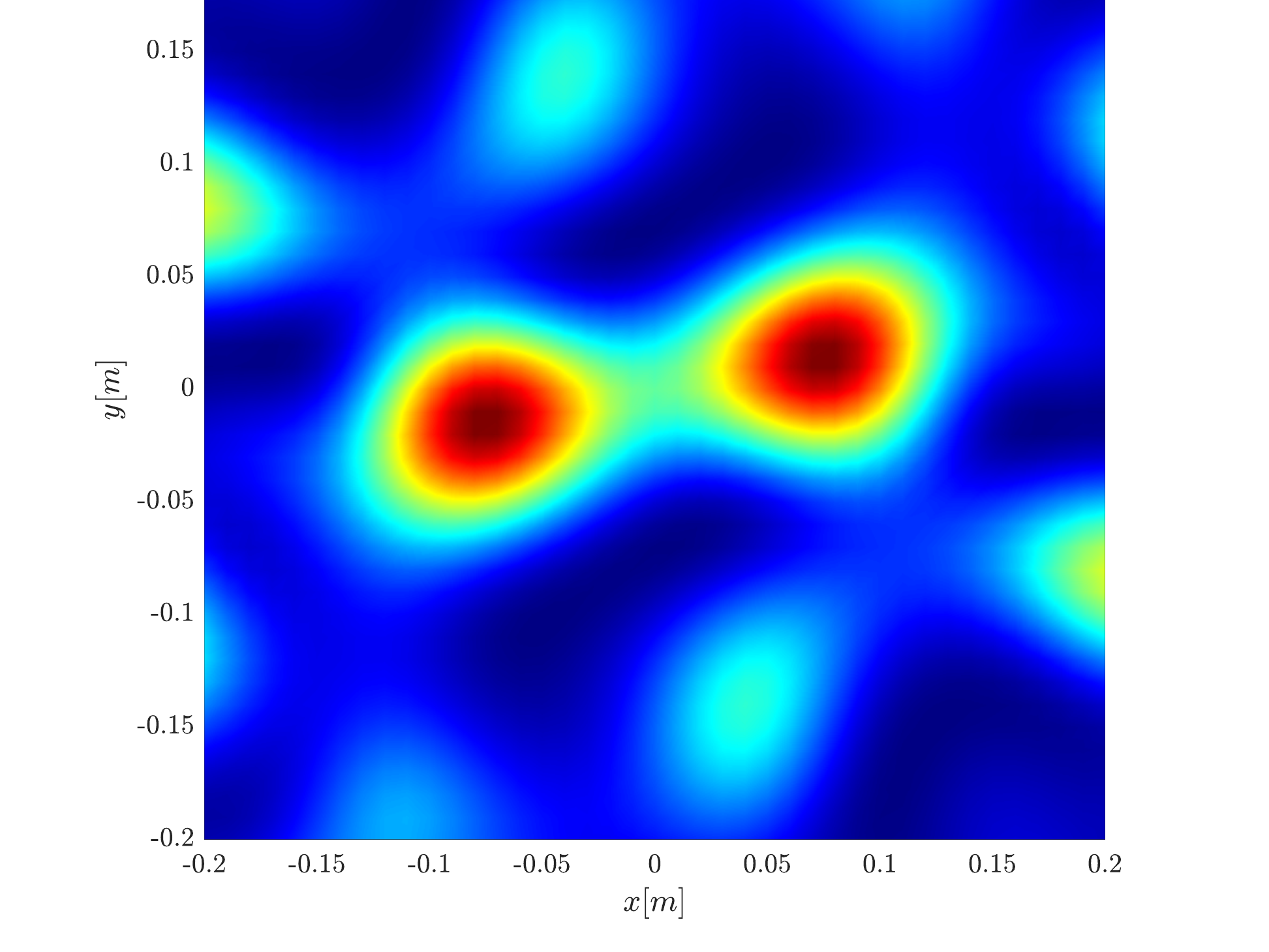}
	\end{subfigure}
	\begin{subfigure}[t]{0.18\textwidth}
		\includegraphics[width=\textwidth]{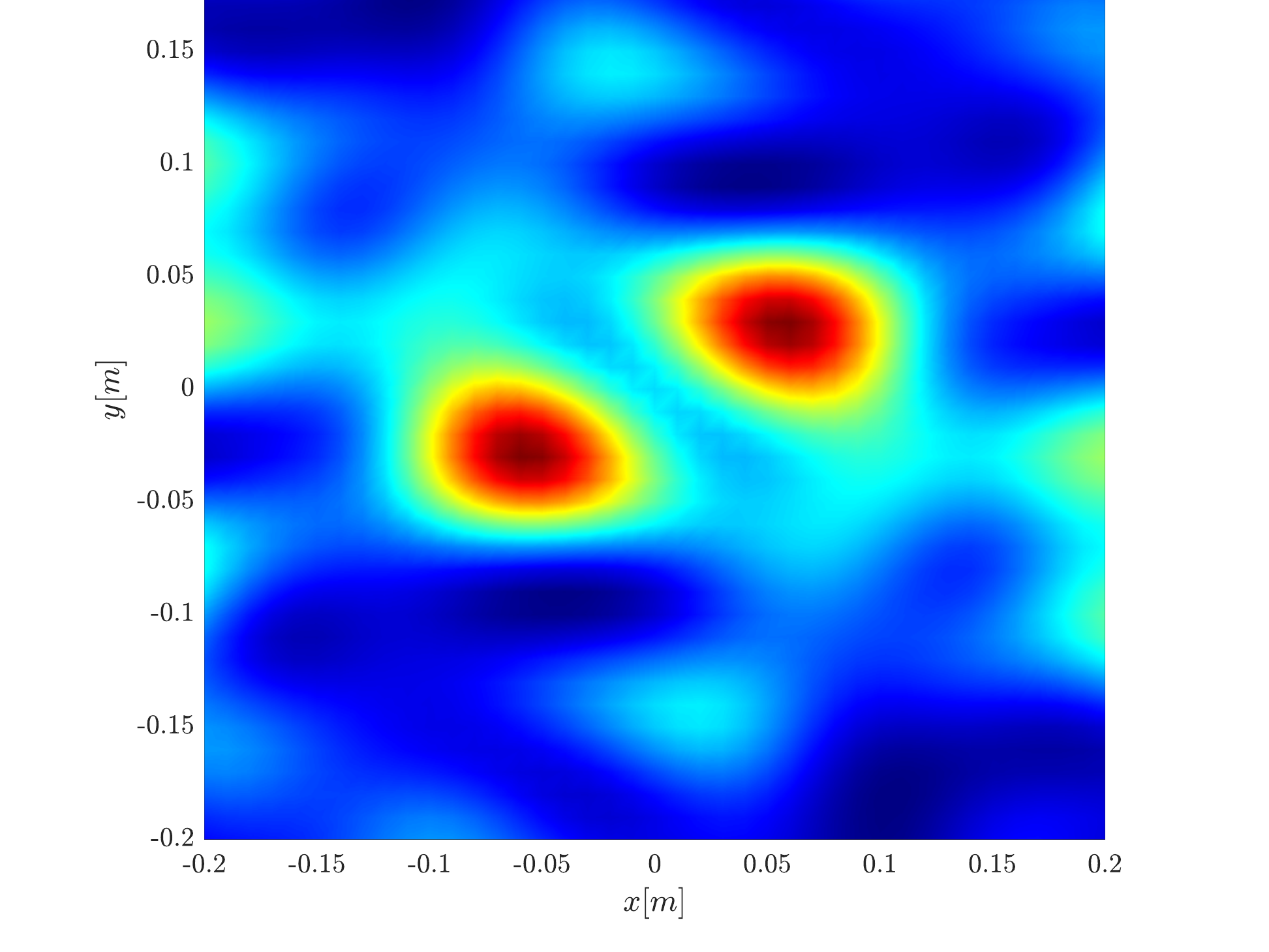}
	\end{subfigure}
	\begin{subfigure}[t]{0.18\textwidth}
		\includegraphics[width=\textwidth]{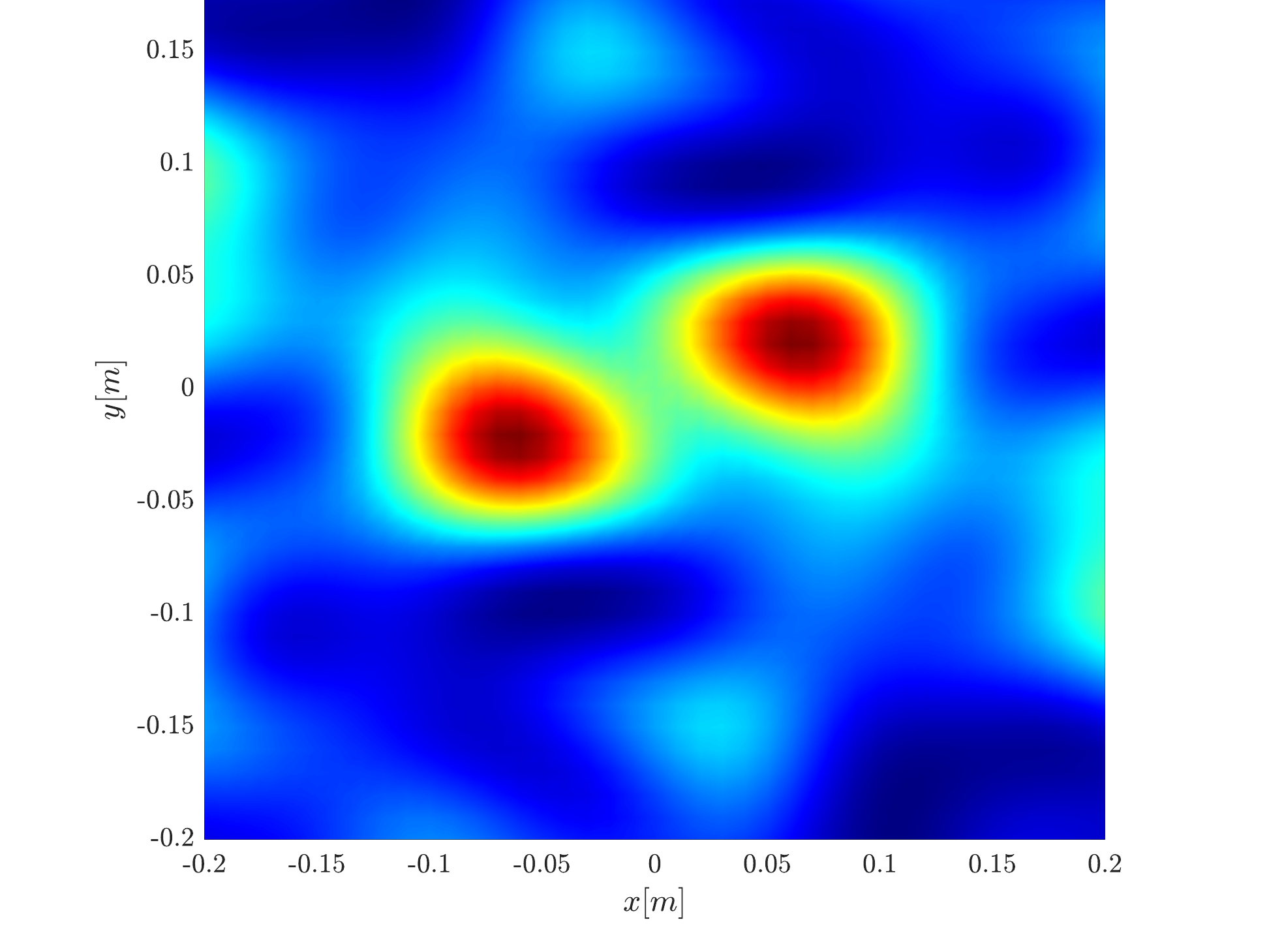}
	\end{subfigure}
	
	\begin{subfigure}[t]{0.18\textwidth}
		\includegraphics[width=\textwidth]{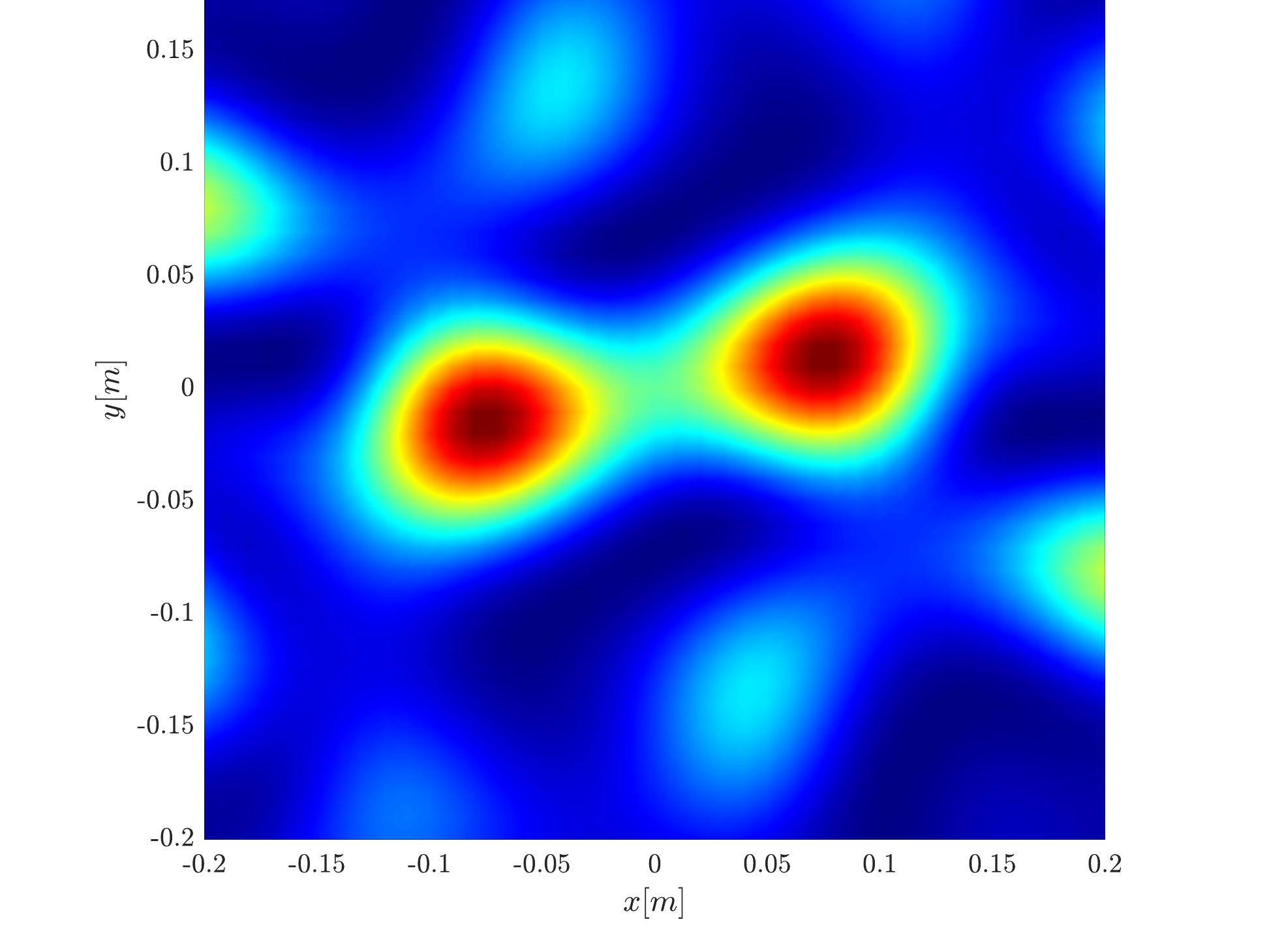}
	\end{subfigure}
	\begin{subfigure}[t]{0.18\textwidth}
		\includegraphics[width=\textwidth]{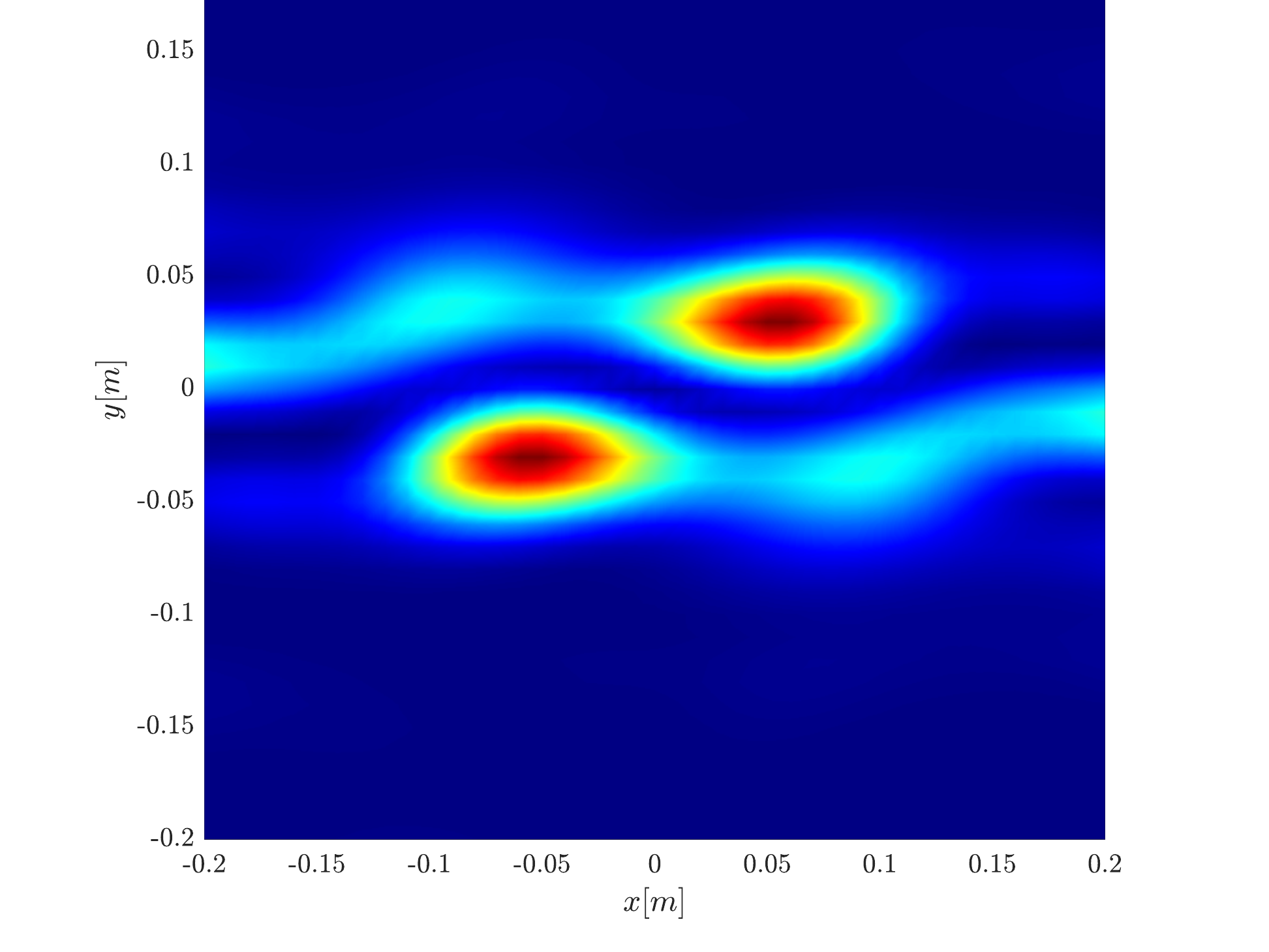}
	\end{subfigure}
	\begin{subfigure}[t]{0.18\textwidth}
		\includegraphics[width=\textwidth]{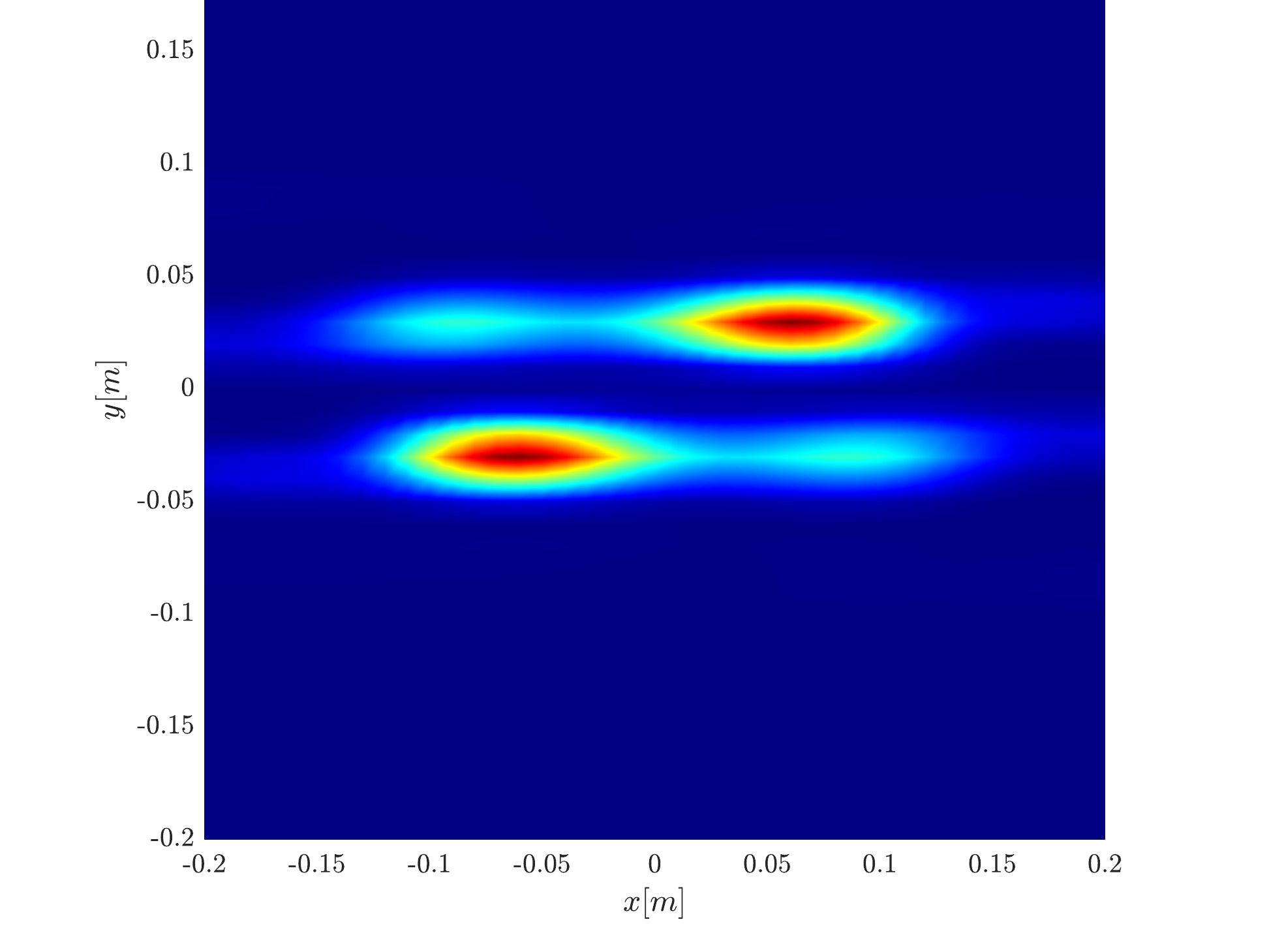}
	\end{subfigure}
	
	\begin{subfigure}[t]{0.18\textwidth}
		\includegraphics[width=\textwidth]{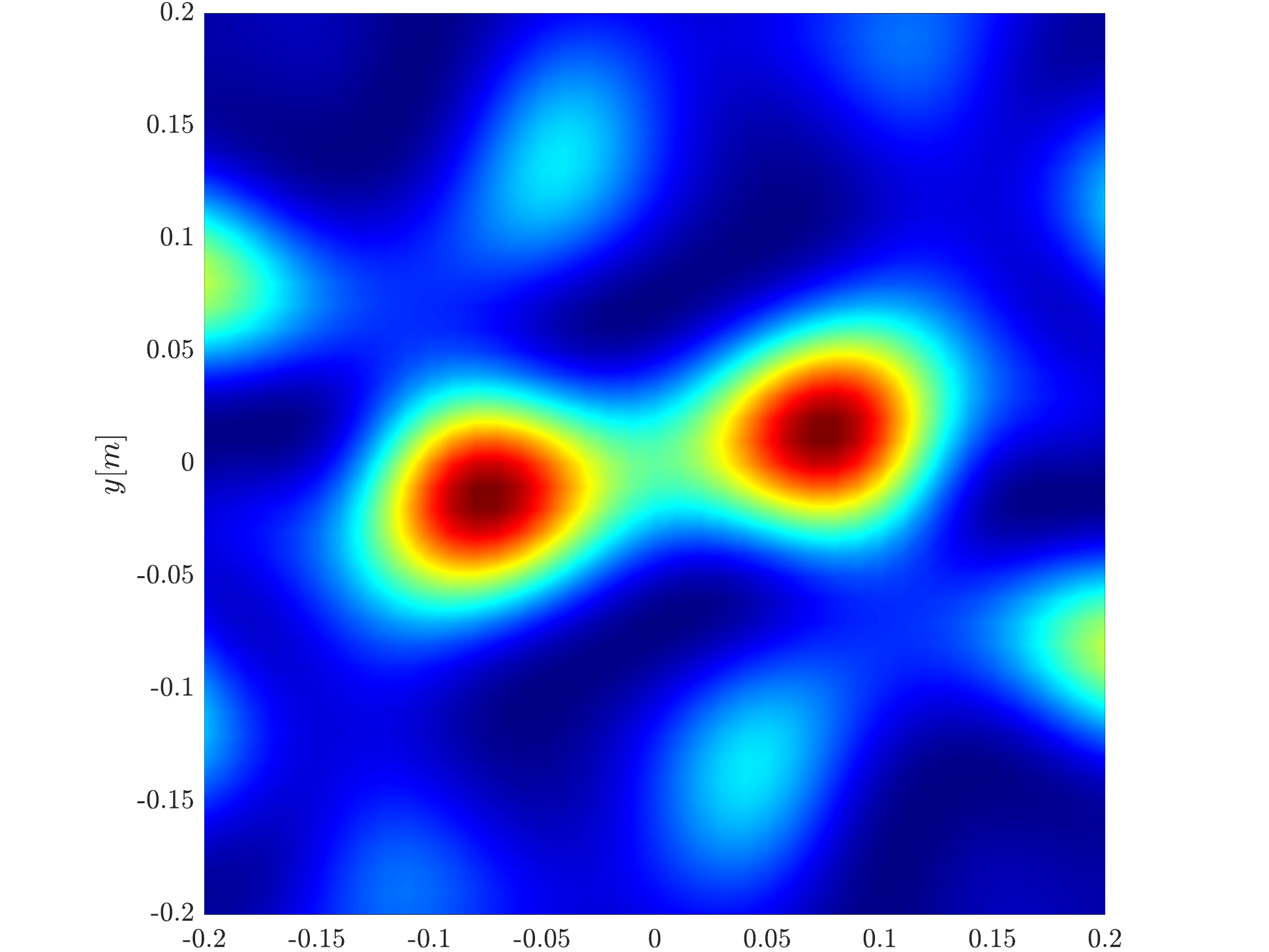}
		\caption{}
	\end{subfigure}
	\begin{subfigure}[t]{0.18\textwidth}
		\includegraphics[width=\textwidth]{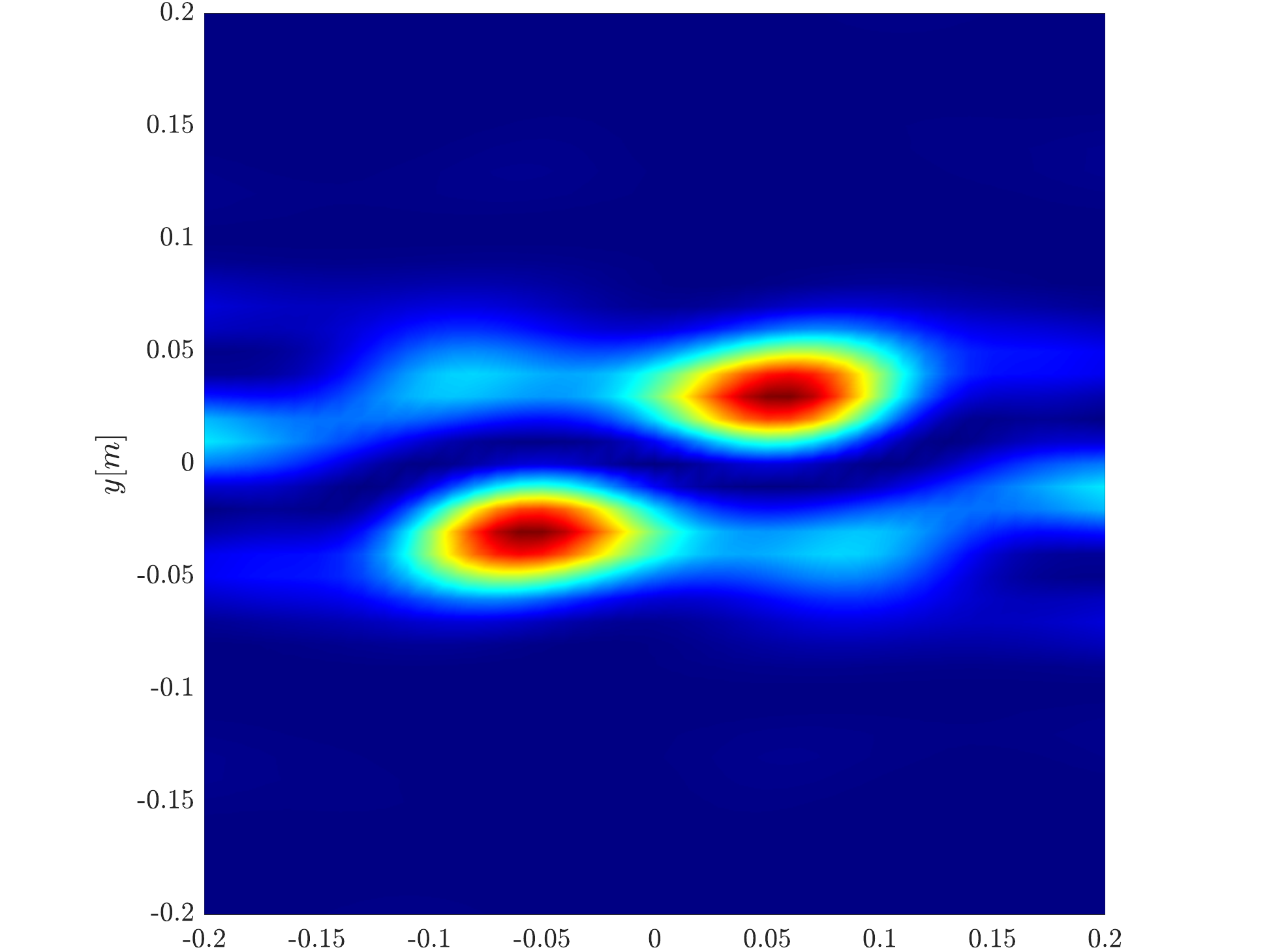}
		\caption{}
	\end{subfigure}
	\begin{subfigure}[t]{0.18\textwidth}
		\includegraphics[width=\textwidth]{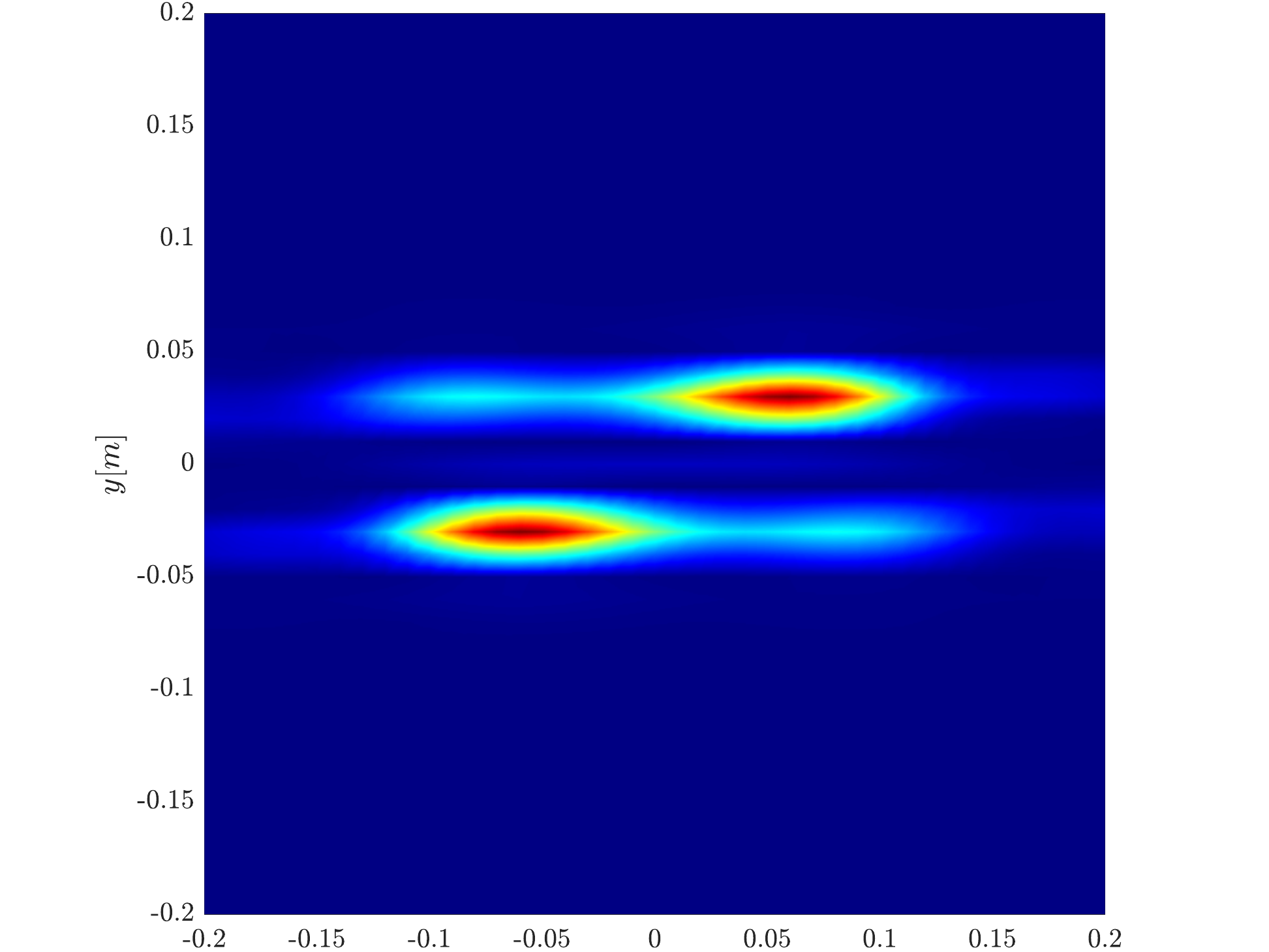}
		\caption{}
	\end{subfigure}
	\caption{Comparison of single point migration (top row), rank-1 image (middle row), and linear Kirchhoff migration (bottom row), for increasing synthetic aperture size of $(a)$ $100\Delta s$,$(b)$ $1000\Delta s$ and $(c)$ $3000\Delta s$. We can see that as the synthetic aperture increases there is a dramatic improvement in the resolution attained by the rank-1 image compared to the single point one, which is only marginally improved by the increased synthetic aperture. }
	\label{fig:eig_v_diag_quad_contrast}
\end{figure}

In Figure~\ref{fig:eig_v_diag_quad_contrast} we consider again the four scatterer cluster but now the scatterers have different reflectivity. Two of the scatterers are weaker having $80$\% of the reflectivity of the other two. This translates to a ratio of $0.64$ for $\rho_i^2$. In this case the single point migration reconstructs only the strong scatterers and those with a resolution that is significantly inferior of that of the rank-1 image. As before rank-1 image and linear KM provide similar results. 


\begin{figure}[htbp]
	\centering
		\begin{subfigure}[t]{0.38\textwidth}
		\includegraphics[width=\textwidth]{Figures_new/quad/1stmode_3000-eps-converted-to.pdf}
		\caption{}
	\end{subfigure}
	\begin{subfigure}[t]{0.38\textwidth}
	\includegraphics[width=\textwidth]{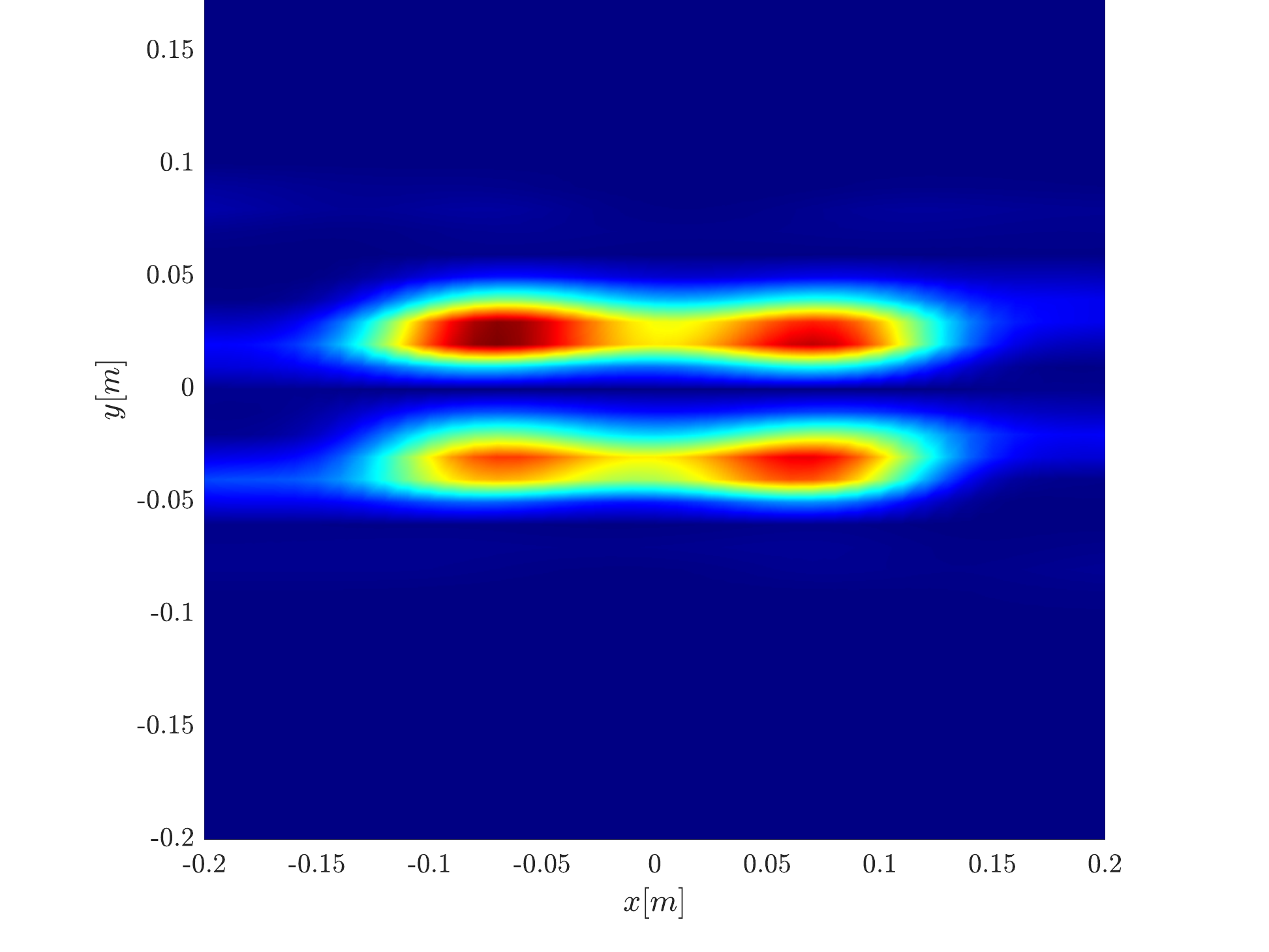}
	\caption{}
\end{subfigure}
\caption{Effect of downsampling: $(a)$ rank-1 image using from a square matrix $\tilde{X}$ $(b)$ rank-1 image from a rectangular matrix with 10\% of the columns chosen at random. We can see that the resolution remains the same.}
\label{fig:quad_downsample}
\end{figure}

To reduce the computational complexity of the rank-1 image, we can downsample the two-point migration with respect to one of its dimensions. In this case instead of computing the first eigenvector of the interference pattern $\tilde{\mb X}$, we compute its first singular vector. This downsampling does not affect the resolution of the rank-1 image. This is illustrated in Figure~\ref{fig:quad_downsample} where we see that the image retains its resolution even if one of the dimensions is downsampled by a factor of 10 (keeping 10\% of the columns chosen at random), reducing the computational complexity by the same factor. 

\begin{figure}
	\centering
	\begin{subfigure}[t]{0.22\textwidth}
	\includegraphics[width=\textwidth]{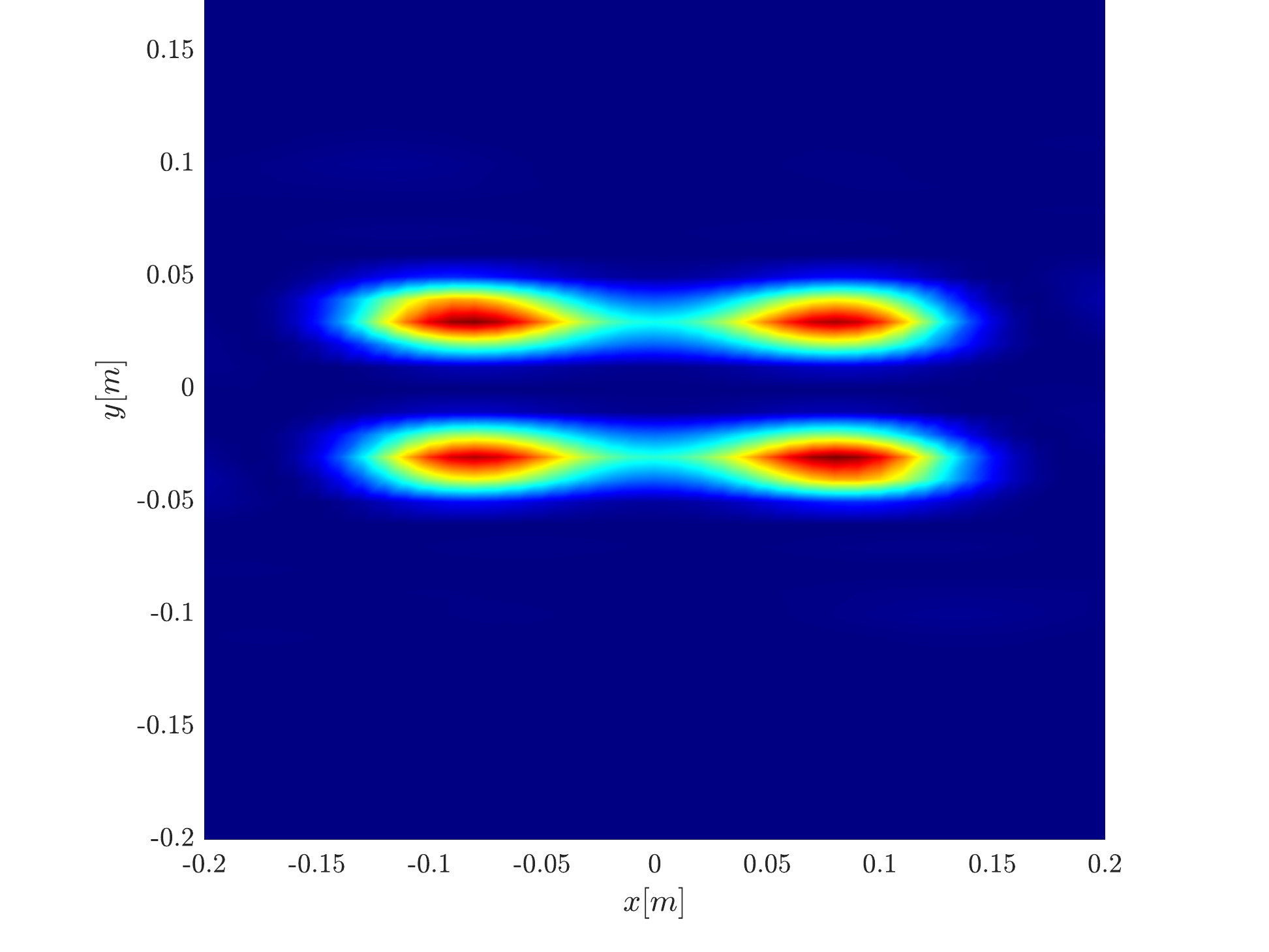}
	\caption{}
\end{subfigure}
\begin{subfigure}[t]{0.22\textwidth}
	\includegraphics[width=\textwidth]{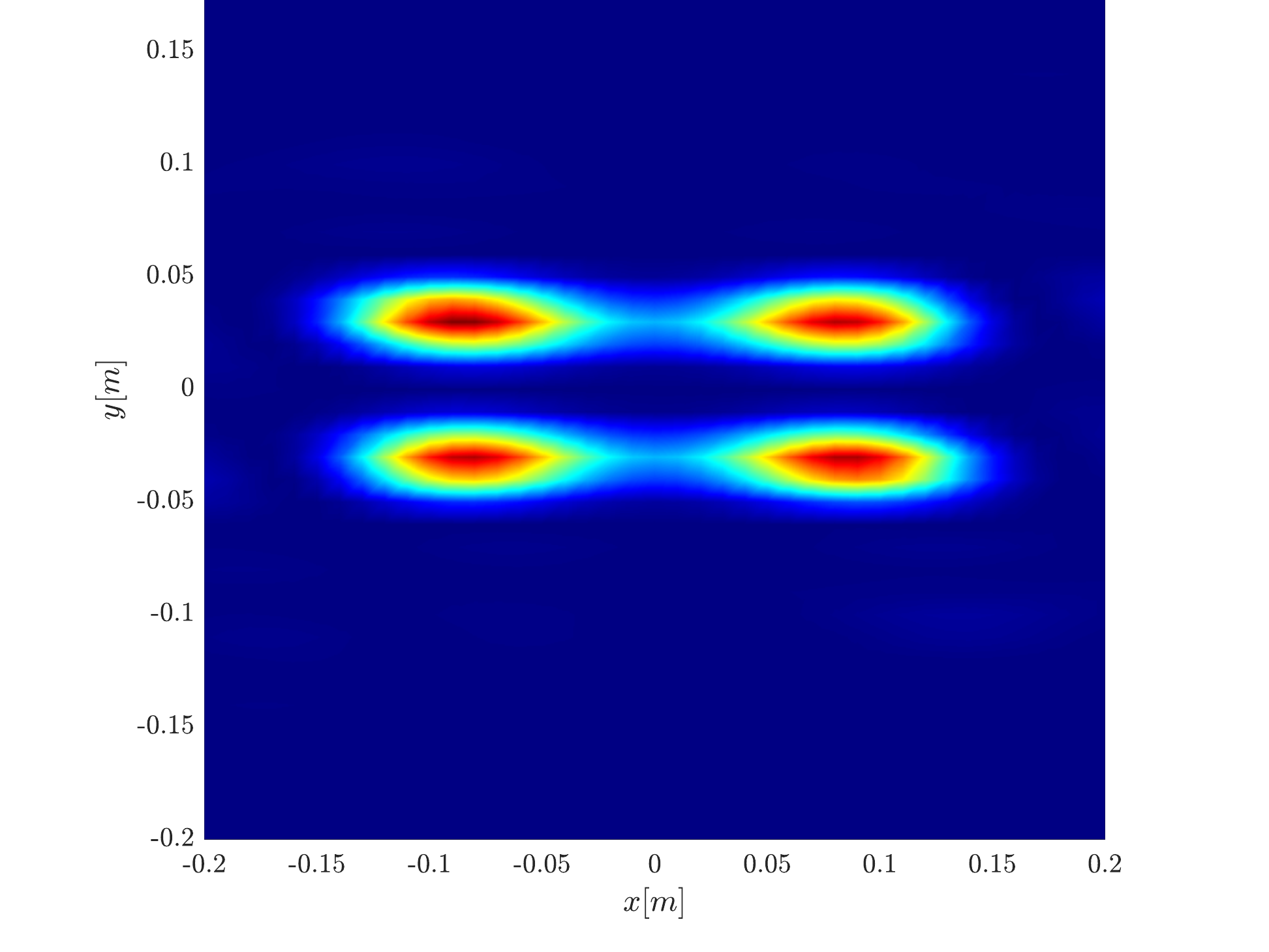}
	\caption{}
\end{subfigure}
\begin{subfigure}[t]{0.22\textwidth}
	\includegraphics[width=\textwidth]{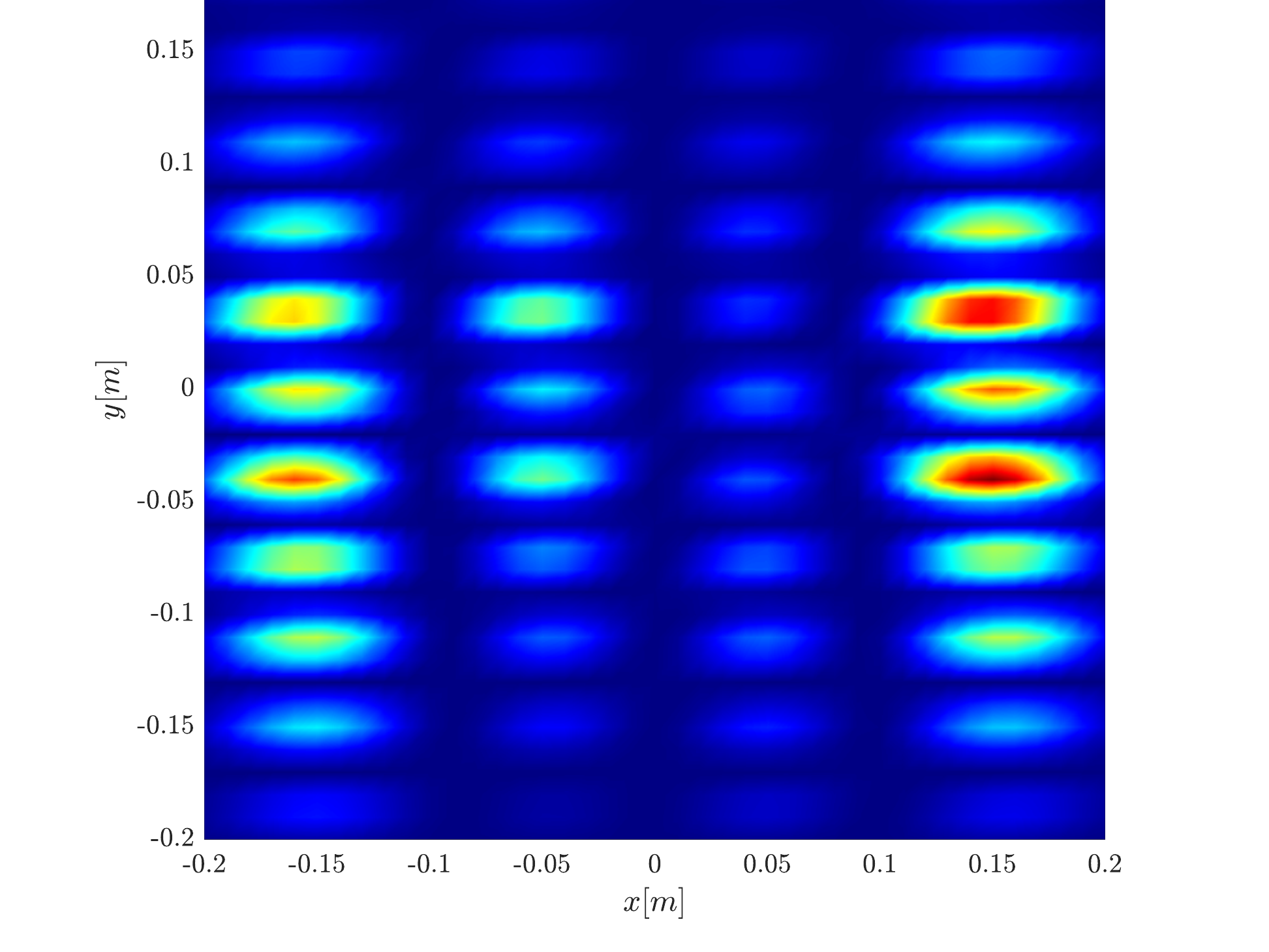}
	\caption{}
\end{subfigure}
	\caption{Effect of signal to noise ratio (SNR): We assume the measurements are corrupted by additive white Gaussian noise with varying noise level, and look at the rank-1 image. $(a)$ SNR $=-14$dB $(b)$ SNR $=-15.5$dB. $(c)$ SNR $=-17$dB. As the SNR reaches a critical value the top eigenvector is no longer robust to noise.}
	\label{fig:quad_SNR}
\end{figure}
\begin{figure}
\centering
\begin{subfigure}[t]{0.82\textwidth}
	\includegraphics[width=\textwidth]{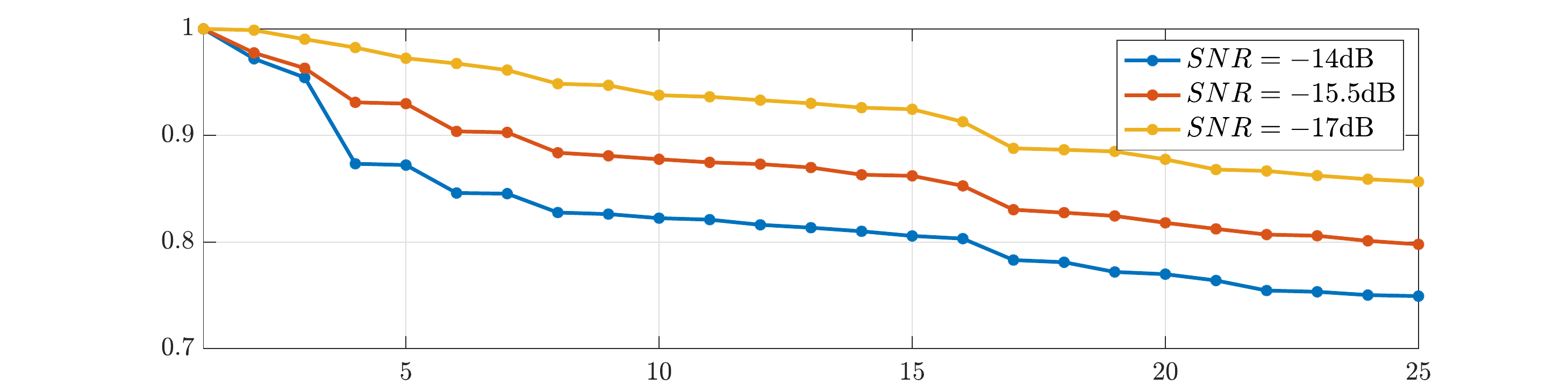}
\end{subfigure}
\caption{Top 25 eigenvalues of $\tilde{\mb X}$ for the reconstructions shown in Figure~\ref{fig:quad_SNR}. The eigenvalues are normalized such that the top eigenvalue is always 1. We can see that as the noise level increase, the noise distorts the form of the spectrum.}
\label{fig:SNR_svd}
\end{figure}

To investigate the robustness of the rank-1 image to noise we corrupt the data corresponding to Figure~\ref{fig:eig_v_diag_quad} with additive white Gaussian noise with varying noise levels. 
We see in Figure~\ref{fig:quad_SNR} that the rank-1 image remains stable down to a low signal to noise ratio (SNR) of $-17$dB. Below that point, as Figure~\ref{fig:SNR_svd} illustrates, the spectrum of the noise distorts that of the data.

\section{Properties of the rank-1 imaging function}
\label{sec:prop_filt}
The results presented in Section~\ref{sec:numerical_simulations} show that the performance of the rank-1 image is superior to the single point migration and comparable to the linear Kirchhoff migration. In order to better understand these results we consider in this section a simplified model problem with a single point scatterer. The results for this simplified case allow us to explain the improved performance of the rank-1 image and are consistent with our theoretical analysis in Appendix~\ref{app:stat_phase}. 

\begin{figure}[htbp]
	\centering
	\begin{subfigure}[t]{0.18\textwidth}
		\includegraphics[width=\textwidth]{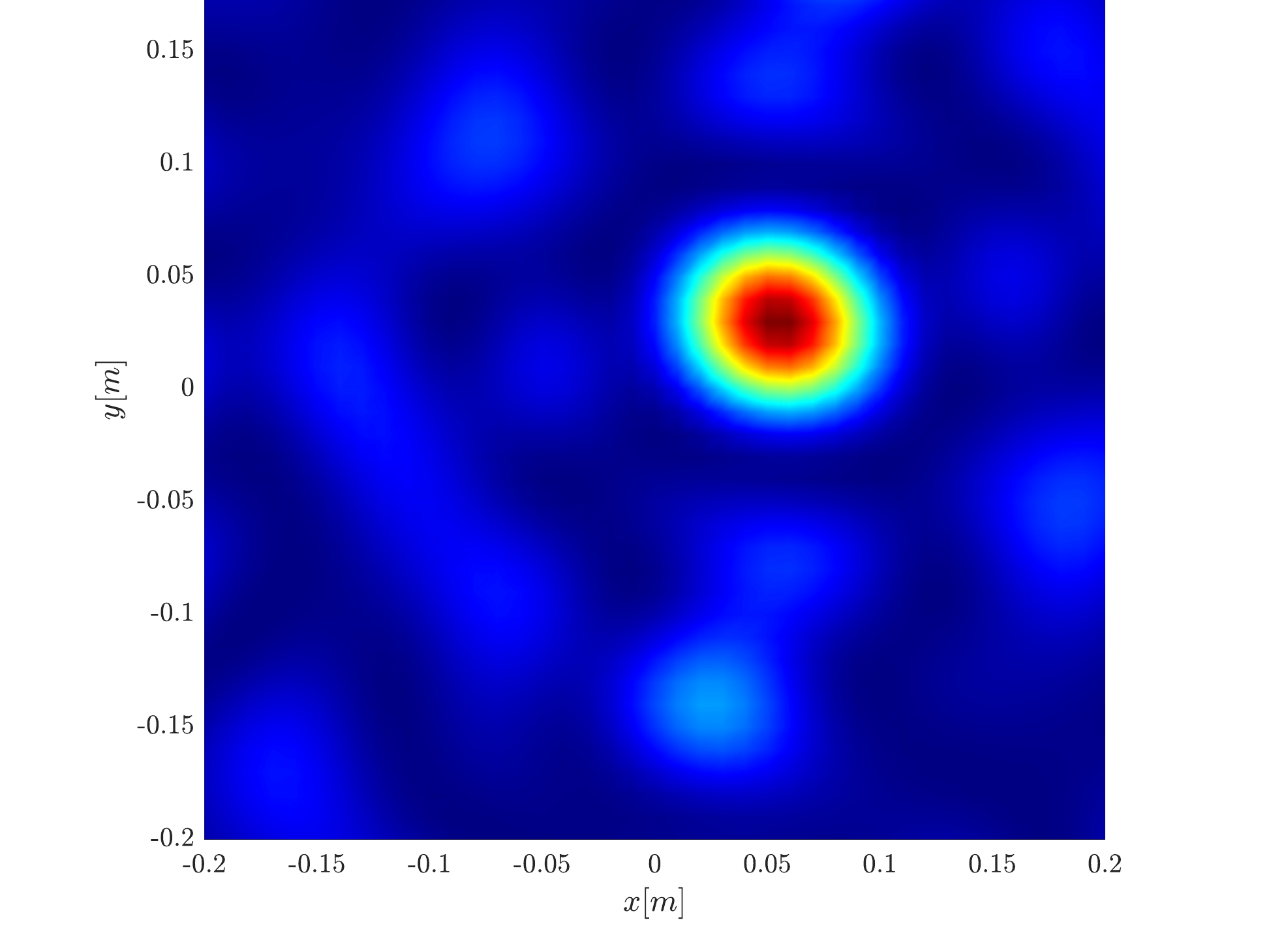}
	\end{subfigure}
	\begin{subfigure}[t]{0.18\textwidth}
		\includegraphics[width=\textwidth]{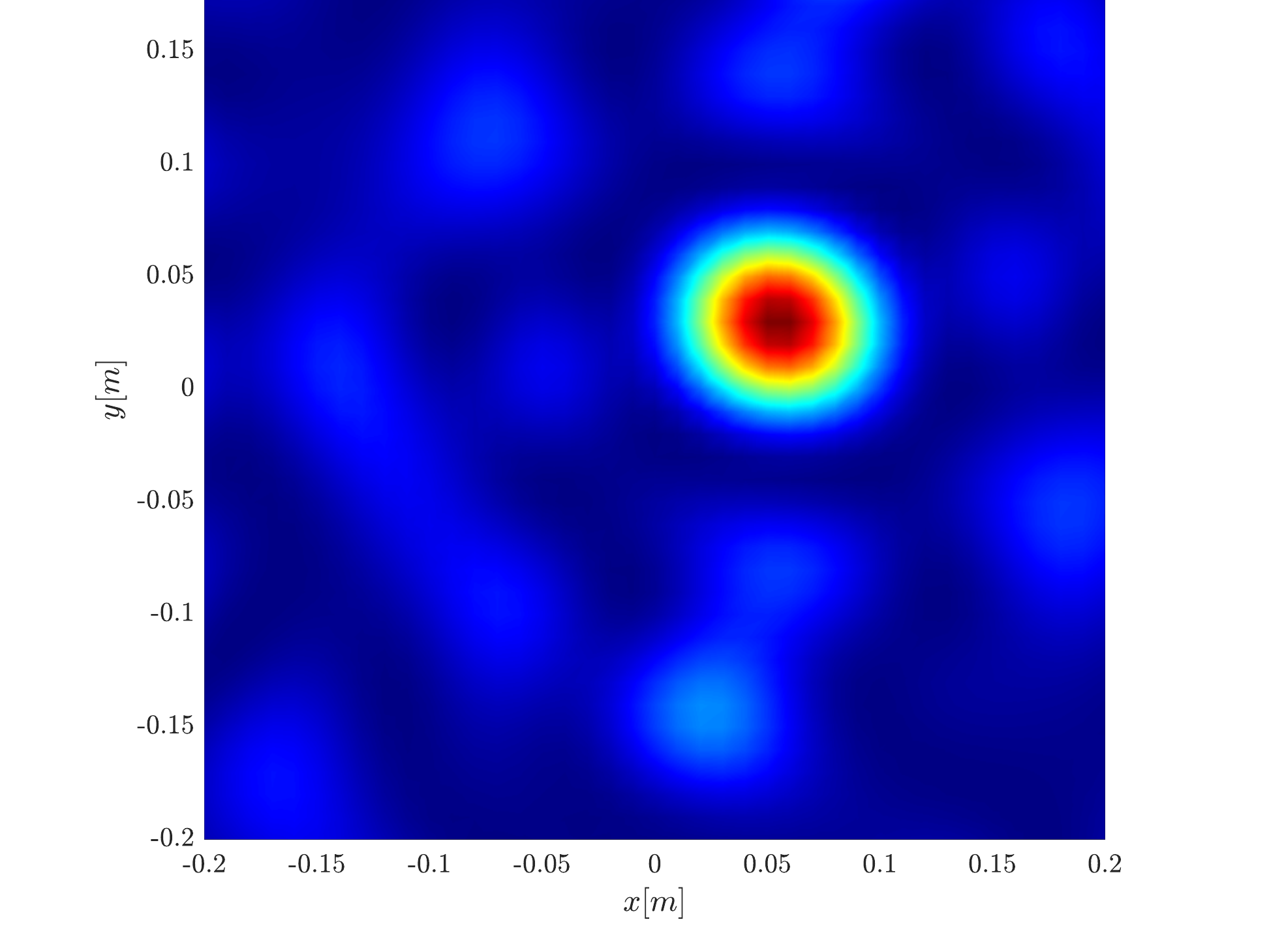}
	\end{subfigure}
	\begin{subfigure}[t]{0.18\textwidth}
		\includegraphics[width=\textwidth]{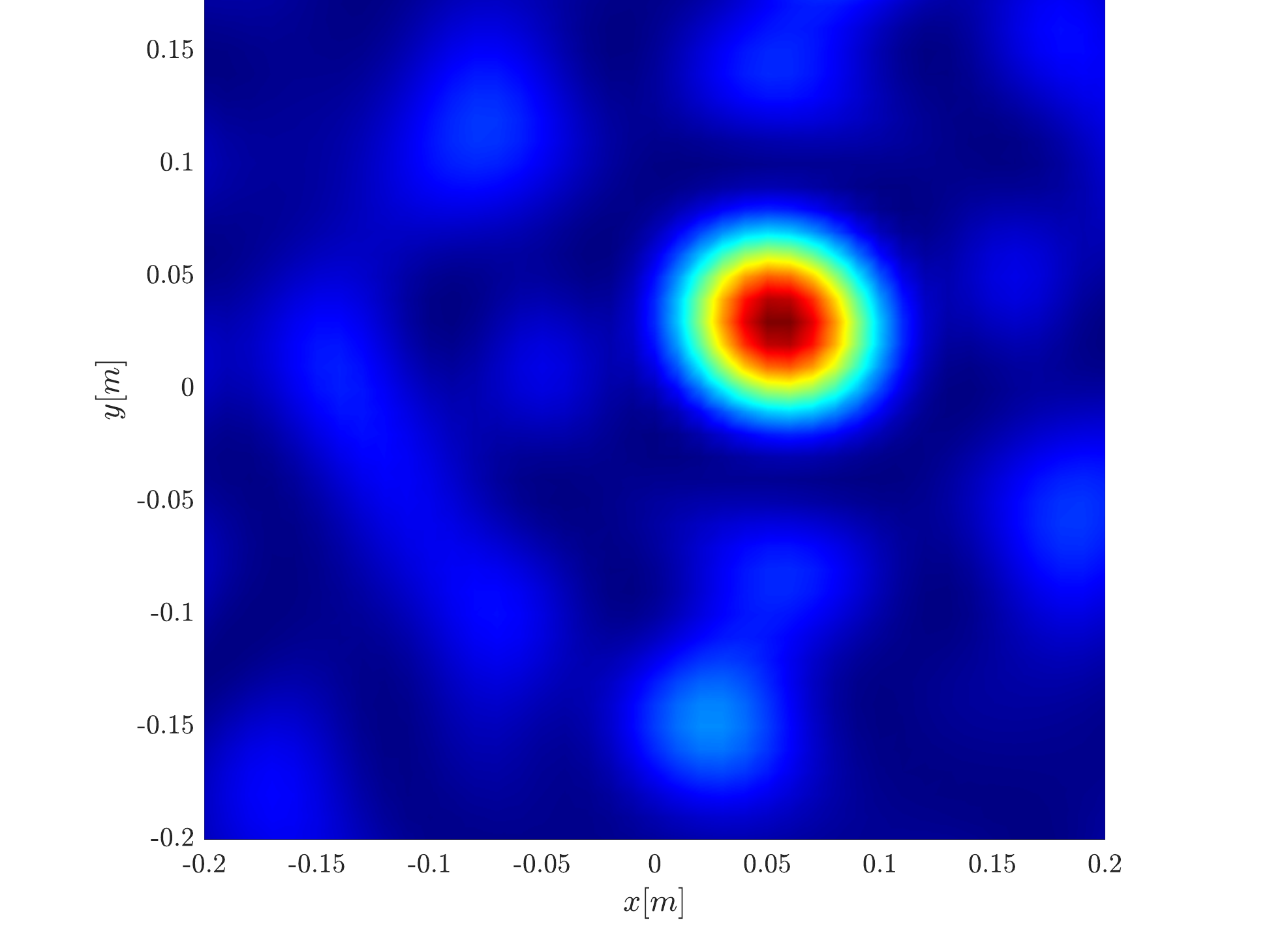}
	\end{subfigure}
	
	\begin{subfigure}[t]{0.18\textwidth}
		\includegraphics[width=\textwidth]{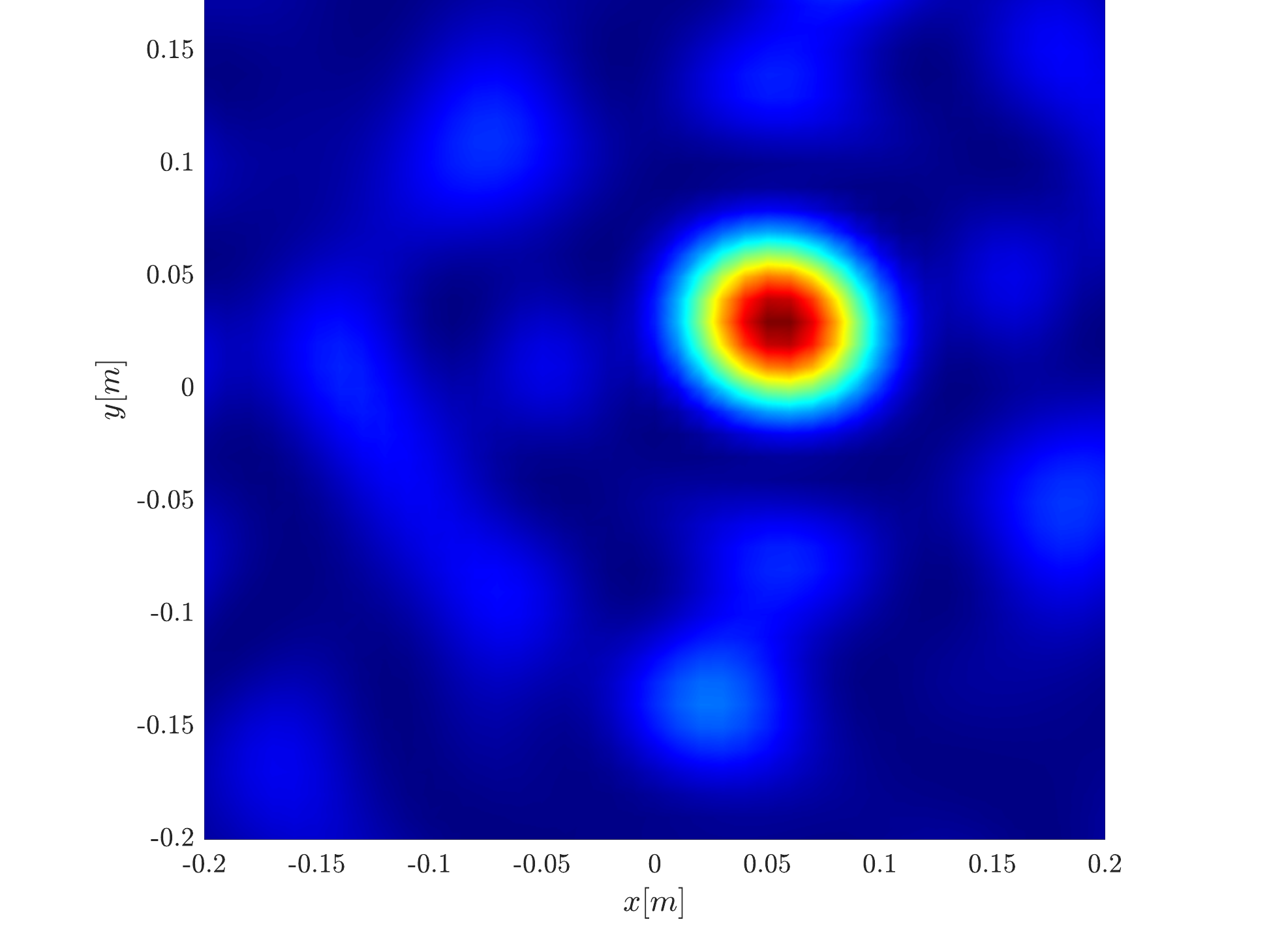}
	\end{subfigure}
	\begin{subfigure}[t]{0.18\textwidth}
		\includegraphics[width=\textwidth]{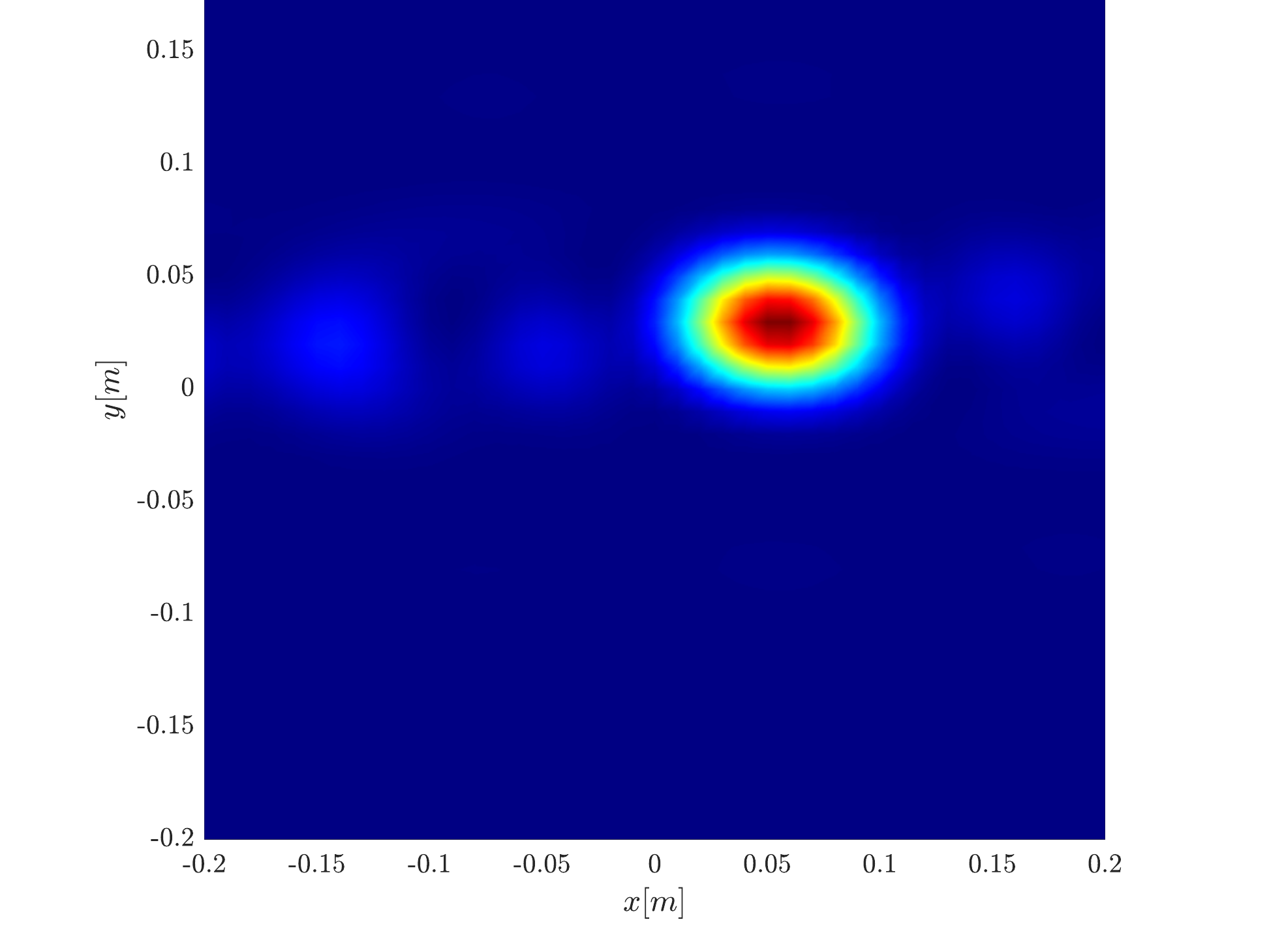}
	\end{subfigure}
	\begin{subfigure}[t]{0.18\textwidth}
		\includegraphics[width=\textwidth]{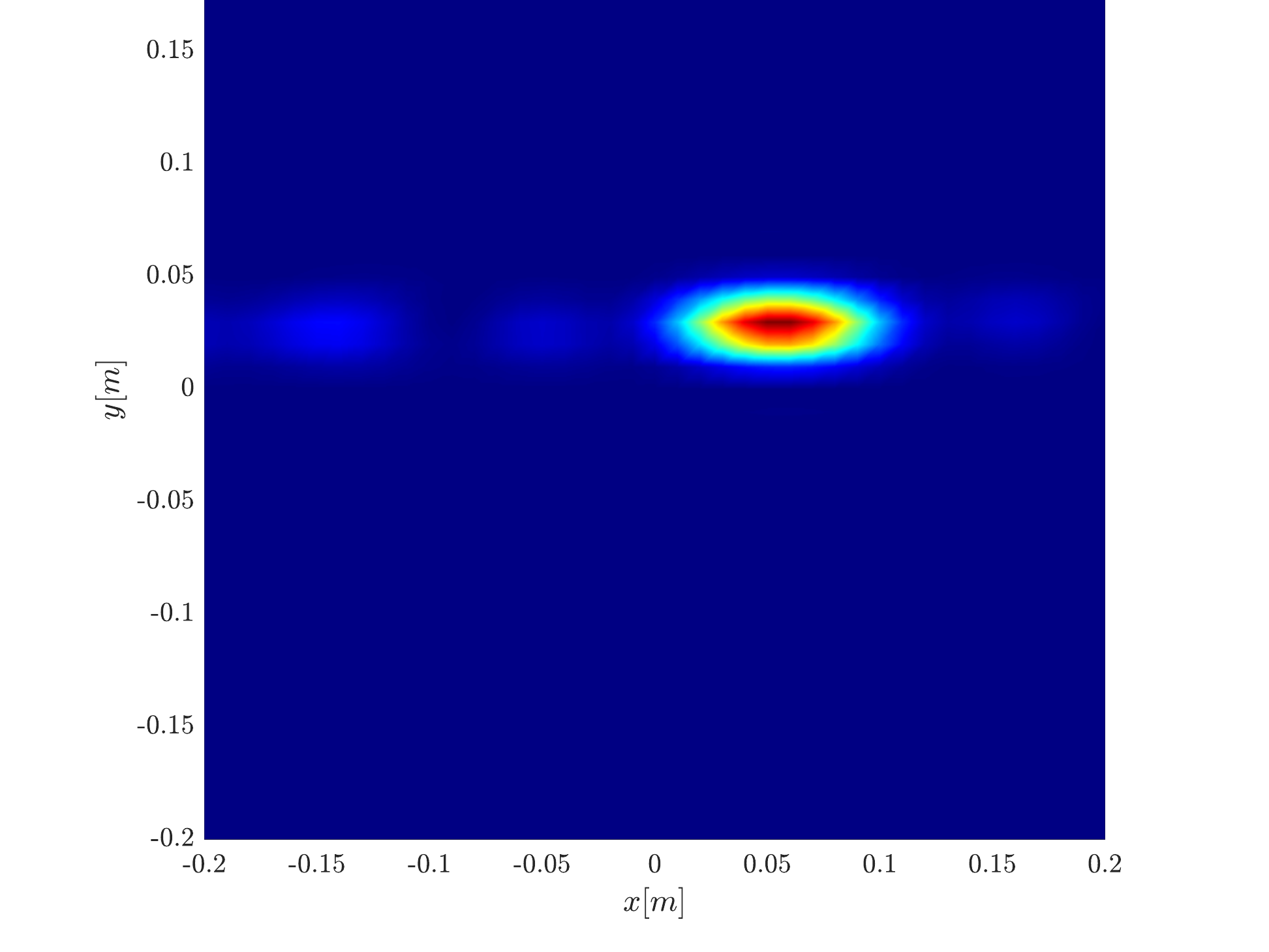}
	\end{subfigure}
	
	\begin{subfigure}[t]{0.18\textwidth}
		\includegraphics[width=\textwidth]{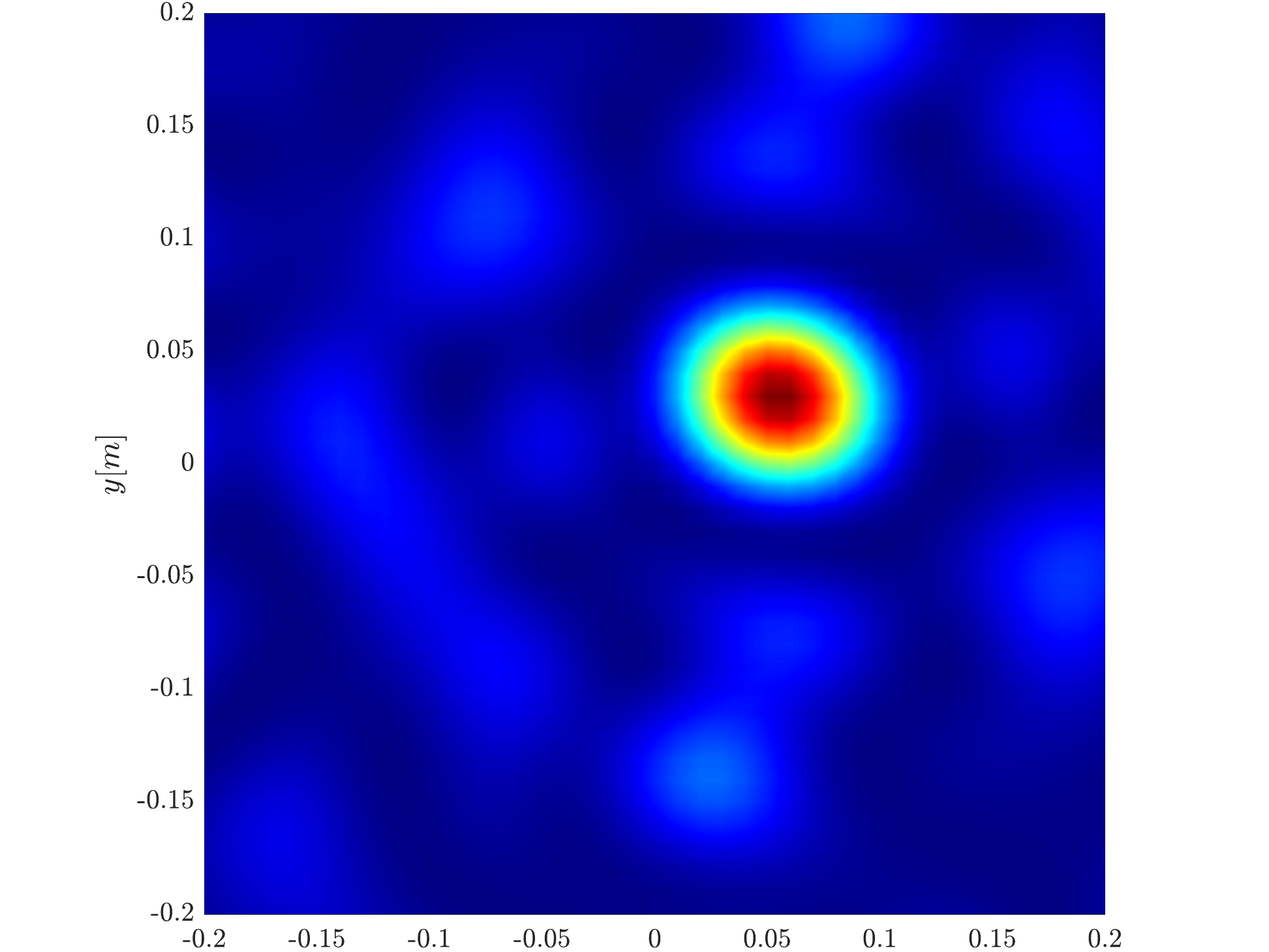}
		\caption{}
	\end{subfigure}
	\begin{subfigure}[t]{0.18\textwidth}
		\includegraphics[width=\textwidth]{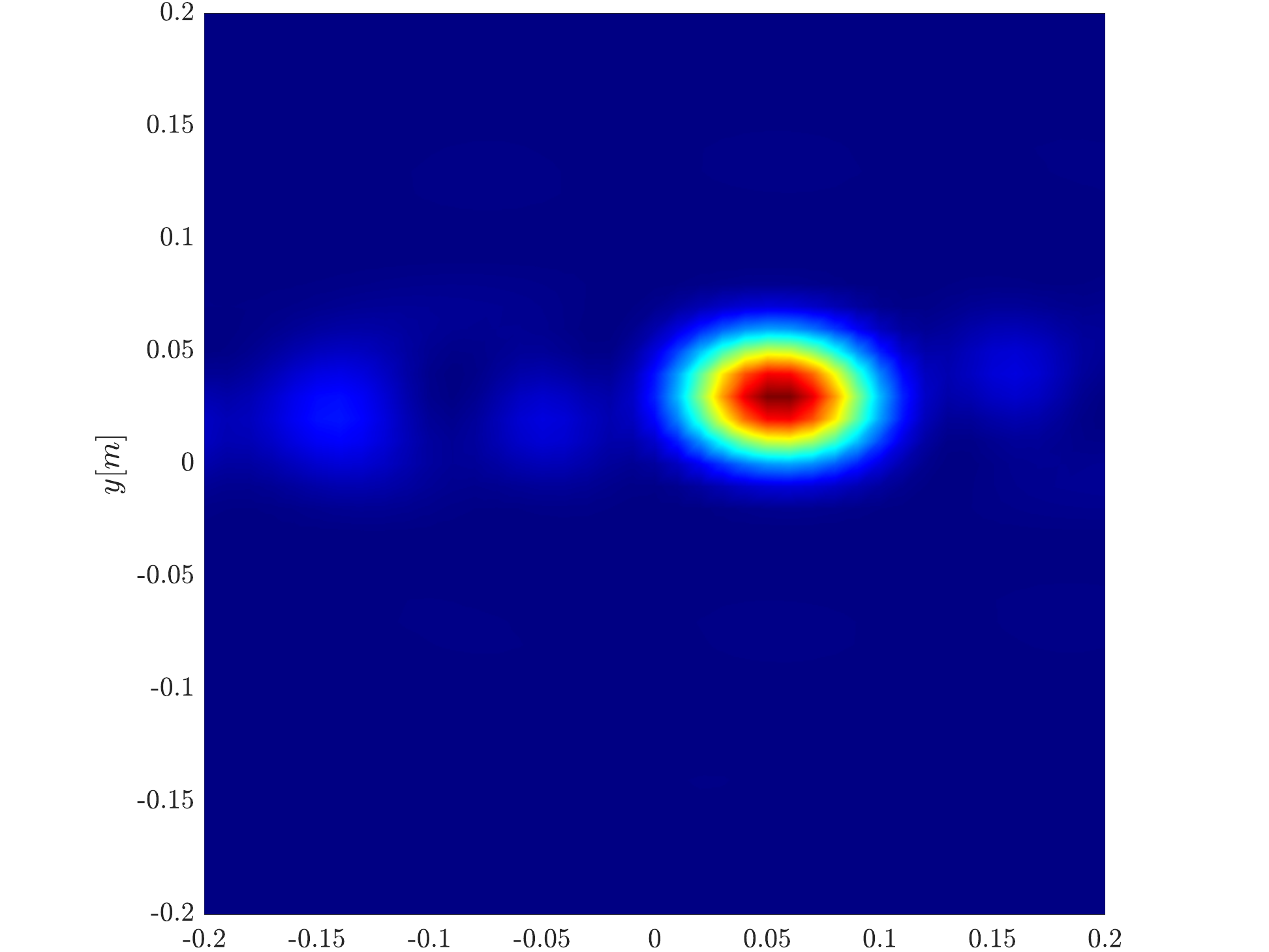}
		\caption{}
	\end{subfigure}
	\begin{subfigure}[t]{0.18\textwidth}
		\includegraphics[width=\textwidth]{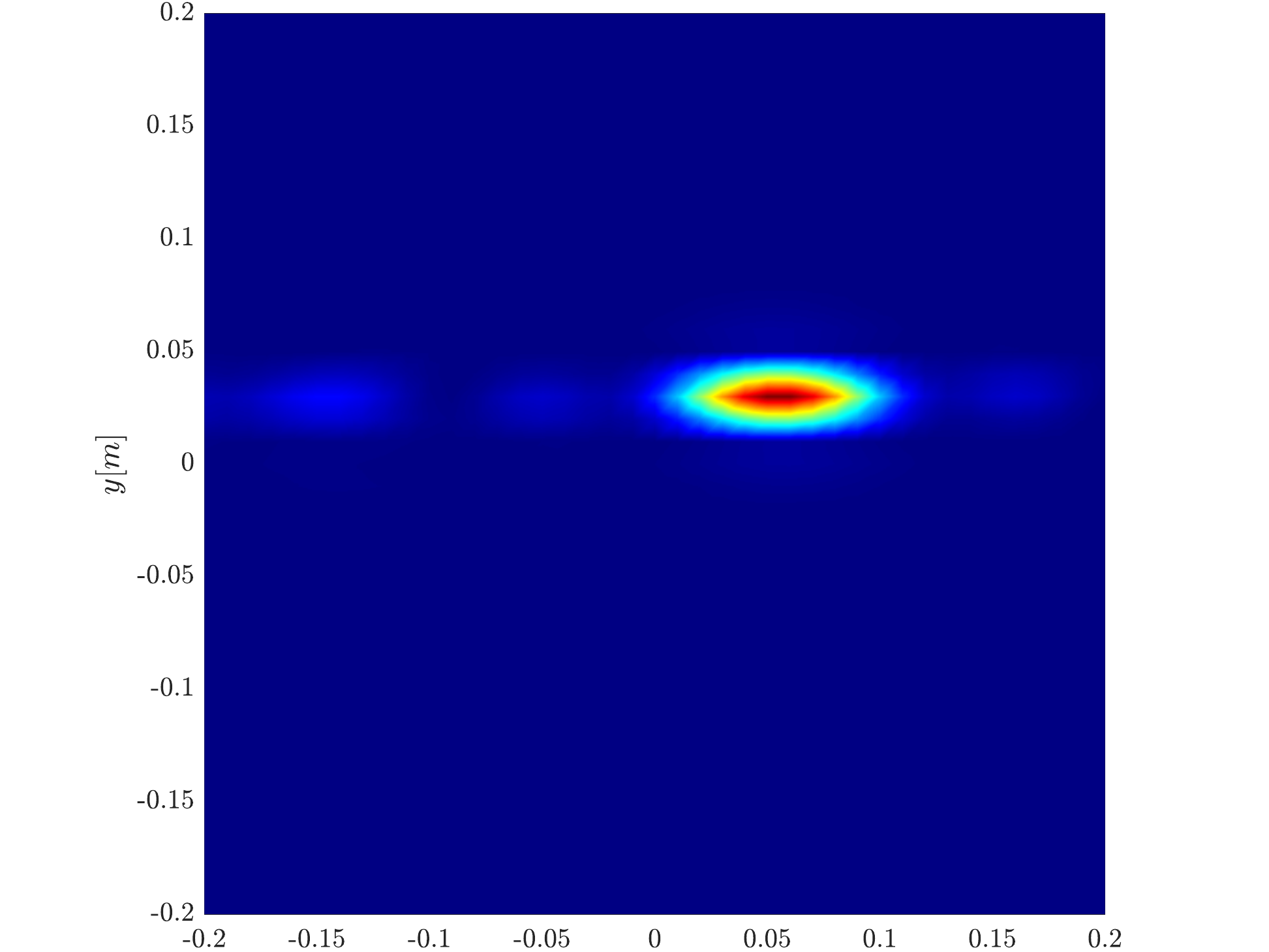}
		\caption{}
	\end{subfigure}
	\caption{Comparison of single point migration (top row)  rank-1 image (middle row) and linear Kirchhoff migration (bottom row), for increasing synthetic aperture size of $(a)$ $100\Delta s$,$(b)$ $1000\Delta s$ and $(c)$ $3000\Delta s$. We can see that as the synthetic aperture increases there is a dramatic improvement in the resolution attained by the rank-1 image compared to the single point one, which is only marginally improved by the increased synthetic aperture. The rank-1 image resolution is comparable to the Kirchhoff migration.}
	\label{fig:eig_v_diag}
\end{figure}

We first repeat the results of Figure~\ref{fig:eig_v_diag_double} for a single scatterer, illustrated in Figure~\ref{fig:eig_v_diag}. We observe the same behavior, namely the resolution increases with the synthetic aperture size for the rank-1 image, greatly improving on the classical migration when the synthetic aperture is large enough.

\begin{figure}[htbp]
	\centering
	\begin{subfigure}[t]{0.68\textwidth}
		\includegraphics[width=\textwidth]{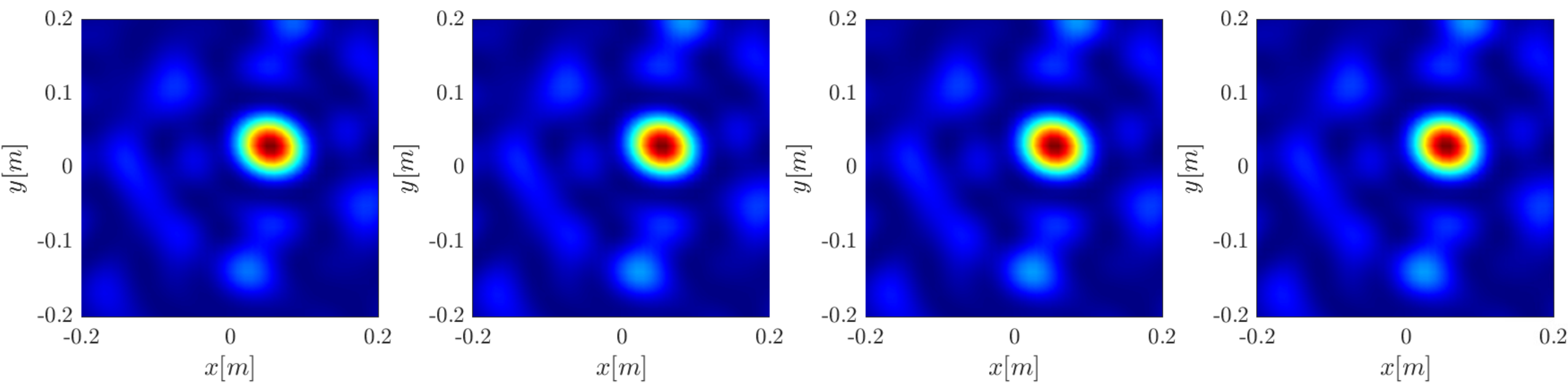}
		\caption{}
	\end{subfigure}
	%
	
	\begin{subfigure}[t]{0.68\textwidth}
		\includegraphics[width=\textwidth]{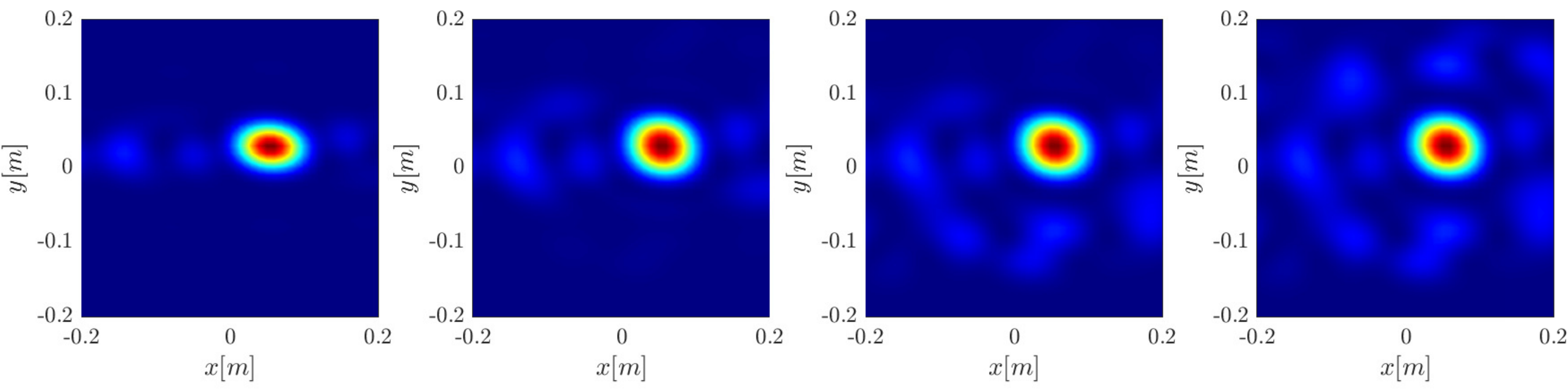}
		\caption{}
	\end{subfigure}
	%
	
	\begin{subfigure}[t]{0.68\textwidth}
		\includegraphics[width=\textwidth]{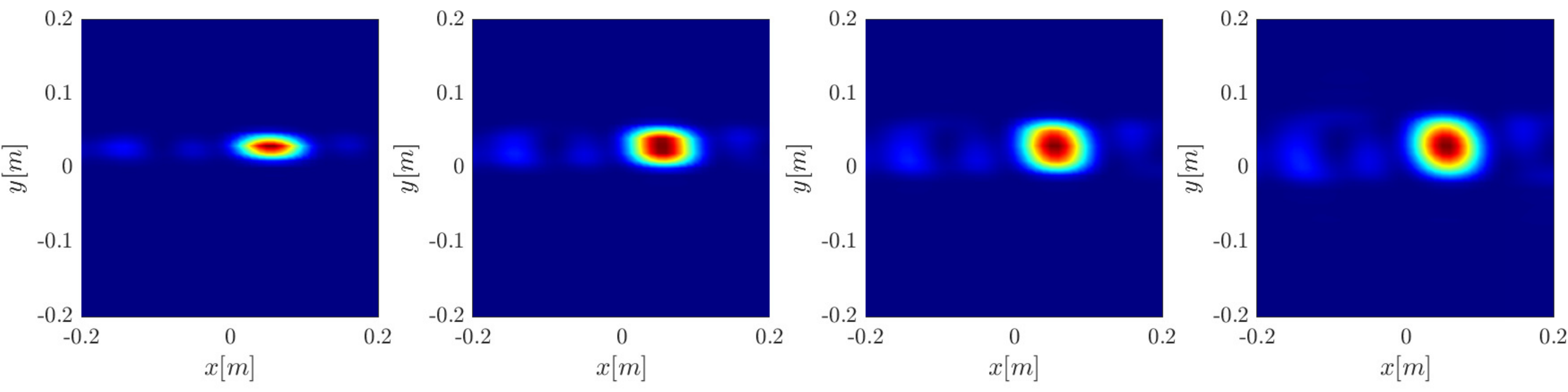}
		\caption{}
	\end{subfigure}
	
	\caption{Comparison of the cumulative sum of weighted eigenvectors, $\sum\limits_{i=1}^m\lambda_i |\mb v_i(\tilde{\mb X})\hspace{0.01em}_k|^2$, for $m=1,\dots,4$, for increasing synthetic aperture size of $(a)$ $100\Delta s$,$(b)$ $1000\Delta s$ and $(c)$ $3000\Delta s$. We can see that for small synthetic aperture size the image converges quickly to a stationary image which greatly resembles the single point migration result of Figure \ref{fig:eig_v_diag}. This suggests a rapid decay in eigenvalues. As the inverse aperture size increases, the convergence is much slower, which indicates $\tilde{\mb X}$ is far from being rank-1, suggesting a greater difference between the single point and the two-point migration schemes.}
	\label{fig:single_comp}
\end{figure}

\begin{figure}
	\centering
	\begin{subfigure}[t]{0.75\textwidth}
		\includegraphics[width=\textwidth]{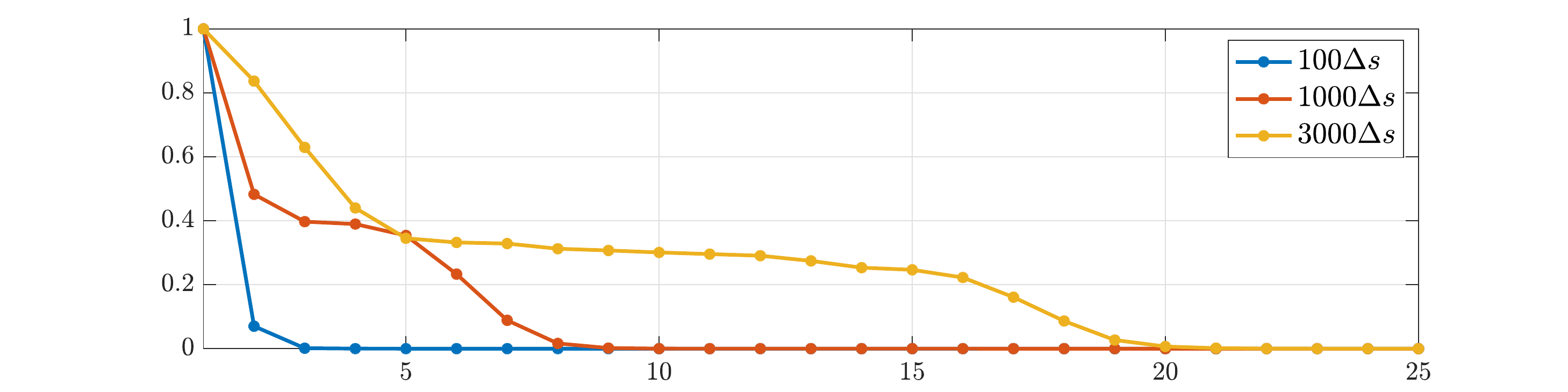}
	\end{subfigure}
	\caption{Top 25 eigenvalues of $\tilde{\mb X}$ for a single scatterer, for different synthetic aperture sizes, normalized such that the top eigenvalue is always 1. We can see that as the synthetic aperture size increases, so does the number of significant eigenvalues.}
	\label{fig:single_svd}
\end{figure}

In order to further investigate the cause for the improved resolution we use \eqref{eq:eig_to_diag}, that is the single point migration is the weighted sum of all the eigenvectors of $\tilde{\mb X}$ squared, weighted by their eigenvalue. We plot in Figure~\ref{fig:single_comp} the cumulative sum $\sum\limits_{i=1}^m\lambda_i |\mb v_i(\tilde{\mb X})\hspace{0.01em}_k|^2$, for $m=1,\dots,4$ for the different inverse aperture sizes. The results motivate us to look into the eigenvalues 
of $\tilde{\mb X}$ illustrated in Figure~\ref{fig:single_svd}. We indeed see that as the inverse aperture increases, $\tilde{\mb X}$, has more significant eigenvalues.

Stationary phase analysis of the two point migration function, carried out in Appendix~\ref{app:stat_phase}, suggest that the stationary points of $\mathcal{I}(\mb y_k,\mb y_{k'})$ are all the points $(\mb y_k,\mb y_{k'})=(\mb y^t_i,\mb y^t_j)$, where $\mb y^t$ is the collection of all scatterer locations. In the single scatterer case we would have only one stationary point. We analyze the behavior of $\tilde{\mb X}$ around the stationary point, to see in greater detail the effect of the synthetic aperture. Since for a 2D imaging domain $\mathcal{I}(\mb y_k,\mb y_{k'})$ would be four dimensional, we look at all the possible planar cross sections of $\mathcal{I}(\mb y_k,\mb y_{k'})$ at the stationary point- 6 in total. Since $\tilde{\mb X}$ is Hermitian, there are only 4 distinct cross sections (up to conjugation). 

\begin{figure}[htbp]
	\centering
	\begin{subfigure}[t]{0.72\textwidth}
		\includegraphics[width=\textwidth]{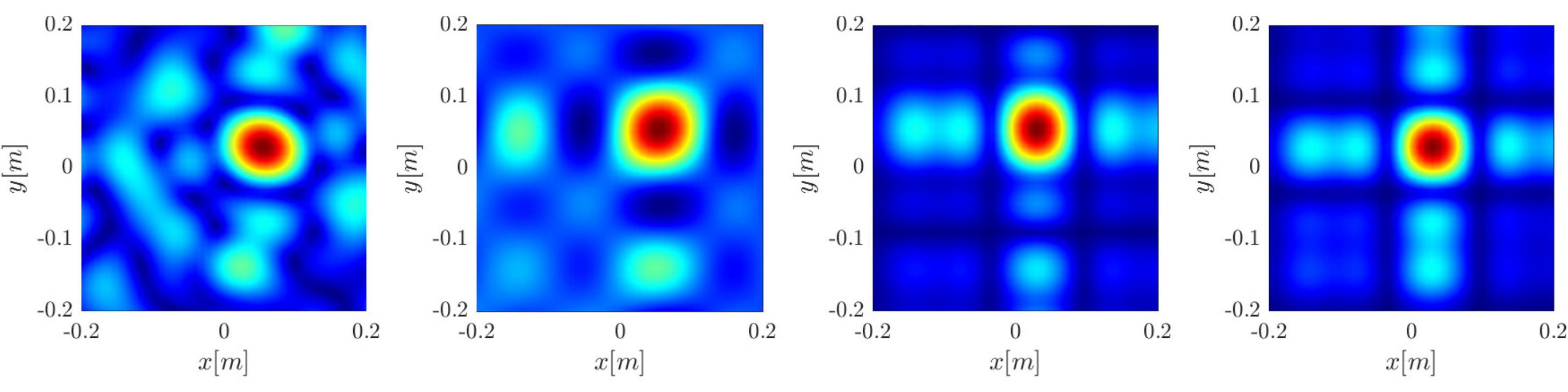}
		\caption{}
	\end{subfigure}
	%
	
	\begin{subfigure}[t]{0.72\textwidth}
		\includegraphics[width=\textwidth]{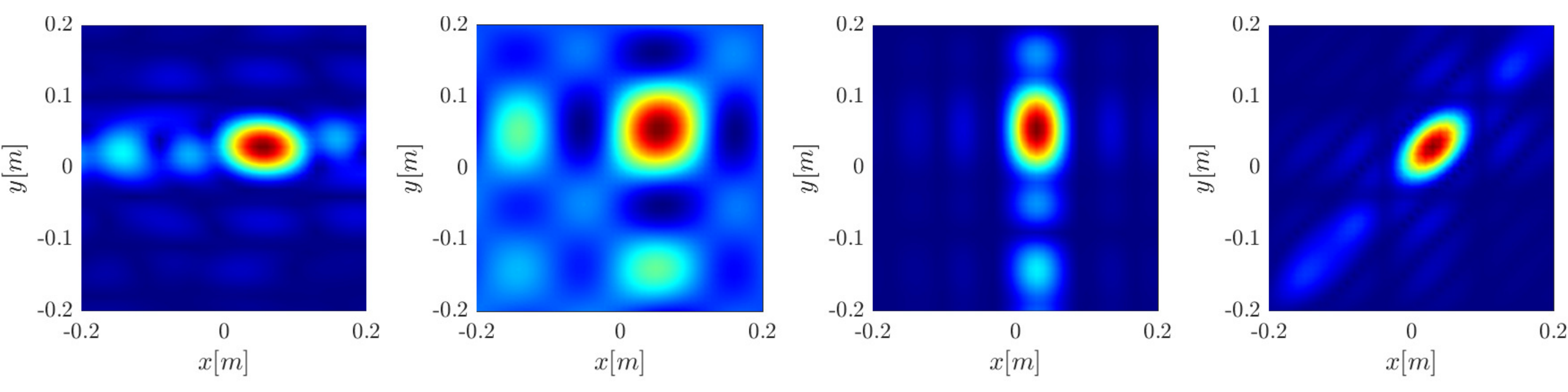}
		\caption{}
	\end{subfigure}
	%
	
	\begin{subfigure}[t]{0.72\textwidth}
		\includegraphics[width=\textwidth]{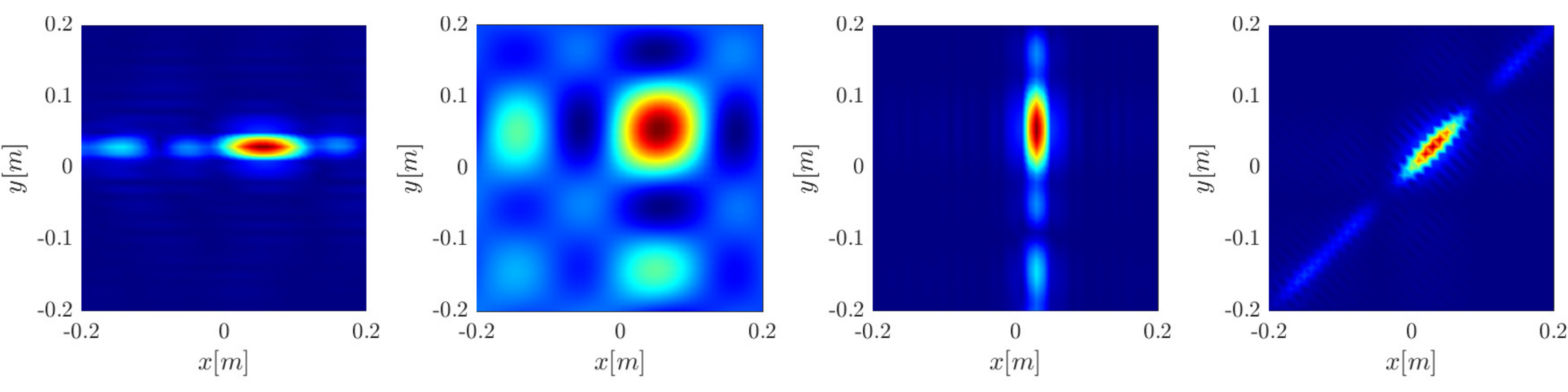}
		\caption{}
	\end{subfigure}
	\caption{Cross -sections of $\tilde{\mb X}$ around the stationary phase point $\mb y$ for different inverse aperture sizes. We use a verbose notation $\mathcal{I}(y_{k,1},y_{k,2}, y_{k',1}, y_{k',2})$. They are ordered (left to right): $\mathcal{I}(y_1,y_2,\cdot,\cdot ),\mathcal{I}(\cdot,y_2,\cdot,y_2 ),\mathcal{I}(y_1,\cdot,\cdot,y_2 ),\mathcal{I}(y_1,\cdot,y_1,\cdot )$. The synthetic aperture size is $(a)$ $100\Delta s$,$(b)$ $1000\Delta s$, $(c)$ $3000\Delta s$. We can see that as the synthetic aperture size increases the stationary point becomes anisotropic in the sense the the main lobe has different width with respect to different directions. Specifically the directions with the weakest decay are the ones for which $\hat{\mb y}_k=\hat{\mb y}_{k'}$- see the diagonal of the second and third columns. This is the direction along which the single point migration is computed.}
	\label{fig:single_cross}
\end{figure}

As illustrated in Figure~\ref{fig:single_cross}, as the synthetic aperture size increases the stationary point becomes anisotropic in the sense the the main lobe has different width with respect to different directions. Specifically, the directions with the weakest decay are the ones for which $\hat{\mb y}_k=\hat{\mb y}_{k'}$, that is also the direction along which the single point migration is computed.  In Appendix~\ref{app:stat_phase}, we show by stationary phase analysis that this anisotropy is the result of the synthetic aperture, and indeed we expect anisotropy to increase with the size of the synthetic aperture. By taking the top eigenvector rather than the diagonal of $\tilde{\mb X}$ we can gain from the anisotropy and reduce the spot size as we show in Appendix~\ref{app:kernel}. 

Anisotropy is also related to the observed increase in rank. If the stationary point is isotropic, it is close to being separable with respect to the two indices, and hence having rank 1. Anisotropy ensures this is no longer the case.

\begin{figure}[htbp]
	\centering
	\begin{subfigure}[t]{0.68\textwidth}
		\includegraphics[width=\textwidth]{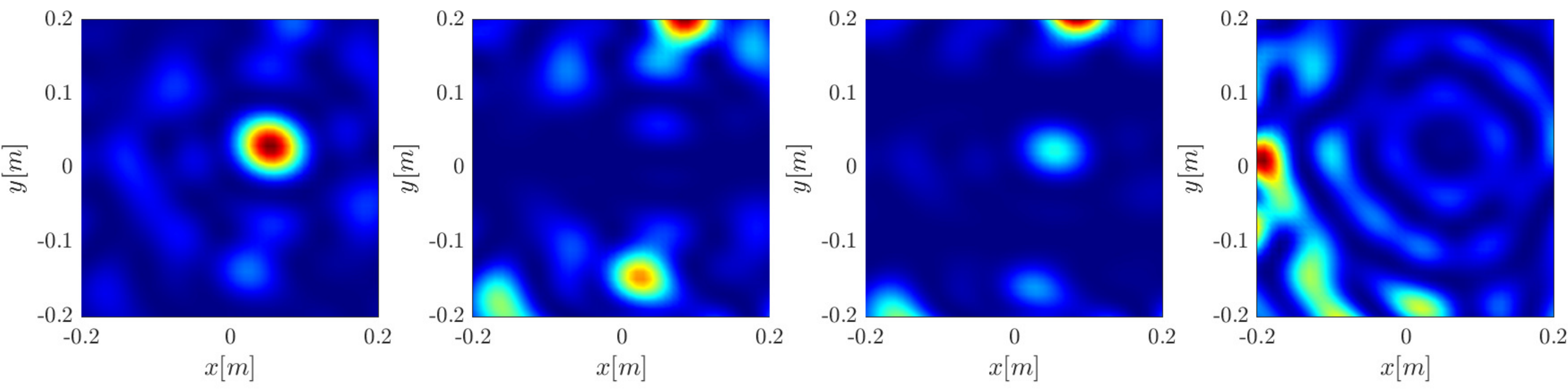}
		\caption{}
	\end{subfigure}
	%
	
	\begin{subfigure}[t]{0.68\textwidth}
		\includegraphics[width=\textwidth]{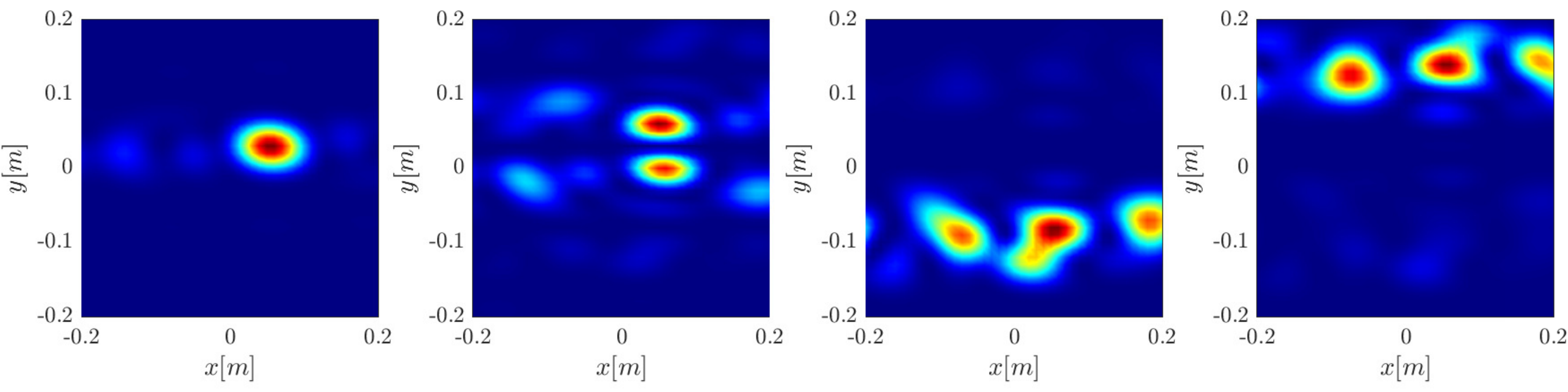}
		\caption{}
	\end{subfigure}
	%
	
	\begin{subfigure}[t]{0.68\textwidth}
		\includegraphics[width=\textwidth]{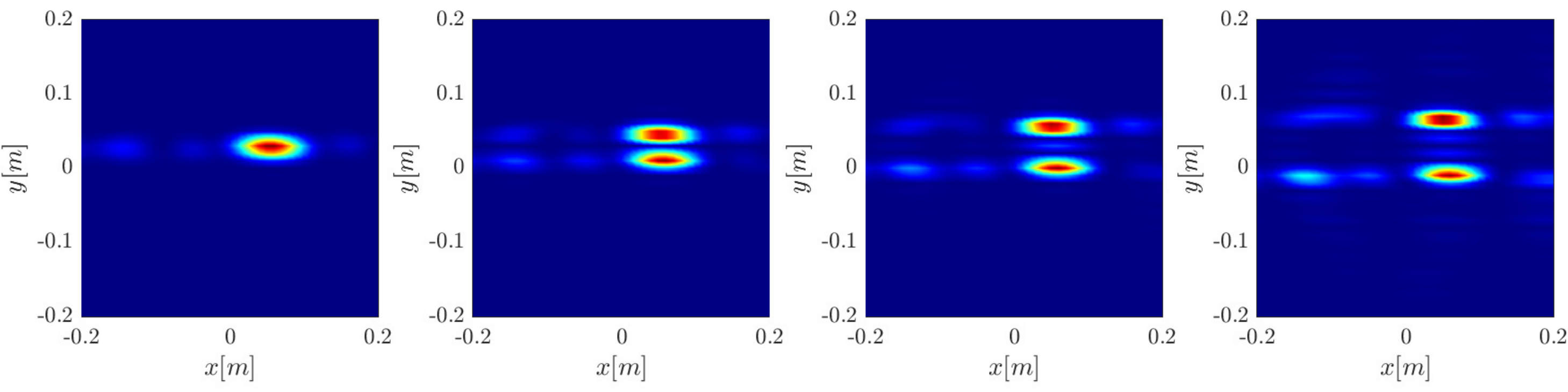}
		\caption{}
	\end{subfigure}
	\caption{Comparison of the first four eigenvectors, $|\mb v_i(\tilde{\mb X})\hspace{0.01em}_k|^2$, for $i=1,\dots,4$, for increasing synthetic aperture  $(a)$ $100\Delta s$,$(b)$ $1000\Delta s$, $(c)$ $3000\Delta s$. We can see that as the first eigenvector becomes more localized, other localized  eigenvectors appear, in tandem with a slower decay in the eigenvalues as illustrated in Figure~\ref{fig:single_svd}. Thus, there is a greater difference in the resolutions achieved by single and two point migrations.}
	\label{fig:single_modes_separate}
\end{figure}

These results can also be interpreted in light of \eqref{eq:svd_opt}. We saw that the top eigenvector solves an optimization problem. The second eigenvector would solve the same optimization problem under the added constraint of orthogonality to the first eigenvector. As the first eigenvector becomes more localized, there might be higher modes who are localized as well, and whose eigenvalues do not decay rapidly. The orthogonality constraint means that they cannot have the same support as the first eigenvector. As a result summing them all would increase the spot size, which is the observed effect in the single point migration limit. This is indeed the case, as illustrated in Figure~\ref{fig:single_modes_separate}.

We close this section by summarizing our observations 
\begin{enumerate}
	\item For a point scatterer model, the interference pattern  $\tilde{\mb X}$ has localized stationary points at all the possible combinations $(\mb y_i,\mb y_j)$ where $\mb y_i,\mb y_j$ are positions of the scatterers.
	\item As the synthetic aperture increases the spot size around the stationary point develops an anisotropy. The directions of weakest decay are the ones that match the single point migration. This is shown both numerically and analytically in Appendix~\ref{app:stat_phase}.
	\item The anisotropy causes the top eigenvector to be more localized, as the spot size is some mean of the spot width in the different directions, as shown in Appendix~\ref{app:kernel}.
	\item Anisotropy also increases the rank, generating more significant localized eigenvectors, and hence accentuating the difference between the two point and single point migrations.
\end{enumerate}
\section{Summary and Conclusions}
We have considered the problem of imaging small fast moving objects, such as satellites in low earth orbit, using several receivers flying above the turbulent atmosphere, and asynchronous sources located on the ground. The resolution of the imaging system depends on the aperture spanned by the receivers, the bandwidth and the central frequency of the probing signals, as well as the inverse synthetic aperture created by the fast moving objects. In Section~\ref{sec:corr_imaging} we have shown that by forming the cross-correlations of the recorded signals and compensating for the Doppler effect in small imaging windows, we can create a cross-correlation data structure $C_{\mb R\mb R'}(s,\tau)$ that can be used for imaging. This data structure suggests a natural extension to linear Kirchhoff migration imaging, which is a two-point interference pattern $\mathcal{I}^{GCC}(\mb y_k,\mb y_{k'})$ defined by \eqref{eq:2_point_mat_imag}. 

In Section~\ref{sec:generalized_migration} we have considered several ways of extracting an image from the interference pattern. The usual single point migration image $\mathcal{I}^{CC}(\mb y_k)$ is the diagonal $\mathcal{I}^{GCC}(\mb y_k,\mb y_{k})$. The alternative that we have introduced here
is the rank-1 image $\mathcal{I}^{R1CC}(\mb y_k)$, which is the first eigenvector of the interference pattern. This rank-1 image has superior resolution compared to the single point migration. 
The improvement in resolution can be explained by studying the structure of the stationary points of the interference pattern. We have shown in Appendix~\ref{app:stat_phase} and Appendix~\ref{app:kernel} that the inverse synthetic aperture induces anisotropy in the profile of the point spread function.  
The effect is weakest in the direction that corresponds to the single point migration image, causing the image resolution to depend weakly on the synthetic aperture size. Taking the leading eigenvector as the image exploits the anisotropy and provides a resolution comparable to the linear KM system, even when considering multiple closely spaced targets. This is supported by extensive numerical simulations in Sections~\ref{sec:numerical_simulations} and \ref{sec:prop_filt}. We have noted in Section~\ref{subsec:rank_1_image} that the interference pattern $\mathcal{I}^{GCC}(\mb y_k,\mb y_{k'})$ need not be a square matrix, with numerical simulations suggesting that one dimension can be downsampled by a up to a factor of 10 while still retaining the same resolution. Instead of the eigenvector, one then uses the top singular vector as the rank-1 image. We have also shown in Section~\ref{sec:numerical_simulations} the robustness of the algorithm to additive noise.

An important effect to be considered in future work is the rotation of the object, which is a more realistic scenario for satellites. Further work will involve a systematic study of the robustness of the proposed rank-1 image to measurement noise, as well as to the effects of atmospheric turbulence. 
\label{sec:summary}

\section{Acknowledgements}
The work of M. Leibovich and G. Papanicolaou was partially supported by AFOSR FA9550-18-1-0519. The work of C. Tsogka was partially supported by AFOSR FA9550-17-1-0238 and AFOSR FA9550-18-1-0519.

\appendix
\setcounter{equation}{0}
\renewcommand{\theequation}{\Alph{section}.\arabic{equation}}
\section{The direct and scattered waves}
\label{app:forward_wave}
The derivation follows the one in \cite{fournier2017matched}. We give it here for completeness.
We assume a source located at  $\mb x_\mb E$ emits a short pulse $f(t)$ whose compact support is in $(0,\infty)$ (or any interval of the form $[-T,\infty], T<\infty$)

Even though in  electromagnetic waves solve a Maxwell's equation rather than the wave equation, the scalar wave equation captures the main propagation and scattering effects which are of interest, neglecting polarization effects \cite{cheney}.   

The scalar wave $u(t,\mb x)$  solves the wave equation excited by the source
	\begin{equation}
	\frac{1}{c^2(t,\mb x)} \frac{\partial^2 u}{\partial t^2}-\Delta u = f(t)\delta(\mb x-\mb x_\mb E).
	\label{eq:waves}
	\end{equation}
	
	The velocity model assumes a uniform background and a localized perturbation $\varrho_\mb T$ centered at $\mb x_\mb T(t)$. We use here the slow/fast time representation so that $\mb x_\mb T(t)=\mb x _\mb T +s\mb v_\mb T+t s\mb v_\mb T$, and the perturbation is
	$$
	\frac{1}{c^2(t,\mb x)} = \frac{1}{c_0^2} \Big(1 + \varrho_T \big(\mb x-\mb x_\mb T(t)\big) \Big).
	$$

	The scattered field, $u^{(1)}(t,\mb x)$, is defined as the difference between the total field $u(t,\mb x)$, solution of \eqref{eq:waves} and the incident field $u^{(0)}(t,\mb x)$. 
	\begin{equation}
	u^{(1)}(t,\mb x)=u(t,\mb x)-u^{(0)}(t,\mb x).
	\end{equation}
$u^{(0)}(t,\mb x)$ solves the free space wave equation
	\begin{equation}
	\frac{1}{c^2_0} \frac{\partial^2 u^{(0)}}{\partial t^2}-\Delta u^{(0)} = f(t)\delta(\mb x-\mb x_E).
	\label{eq:waves0}
	\end{equation}
	The incident field has the form
	\begin{equation}
	u^{(0)}(t,\mb x) = \int_{0}^t d\tau f(\tau) G(t-\tau,\mb x,\mb x_\mb E),
	\label{eq:ud1} 
	\end{equation}
	where $G(t,\mb x,\mb x')$ is the free space Green's function, given by
	\begin{equation}
	G(t,\mb x,\mb x') = \frac{\delta \Big( t - \frac{ |\mb x -\mb x'| }{c_0}\Big)}{4 \pi \left| \mb x -\mb x'\right|} .
	\label{eq:green} 
	\end{equation}
	Plugging in Green's function in \eqref{eq:ud1} we obtain,
	\begin{equation}
	u^{(0)}(t,\mb x)  = \frac{1}{4 \pi \left| \mb x -\mb x_\mb E \right|} f\Big( t - \frac{ |\mb x -\mb x_\mb E|}{c_0}\Big).
	\label{eq:ud2} 
	\end{equation}
	Thus, the scattered field solves,
	$$
	\frac{1}{c^2_0} \frac{\partial^2 u^{(1)}}{\partial t^2}-\Delta u^{(1)} = 
	-\frac{1}{c^2_0}\varrho_T (\mb x-\mb x_\mb T(t)) \frac{\partial^2 (u^{(0)}+u^{(1)})}{\partial t^2} (t,\mb x),
	$$
	so that
	$$
	u^{(1)}(t,\mb x) = - \frac{1}{c_0^2} \int_0^t d\tau  \int d\mb y  G(t-\tau,\mb x,\mb y) \varrho_T(\mb y-\mb x_\mb T(\tau))
	\frac{\partial^2}{\partial \tau^2}  (u^{(0)}+u^{(1)})(\tau,\mb y) .
	$$
	In the Born approximation, we can neglect $u^{(1)}$ as an effective source term in the integral, as its contributions would be quadratic in $\varrho_T$, assumed to be small, so that
	$$
	u^{(1)}(t,\mb x) = - \frac{1}{c_0^2} \int_0^t d\tau  \int d\mb y  G(t-\tau,\mb x,\mb y) \varrho_T(\mb y-\mb x_\mb T(\tau))
	\frac{\partial^2}{\partial \tau^2} u^{(0)}(\tau,\mb y) .
	$$
	If $\varrho_T(\mb x)= \rho \delta(\mb x)$, i.e. a point- like scatterer, we get
	$$
	u^{(1)}(t,\mb x) = - \frac{\rho}{c_0^2} \int_0^t d\tau   G(t-\tau,\mb x, \mb x_\mb T(\tau))
	\frac{\partial^2}{\partial \tau^2} u^{(0)}(\tau,\mb y) \mid_{\mb y=\mb x_\mb T(\tau)}  .
	$$
	Substituting the expression of $u^{(0)}$ and integrating by parts twice, we obtain:
	$$
	u^{(1)}(t,\mb x) = - \frac{\rho}{c_0^2} \int_0^t d\tau  \int_0^\tau d\tau' f''(\tau')  
	G(\tau-\tau',\mb x_\mb T(\tau),\mb x_\mb E) G(t-\tau,\mb x,\mb x_\mb T(\tau))  .
	$$
	Therefore the scattered field recorded by the receiver at $\mb x=\mb x_\mb R$, when the target is at $\mb x _\mb T(s)$, $u_\mb R (s,t)= u^{(1)}(s+t,\mb x_\mb R)$ has the form:
\begin{equation}
	u_{\mb R}(s,t)  =-\frac{\rho}{c_0^2} \int_0^{t} d\tau 
	\frac{1}{4 \pi |\mb x_\mb T(s+\tau) -\mb x_\mb E|}  {f}'' \Big( \tau- \frac{|\mb x_\mb T(s+\tau) - \mb x_\mb E|}{c_0}\Big) \frac{1}{ 4 \pi  |  \mb x_\mb R- \mb x_\mb T(s+\tau) |} 
	\delta \Big( t- \tau - \frac{| \mb x_\mb R -\mb x_\mb T(s+\tau) |}{c_o}\Big) .
\end{equation}
Introduce
$$
\Phi(\tau;t) =  t -\tau - \frac{ \left| \mb x_\mb T+s \mb v_\mb T+\tau \mb v_\mb T -\mb x_\mb R  \right|}{c_0} ,
$$
then  we have
$$
\delta\left[ \Phi(\tau;t) \right] = \sum\limits_i \frac{\delta[ \tau-\tau_i(t)]}{\left| \partial_\tau \Phi(\tau_i(t);t) \right|}, \quad \Phi(\tau_i(t);t)=0 ,\tau_i(t)\le t .
$$
Denoting
\begin{equation}
\bD(t) = \mb x_\mb T+s\mb v_\mb T - \mb x_\mb R + t \mb v _\mb T , 
\label{eq:by} 
\end{equation}
then we can write
	\begin{equation}
	\Phi(\tau;t) =  t -\tau - \frac{ \left| \bD(t) -(t- \tau) \mb v_\mb T  \right|}{c_0} .
	\end{equation}

	The unique zero of $\Phi(\tau;t)$ in $(0,t)$, can  is a root of the quadratic equation 
	$$\dsp ( t -\tau)^2 \left( 1 - \frac{| \mb v_\mb T|^2}{c_0^2} \right) + 2(t-\tau) \frac{\mb v_\mb T}{c_0} \cdot \frac{\bD(t)}{c_0}-\frac{|\bD(t)|^2}{c_0^2} =0 .$$

Which is given by
\begin{equation}
\tau(t) = t - \frac{|\bD(t)|}{c_0 \big(1- \big| \frac{\mb v_\mb T}{c_0}\big|^2 \big) } \left[ \sqrt{ 1 - \left| \frac{\mb v_\mb T}{c_0}\right|^2 + \left(\frac{\mb v_\mb T}{c_0} \cdot \frac{\bD(t)}{|\bD(t)|} \right)^2} - \frac{\mb v_\mb T}{c_0} \cdot \frac{\bD(t)}{|\bD(t)|} \right].
\label{eq:tau} 
\end{equation}

We also have
$$
\left| \partial_\tau \Phi(\tau(t);t) \right| = \left| 1 + \frac{\mb v_\mb T}{c_0} \cdot  \frac{\bD( \tau(t))} {| \bD( \tau(t))  |} \right|   .
$$
Hence, we obtain
\begin{equation}
u_\mb R(s,t) = - \frac{\rho    f''\Big( s+\tau(t) - \frac{|\mb x_\mb T(\tau(t)) - \mb x_\mb E|}{c_0} \Big) }{(4\pi)^2 c_0^2 |\mb x_\mb T(\tau(t)) - \mb x_\mb E| |\mb x_\mb R - \mb x_\mb T(\tau(t))| \left| 1 + \frac{\mb v_\mb T}{c_0} \cdot \frac{ \bD(\tau(t)) }{| \bD( \tau(t))   | } \right|}  .
\label{eq:ur} 
\end{equation}
This is the Born approximation for the scattered field, which is essentially exact for well separated, localized and weak reflectors.  The expressions in \eqref{eq:scattered_field}, \eqref{eq:gamma_t} are derived by expanding $\tau(t)$ to first order, in $|\mb v_{\mb T}|/c_0$. Thus
\begin{equation}
\tau(t) - \frac{|\mb x_\mb T(\tau(t)) - \mb x_\mb E|}{c_0} \approx \gamma_\mb R (\mb x_\mb T(s),\mb x_\mb E, \mb v_\mb T)t -t_\mb R (\mb x_\mb T(s),\mb x_\mb E, \mb v_\mb T).
\end{equation}
Here $\gamma_\mb R (\mb x_\mb T(s),\mb x_\mb E, \mb v_\mb T)$ and $t_\mb R (\mb x_\mb T(s),\mb x_\mb E, \mb v_\mb T)$ are given by  \eqref{eq:gamma_t}. 

\section{Approximate evaluation of the two point interference pattern}
\label{app:stat_phase}
\setcounter{equation}{0}
In this appendix we analyze the expression \eqref{eq:M_tilde} of the two point interference pattern. The structure of the interference pattern plays a determining role in the resolution of the rank-1 image.

Assume the scatterers are located at $\mb y_i, i=1,\dots,M$ with respect to the center of the image window $\mb x_\mb T(s)$, that is, 
$$ \mb y_i(s)=\mb x_\mb T(s)+\mb y_i$$
and the reflectivities are $\rho_i$. Then, 
\eqref{eq:C_form_element} takes the form
\begin{equation}
\hat{C}_{\mb R \mb R'}(s,\omega)\approx |\xi(\omega,s)|^2\sum\limits_{i,j=1}^M\rho_i\rho_je^{i\omega (t_\mb R^i(s)-t_{\mb R'}^j(s)-(t_\mb R(s)-t_{\mb R'}(s)))} 
\end{equation}
with 
$$
\begin{array}{ll} 
t_\mathbf{R}^{i}(s)& \displaystyle =\frac{|\mb y_i(s) -\mb x_\mathbf{E}|}{c_0}+\frac{|\mb y_i(s) -\mb x_\mathbf{R}|}{c_0}\gamma_{\mathbf{R}}(\mb y_i(s),\mb x_\mathbf{E},\mathbf{v}_T), \\[12pt]
t_\mathbf{R}(s) & \displaystyle=\frac{|\mb  x_\mb T(s) -\mb x_\mathbf{E}|}{c_0}+\frac{|\mb x_\mb T(s) -\mb x_\mathbf{R}|}{c_0}\gamma_{\mathbf{R}}(\mb x_\mb T(s),\mb x_\mathbf{E},\mathbf{v}_T). 
\end{array}
$$
Let us define, 
\begin{equation}
\begin{split}
t_\mathbf{R}^{\mb x}(s)=&\frac{|\mb x_\mb T(s)+\mb x-\mb x_\mathbf{E}|}{c_0}+\frac{| \mb x_\mb T(s)+\mb x-\mb x_\mathbf{R}|}{c_0}\gamma_{\mathbf{R}}(\mb x_\mb T(s) ,\mb x_\mathbf{E},\mathbf{v}_T).\\
\end{split}
\end{equation}
The two point migration translates the cross correlation data to a pair of points $(\mb x,\mb y)$ in the image window, and then sums over all receiver pairs $\mb R,\mb R'$, pulses $s$, and frequencies $\omega$. 
The interference pattern $\tilde{\mb X}$ in \eqref{eq:M_tilde} then has the form
\begin{equation}
\begin{split}
\tilde{\mb X}_{\mb x,\mb y}&=\sum\limits_{s,\omega,\mb R,\mb R'}\hat{\mb C}_{\mb R\mb R'}(s,\omega)e^{i\omega (t_\mb R^\mb x(s)-t_\mb R(s))}e^{-i\omega (t_{\mb R'}^\mb y(s)- t_{\mb R'}(s))}\\
&\approx\sum\limits_{s,\omega,\mb R,\mb R'}  |\xi(\omega,s)|^2\sum\limits_{i,j=1}^M\rho_i\rho_je^{i\omega (t_\mb R^{\mb x}(s)-t_\mb R^i(s)-(t_{\mb R'}^\mb y(s)-t_{\mb R'}^j(s)))}
\end{split}
\end{equation}

\begin{figure}[htbp]
	\centering
	\includegraphics[width=0.5\textwidth]{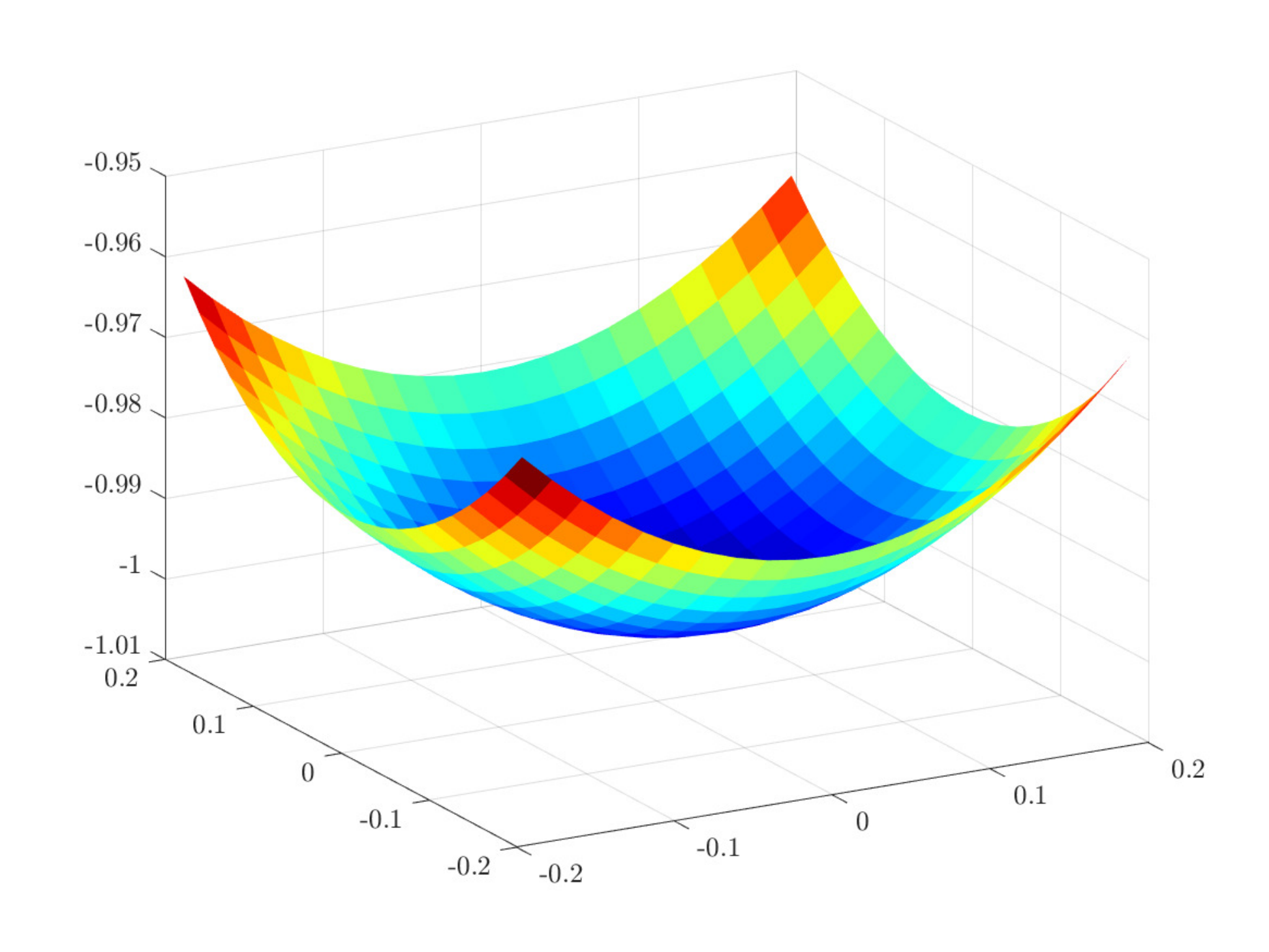}
	\caption{Example of $\mb k_\mb R$ - a sub domain of the unit sphere induced by the positions of the receivers on the ground.}
	\label{fig:k_w}
\end{figure}

We approximate the sum over pulses and frequencies with integrals. The sum over receiver pairs is replaced by a double integral over the physical aperture spanned by the receivers. This is well justified if we assume that the receiver positions are uniformly distributed over an area of $200 \text{km} \times 200\text{km}$ as we do here. By making the change $\sum\limits_{s,\mb R,\mb R'}\rightarrow \int ds \int d\mb x_\mb R\int d\mb x_{\mb R'}$, the interference pattern $\tilde{\mb X}$ has the form
\begin{equation}
\begin{split}
\tilde{\mb X}_{\mb x,\mb y}&\approx\int d\mb x_\mb R d\mb x_{\mb R'} ds d\omega |\xi(\omega,s)|^2\sum\limits_{i,j=1}^M\rho_i\rho_je^{i\omega (t_\mb R^{\mb x}(s)-t_\mb R^i(s)-(t_{\mb R'}^\mb y(s)-t_{\mb R'}^j(s)))}\\&=\sum\limits_{i,j=1}^M\rho_i\rho_j\int d\mb x_\mb R d\mb x_{\mb R'} ds d\omega |\xi(\omega,s)|^2 e^{i\omega (t_\mb R^{\mb x}(s)-t_\mb R^i(s)-(t_{\mb R'}^\mb y(s)-t_{\mb R'}^j(s)))}\\
&=\sum\limits_{i,j=1}^M\rho_i\rho_j\int d\omega \int ds |\xi(\omega,s)|^2 \int d\mb x_\mb R e^{i\omega (t_\mb R^{\mb x}(s)-t_\mb R^i(s))}\int d\mb x_{\mb R'}e^{-i\omega (t_{\mb R'}^\mb y(s)-t_{\mb R'}^j(s))}
\end{split}
\label{eq:M_w}
\end{equation}
We see that the integrals over the physical receiver aperture separate.
We can approximate the travel time difference in the exponent using $|\mb x_\mb T(s)-\mb x_\mb E|,|\mb x_\mb T(s)-\mb x_\mb R|\gg |\mb x|,|\mb y|$. We have to first order
$$|\mb z+ \mb w|\approx |\mb z|+\frac{\mb z}{|\mb z|} \cdot \mb w +O(|\mb w|^2) .$$
Thus, also taking $\gamma_{\mathbf{R}}\approx 1$, we can approximate the argument of the exponent as
\begin{equation}
\begin{split}
t_\mathbf{R}^{\mb x}(s)-t_\mathbf{R}^{i}(s)=&\frac{|\mb x_\mb T(s)+\mb x -\mb x_\mathbf{E}|}{c_0}+\frac{| \mb x_\mb T(s)+\mb x -\mb x_\mathbf{R}|}{c_0}\gamma_{\mathbf{R}}(\mb x_\mb T(s)+\mb x ,\mb x_\mathbf{E},\mathbf{v}_T)\\
-&\frac{|\mb x_\mb T(s)+\mb y_i -\mb x_\mathbf{E}|}{c_0}-\frac{| \mb x_\mb T(s)+\mb y_i -\mb x_\mathbf{R}|}{c_0}\gamma_{\mathbf{R}}(\mb x_\mb T(s)+\mb y_i ,\mb x_\mathbf{E},\mathbf{v}_T)\\
&\approx \frac{1}{c_0}(\mb x-\mb y_i)\cdot \frac{\mb x_\mb T(s)-\mb x_\mb E}{|\mb x_\mb T(s)-\mb x_\mb E|}+ \frac{1}{c_0}(\mb x-\mb y_i)\cdot \frac{\mb x_\mb T(s)-\mb x_\mb R}{|\mb x_\mb T(s)-\mb x_\mb R|}
\end{split}
\end{equation}
We can then write the integral over the physical aperture as
\begin{equation}
\begin{split}
\int d\mb x_\mb R e^{i\omega (t_\mb R^{\mb x}(s)-t_\mb R^i(s))}&\approx \int d\mb x_\mb R e^{i \omega \left(\frac{1}{c_0}(\mb x-\mb y_i)\cdot \frac{\mb x_\mb T(s)-\mb x_\mb E}{|\mb x_\mb T(s)-\mb x_\mb E|}+ \frac{1}{c_0}(\mb x-\mb y_i)\cdot \frac{\mb x_\mb T(s)-\mb x_\mb R}{|\mb x_\mb T (s)-\mb x_\mb R|}\right)}\\
&= e^{i \omega \left(\frac{1}{c_0}(\mb x-\mb y_i)\cdot \frac{\mb x_\mb T(s)-\mb x_\mb E}{|\mb x_\mb T(s)-\mb x_\mb E|}\right)}\int d\mb x_\mb R 
e^{i \omega \left(\frac{1}{c_0}(\mb x-\mb y_i)\cdot \frac{\mb x_\mb T(s)-\mb x_\mb R}{|\mb x_\mb T(s)-\mb x_\mb R|}\right)}
\end{split}
\label{eq:k_w_ex}
\end{equation}
This is a classical result in array imaging \cite{bleistein}. The physical receiver array induces an integral over phases linear in the difference between the positions of the scatterer and the search point $\mb x- \mb y_i$. 
If we further define $$\mb k_\mb R=\frac{\omega}{c_0}\frac{\mb x_\mb T(s)-\mb x_\mb R}{|\mb x_\mb T(s)-\mb x_\mb R|},\quad  |\mb k _{\mb R }|=\frac{\omega}{c_0},$$ 
the integral is in fact one over a domain $\Omega$ in $\mb k$ space, which is a subdomain of the 2-sphere with radius $\omega/c_0$, as illustrated in Figure~\ref{fig:k_w},
\begin{equation}
\begin{split}
e^{i \omega \left(\frac{1}{c_0}(\mb x-\mb y_i)\cdot \frac{\mb x_\mb T(s)-\mb x_\mb E}{|\mb x_\mb T(s)-\mb x_\mb E|}\right)}\int\limits_{\Omega} d \mb k_{\mb R}|J|
e^{i \mb k_{\mb R}\cdot(\mb x-\mb y_i)},\quad \Omega\subset S^2(\omega/c_0)
\end{split}
\end{equation}
The Jacobian $J$ and the integral can be evaluated in a more general setting using the Green-Helmholtz identity \cite{garnier2016passive}.
We can then  write \eqref{eq:M_w} as
\begin{equation}
\sum\limits_{i,j=1}^M\rho_i\rho_j\int d\omega \int ds|\xi(\omega,s)|^2e^{i \omega \left(\frac{1}{c_0}((\mb x-\mb y_i)-(\mb y-\mb  y_j))\cdot \frac{\mb x_\mb T(s)-\mb x_\mb E}{|\mb x_\mb T(s)-\mb x_\mb E|}\right)} \int\limits_{\Omega\times \Omega} d \mb k_{\mb R}d \mb k_{\mb R'}|J||J'|
e^{i \left(\mb k_{\mb R}\cdot(\mb x-\mb y_i)- \mb k_{\mb R'}\cdot(\mb y-\mb y_j)\right)}
\label{eq:k_w_int}
\end{equation}
Similar expressions have also been analyzed in \cite{bleistein}. As expected, the resolution is inversely proportional to the physical aperture size, and is bounded by $\lambda/2$. 

\begin{figure}[htbp]
	\centering
	\begin{subfigure}[t]{0.6\textwidth}
	\includegraphics[width=\textwidth]{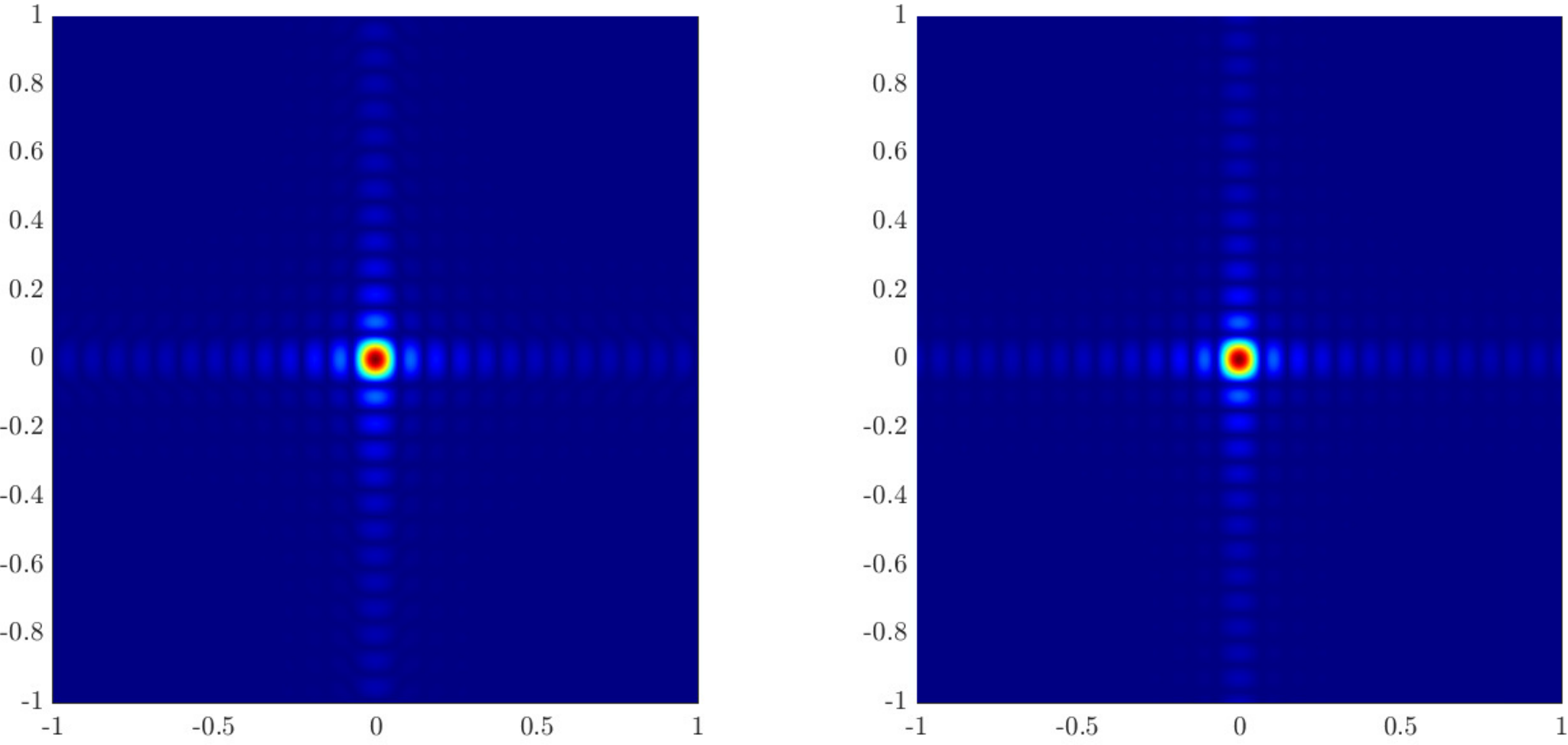}
	\caption{}
	\label{fig:k_w_comp}
\end{subfigure}
\begin{subfigure}[t]{0.4\textwidth}
	\includegraphics[width=\textwidth]{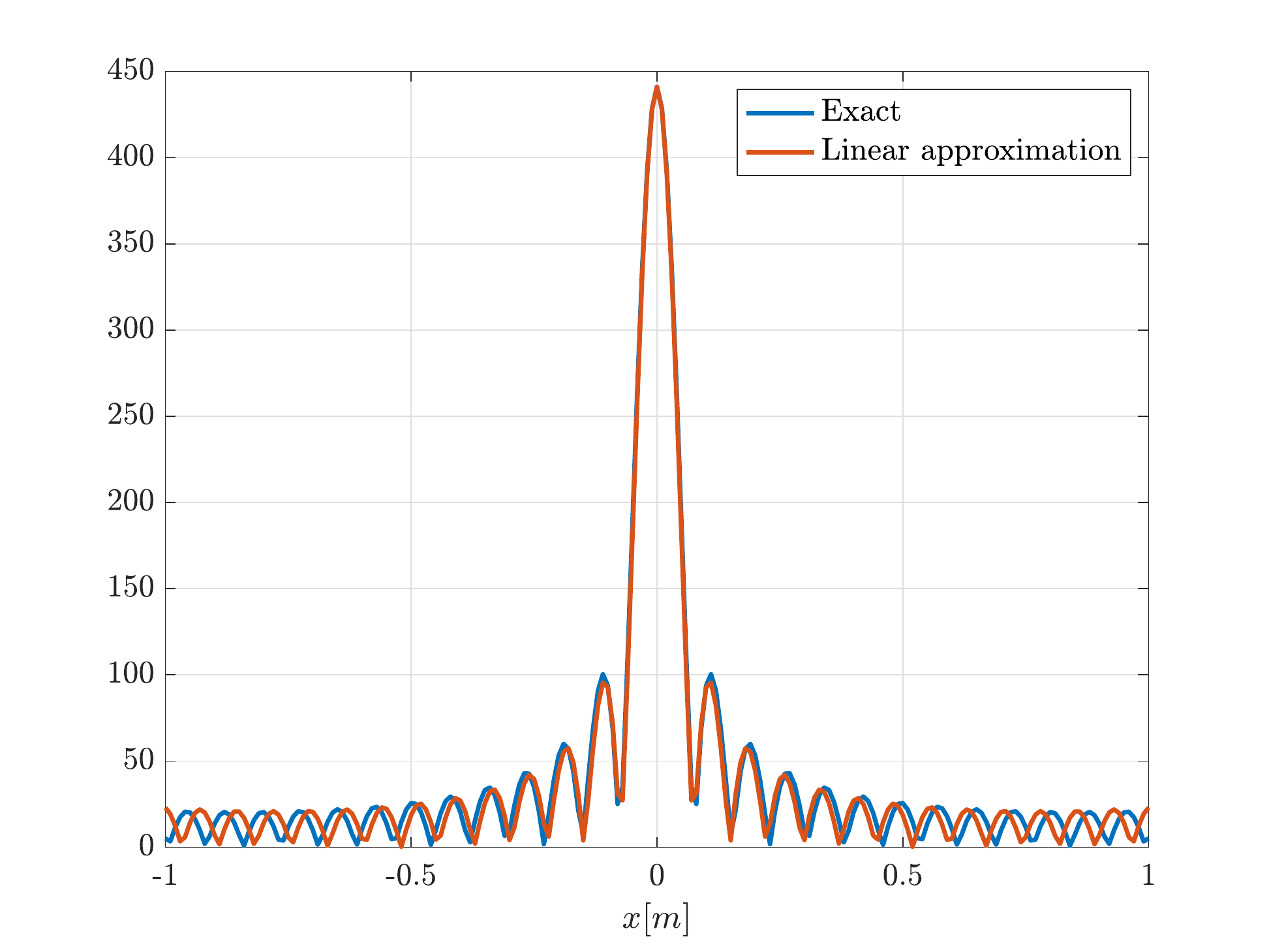}
	\caption{}
\end{subfigure}
\caption{$(a)$ Evaluation of the absolute value of \eqref{eq:k_w_ex}, with exact expressions for $t_{\mb R} ^\mb x$ (left), and calculation using the linear approximation of \eqref{eq:k_w_cart} (right). The 400 receivers are located on a Cartesian grid  of size $200\text{km} \times 200\text{km}$ with 10km spacing $(b)$ 1D cut of Figure~\ref{fig:k_w_comp}. We can see that the linear approximation is well justified.}
	\label{fig:k_w_comp_crossection}
\end{figure}

\begin{figure}[htbp]
	\centering
	\begin{subfigure}[t]{0.6\textwidth}
		\includegraphics[width=\textwidth]{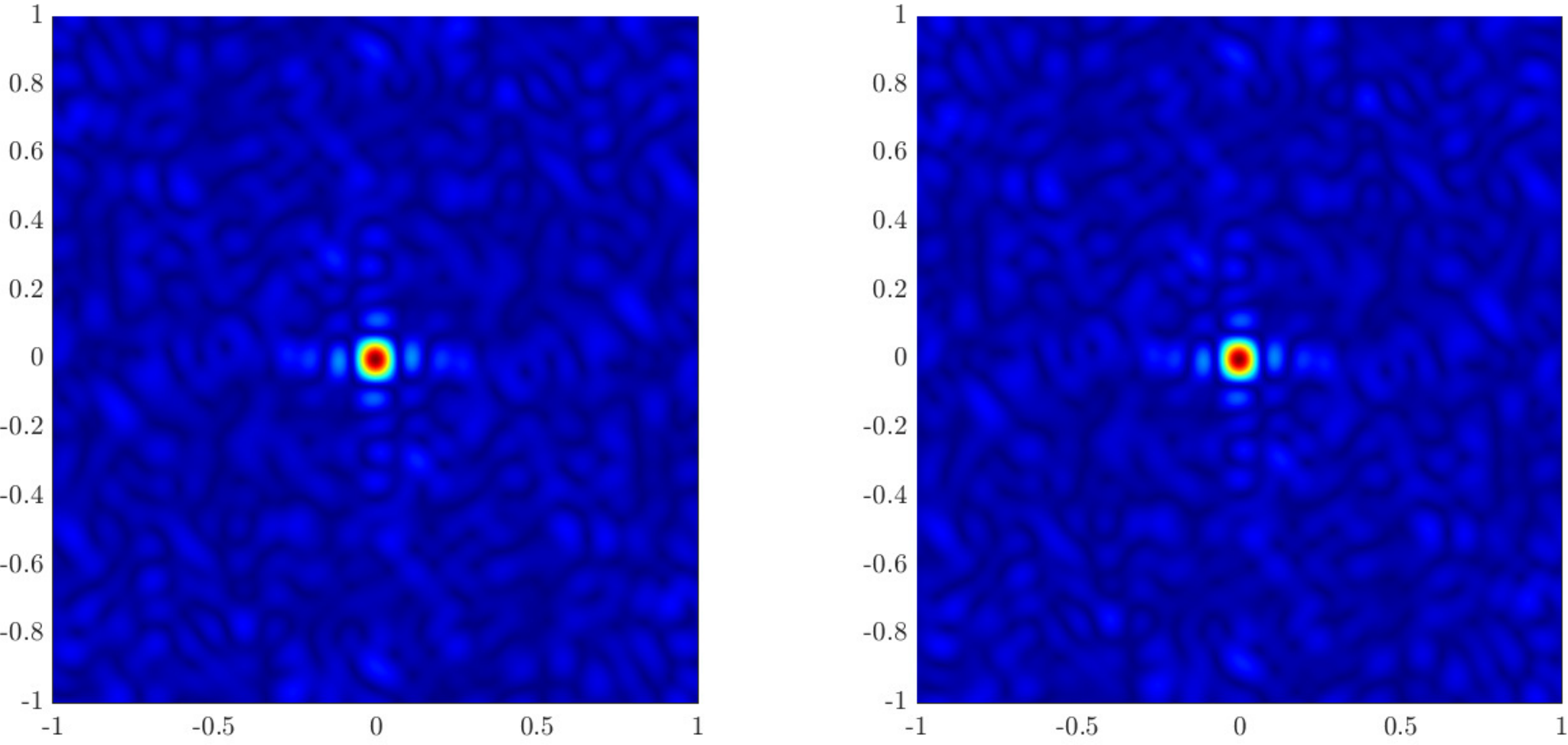}
		\caption{}
		\label{fig:k_w_comp_random}
	\end{subfigure}
	\begin{subfigure}[t]{0.4\textwidth}
		\includegraphics[width=\textwidth]{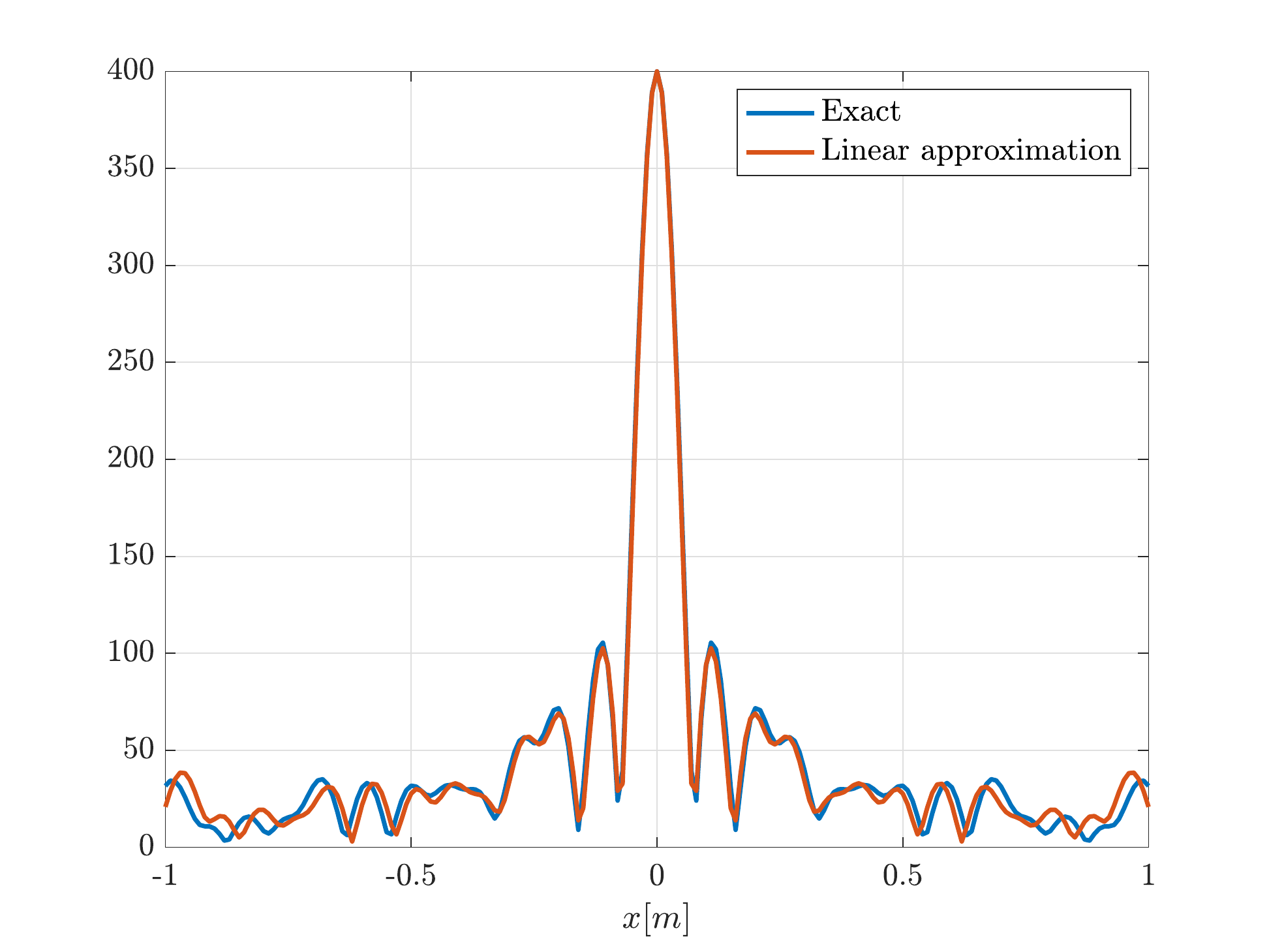}
		\caption{}
	\end{subfigure}
	\caption{$(a)$Evaluation of the absolute value of \eqref{eq:k_w_ex}, with exact expressions for $t_{\mb R} ^\mb x$ (left), and calculation using the linear approximation of \eqref{eq:k_w_cart} (right). The 400 receivers uniformly distributed in a square of size $200\text{km} \times 200\text{km}$ with 10km spacing $(b)$ 1D cut of Figure~\ref{fig:k_w_comp_random}. We can see that the linear approximation is well justified even when receivers are randomly distributed.}
	\label{fig:k_w_comp_crossection_random}
\end{figure}

We can make a further approximation to simplify the calculation in our case. Assume $$|\mb x_\mb T(s)-\mb x_\mb R|,|\mb x_\mb T(s)-\mb x_\mb E|\approx H_\mb T, $$ where $H_\mb T$ is the height of the targets. This is well justified for uniformlly distributed ground receivers and an inverse aperture of a size comparable to the ground array ($|\mb v_T S|\gtrsim a$,  $a$ the diameter of the receiver array). In this approximation we also have that $\xi(\omega,s)\approx \xi(\omega)$. Noting $\mb x_\mb T(s)=\mb x_\mb T+s \mb v_\mb T$, we can rewrite the integral \eqref{eq:k_w_int} as
\begin{equation}
\label{eq:inter_res}
\begin{split}
& \sum\limits_{i,j=1}^M\rho_i\rho_je^{i \omega \frac{1}{c_0}\frac{\mb x_\mb T-\mb x_\mb E}{H_\mb T}\cdot\left((\mb x-\mb y_i)-(\mb y-\mb  y_j)\right)}\int d\omega |\xi(\omega)|^2 \\&\times\int d\mb x_\mb R e^{i \frac{\omega}{c_0}\frac{\mb x_\mb T-\mb x_\mb R}{H_\mb T}\cdot(\mb x-\mb y_i)}\int d\mb x_\mb {R'} e^{-i \frac{\omega}{c_0}\frac{\mb x_\mb T-\mb x_\mb {R'}}{H_\mb T}\cdot(\mb y-\mb y_j)}
\int \limits_{-S/2}^{S/2}ds e^{i2 \frac{\omega}{c_0}\frac{s \mb v_\mb T}{H_\mb T}\cdot\left((\mb x-\mb y_i)-(\mb y-\mb y_j)\right)}\\
&\approx  C\sum\limits_{i,j=1}^M\rho_i\rho_j\int d\omega |\xi(\omega)|^2 \mathcal{B} (\mb x-\mb y_i)\mathcal{B}^*(\mb y-\mb y_j)\hspace{0.1em}\text{sinc} \left( \frac{\omega }{c_0}\frac{S\mb v_\mb T }{H_\mb T }  \cdot \left((\mb x-\mb y_i)-(\mb y-\mb y_j)\right)\right).
\end{split}
\end{equation}
If we assume a rectangular grid $\mb x_\mb R=(x_1,x_2), x_i\in[-a/2,a/2]$ then the point spread function is
\begin{equation}
\mathcal{B}(\mb x-\mb y_i)=\int d\mb x_\mb R e^{i \frac{\omega}{c_0}\frac{\mb x_\mb T-\mb x_\mb R}{H_\mb T}\cdot(\mb x-\mb y_i)}=a^2e^{i \frac{\omega}{c_0}\frac{\mb x_\mb T}{H_\mb T}\cdot(\mb x-\mb y_i)}\text{sinc}\left(\frac{\omega}{c_0}\frac{a}{2H_\mb T}(x_1-y_{i,1})\right)\text{sinc}\left(\frac{\omega}{c_0}\frac{a}{2H_\mb T}(x_2-y_{i,2})\right)
\label{eq:k_w_cart}
\end{equation}
As illustrated in  figures~\ref{fig:k_w}-\ref{fig:k_w_comp_crossection_random}, the approximations made above are well justified in the context of our numerical simulations.

We see that $\mathcal{B}(\mb x-\mb y_i)$ is, as a function of $\mb x$, localized around the scatterer location $\mb y_i$. Hence $\tilde{\mb X}_{\mb x,\mb y}$ has peaks at points $(\mb x,\mb y)=(\mb y_i,\mb y_j)\in \mathbb{R}^4$, where $\mb y_i,\mb y_j$ are scatterer positions. $\mathcal{B}(\mb x)$ has an effective resolution $\lambda \frac{H_\mb T}{a}$. Recall that this is the classical cross range resolution obtained with an array of size $a$, imaging a point at range $H_\mb T$. On the other hand, the synthetic aperture produces the term
$\text{sinc} \left( \frac{\omega }{c_0}\frac{\mb v_\mb T}{H_\mb T} S \cdot \left((\mb x-\mb y_i)-(\mb y-\mb y_j)\right)\right)$ in (\ref{eq:inter_res}), which has resolution $\lambda \frac{H_\mb T}{2\mb v_\mb T S}$. For small $S$ the resolution is dominated by the size $a$ of the receiver array. However as the inverse aperture increases and becomes comparable to $a/2$, anisotropy appears, as illustrated in Figure~\ref{fig:single_cross} in Section ~\ref{sec:prop_filt}. The effect of the synthetic aperture is however not isotropic. Since the argument of the sinc is $\mb x-\mb y_i-(\mb y-\mb y_j)$, the direction in which the argument is constant will be unaffected, while the orthogonal direction experiences the strongest decay. We note again that the single point migration is equivalent to taking $\mb x=\mb y$ with $\mb x-\mb y=0$ and so its resolution does not benefit from the synthetic aperture.

We have thus shown that we can approximate the form of $\tilde{\mb X}$ in \eqref{eq:M_tilde} as a collection of localized peaks, located at pairs of scatterer positions. When considering the extended domain of the interference pattern the peaks exhibit anisotropy with principal widths that are aligned with the directions $\mb x+\mb y=\text{constant}$, $\mb x-\mb y=\text{constant}$. 

The rank-1 image takes as an alternative image the top eigenvector of $\tilde{\mb X}$. In Appendix ~\ref{app:kernel}, we analyze the top eigenfunction of kernels that have a similar anisotropic form, and show that in general the resolution of the eigenfunction is better than the maximal width associated with single point migration.

\section{Eigenfunction of localized kernels}
\label{app:kernel}
\setcounter{equation}{0}

In this appendix we estimate the top eigenfunction of a localized anisotropic kernel in one dimension. This is the continuum limit of the results obtained in Appendix~\ref{app:stat_phase}, where it was shown that the peaks of the two point interference pattern are anisotropic. We evaluate the top eigenfunction, show it is localized, and get an estimate for its width in terms of the principal widths of the kernel. We show results in 1D, but they generalize to higher dimensions.

We saw in (\ref{eq:inter_res}) that the peaks of the two point interference pattern have principal widths in the directions $\mb x+\mb y=\text{constant}$, $\mb x-\mb y=\text{constant}$. Hence, we are interested in analyzing kernels of the form $$\mathcal{K}(x,y)=\sum\limits_{i,j=1}^M \rho_i\rho_j K(x-y_i,y-y_j),\quad K(x,y)=f\left( \frac{x-y}{\alpha}\right)f\left( \frac{x+y}{\beta}\right),$$ where $f(x)$ is localized around $x=0$. This is an intermediate case between two extreme cases. If $\alpha=\beta$, $f(x)$ is approximately separable (indeed separable if $f(x)$ is a Gaussian), and $K(x,y)$ has rank 1 with eigenfunction $f\left(\frac{x}{\alpha}\right)$. On the other hand if either $\alpha,\beta\rightarrow \infty$,  $K(x,y)$ would be a correlation/convolution kernel and will have full rank, with non localized eigenfunctions.  We are thus interested in the intermediate case.

We specifically look at two cases for $f(x)$- we start with a Gaussian kernel for which an analytical solution exists, and then consider a more realistic sinc model. While the Gaussian case is given here for illustration purposes, it could represent a situation where the medium has random fluctuations \cite{borcea2003theory}.

We first analyze the eigenfunctions of $K(x,y)$ in the two cases. We then show how the eigenfunction of $\mathcal{K}(x,y)$ can be expressed using them.
\subsection{Gaussian Kernel Approximation}
Introduce the following Gaussian model  for $K(x,y)$,
\begin{equation}
K(x,y)=e^{-\frac{\alpha}{2}(x+y)^2-\frac{\beta}{2}(x-y)^2}=e^{-\alpha(\frac{x+y}{\sqrt{2}})^2-\beta(\frac{x-y}{\sqrt{2}})^2},\quad x,y\in\mathbb{R}.
\label{eq:kernel_form_1D}
\end{equation}
And plug in a candidate eigenfunction $u(y)=e^{-\gamma y^2}$. For $u$ to be an eigenfunction, it must satisfy
\begin{equation}
\int_{\mathbb{R}} dy K(x,y) u(y)=\lambda_u u(x).
\end{equation}
The exponential argument of $K(x,y)u(y)$ has the form:
\begin{equation}
\begin{split}
&-\frac{\alpha}{2}(x+y)^2-\frac{\beta}{2}(x-y)^2-\gamma y^2
= -\frac{\alpha+\beta}{2} x^2-\frac{\alpha+\beta+2\gamma}{2}y^2+(\alpha-\beta)xy.
\end{split}
\end{equation}
Denote $z=\frac{\alpha+\beta+2\gamma}{2}$, then we can rewrite the argument as
\begin{equation}
\begin{split}
-\frac{\alpha+\beta}{2} x^2-z\left(y-\frac{\alpha-\beta}{2z}x\right)^2+\frac{(\alpha-\beta)^2}{4z}x^2.
\end{split}
\end{equation}
Integrating over $y$ we get
\begin{equation}
\int _\mathbb{R}dy K(x,y) u(y)\propto e^{-\left(\frac{\alpha+\beta}{2}-\frac{(\alpha-\beta)^2}{4z}\right)x^2}
\end{equation}
i.e. plugging in the value for $z$, the equation for $\gamma>0$ is
\begin{equation}
\begin{split}
&\gamma=\frac{\alpha+\beta}{2}-\frac{(\alpha-\beta)^2}{2(\alpha+\beta)+4\gamma}\\
\Rightarrow & 2\gamma(\alpha+\beta+2\gamma)=(\alpha+\beta)^2+2\gamma(\alpha+\beta)-(\alpha-\beta)^2\\
\Rightarrow &4\gamma^2=4\alpha\beta \Rightarrow \gamma =\sqrt{\alpha\beta}
\end{split}
\end{equation}
The eigenfunction has a variance which is the geometric mean of the variances of the two principal axes.

Since $K(x,y)$ has non-negative entries and its support is a non-disjoint set, we know by the Krein-Rutman theorem \cite{lax2002functional} (extension of the Perron-Frobenius theorem to Banach spaces) that its top eigenfunction can be chosen to be strictly non-negative, so we have indeed recovered the top eigenfunction. 
\begin{figure}[htbp]
	\centering
	\includegraphics[width=0.5\textwidth]{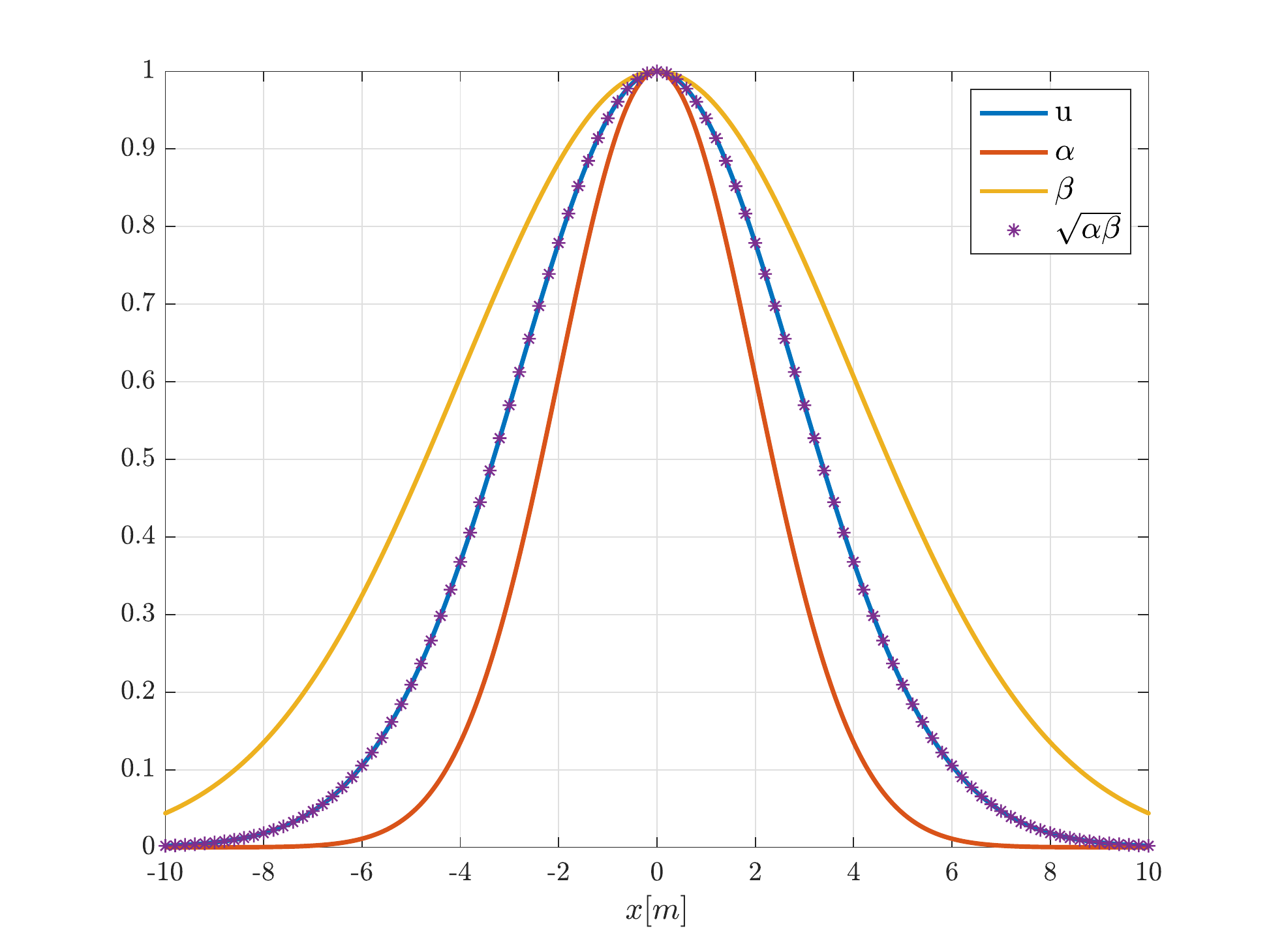}
	\caption{Comparison of the top singular value of the Gaussian kernel of form \eqref{eq:kernel_form_1D}. We can see that indeed the eigenfunction is a Gaussian with variance which is a geometric mean of $\alpha,\beta$.}
	\label{fig:geo_mean}
\end{figure}
\subsection{Sinc kernel approximation}
We now wish to approximate the kernel as a product of 2 sinc functions:
\begin{equation}
K(x,y)=\text{sinc}(\alpha(x-y))\text{sinc}(\beta(x+y)),\quad \alpha>\beta ,\quad x,y\in\mathbb{R}.
\label{eq:ker_sinc}
\end{equation}
This is a more realistic case since we saw that the behavior around the peaks of the two point interferece pattern can be approximated as a product of sinc functions.
We'll look for a candidate eigenfunction of the form $u=\text{sinc}(\gamma y)$.
 Numerical simulations suggest that while this is not an exact eigenfunction, it is close to one. Specifically, $K u\approx \lambda u +\epsilon, \|\epsilon\|_2\ll1$.

We are looking to estimate an integral of the form:
\begin{equation}
\int_\mathbb{R}\frac{\sin(\alpha(x-y))\sin(\beta(x+y))\sin(\gamma y)}{\alpha(x-y)\beta(x+y)\gamma y}dy
\end{equation}
We first drop the constants from the denominator and use the following identity:
\begin{equation}
\frac{1}{(x-y)(x+y )y}=\frac{1}{2x^2}\left[\frac{2}{y}-\frac{1}{x+y}+\frac{1}{x-y}\right].
\label{eq:denom_decomp}
\end{equation}
Next we decompose the numerator using the following identity
\begin{equation}
\sin(a)\sin(b)\sin(c)=\frac{1}{4}\left(\sin(a-b+c)-\sin(a-b-c)-\sin(a+b+c)+\sin(a+b-c)\right).
\end{equation}

Which yields in our case:
\begin{equation}
\begin{split}
\frac{1}{4}& [\sin((\alpha-\beta)x-(\alpha+\beta-\gamma) y)-\sin((\alpha-\beta)x-(\alpha+\beta+\gamma) y)\\
&-\sin((\alpha+\beta)x-(\alpha-\beta-\gamma) y)+
\sin((\alpha+\beta)x-(\alpha-\beta+\gamma) y)].
\end{split}
\end{equation}
We can combine all the terms in \eqref{eq:denom_decomp} to a single denominator $z$ using an appropriate variable transformation
$z=y, z=x+y,z=x-y$.

This will give us the following numerator, up to a constant
\begin{equation}
\begin{split}
& [2\sin((\alpha-\beta)x-(\alpha+\beta-\gamma) z)-2\sin((\alpha-\beta)x-(\alpha+\beta+\gamma) z)\\
&-2\sin((\alpha+\beta)x-(\alpha-\beta-\gamma) z)+
2\sin((\alpha+\beta)x-(\alpha-\beta+\gamma) z)]\\
-& [\sin((\alpha-\beta)x-(\alpha+\beta-\gamma) (z-x))-\sin((\alpha-\beta)x-(\alpha+\beta+\gamma) (z-x))\\
&-\sin((\alpha+\beta)x-(\alpha-\beta-\gamma) (z-x)+
\sin((\alpha+\beta)x-(\alpha-\beta+\gamma) (z-x))]\\
+& [\sin((\alpha-\beta)x-(\alpha+\beta-\gamma) (x-z))-\sin((\alpha-\beta)x-(\alpha+\beta+\gamma) (x-z))\\
&-\sin((\alpha+\beta)x-(\alpha-\beta-\gamma) (x-z)+
\sin((\alpha+\beta)x-(\alpha-\beta+\gamma) (x-z))]
\end{split}
\end{equation}
Which is up to a constant
\begin{equation}
\begin{split}
& [2\sin((\alpha-\beta)x-(\alpha+\beta-\gamma) z)-2\sin((\alpha-\beta)x-(\alpha+\beta+\gamma) z)\\
&-2\sin((\alpha+\beta)x-(\alpha-\beta-\gamma) z)+
2\sin((\alpha+\beta)x-(\alpha-\beta+\gamma) z)]\\
-& [\sin((2\alpha-\gamma)x-(\alpha+\beta-\gamma) z)-\sin((2\alpha+\gamma)x-(\alpha+\beta+\gamma) z)\\
&-\sin((2\alpha-\gamma)x-(\alpha-\beta-\gamma) z)+
\sin((2\alpha+\gamma )x-(\alpha-\beta+\gamma) z)]\\
+& [\sin((-2\beta+\gamma)x+(\alpha+\beta-\gamma) z)-\sin(-2\beta-\gamma )x+(\alpha+\beta+\gamma) z)\\
&-\sin((2\beta+\gamma )x+(\alpha-\beta-\gamma) z+
\sin((2\beta-\gamma )x+(\alpha-\beta+\gamma) z)]
\end{split}
\end{equation}
We will use extensively the following result:
\begin{equation}
\text{PV}\int\limits_{-\infty}^{\infty}\frac{\sin(a z+b)}{z}dz=\pi \cos(b)\text{sign}(a) \end{equation}

The integral then yields:
\begin{equation}
\begin{split}
& [2\cos((\alpha-\beta)x)\text{sign}(-(\alpha+\beta-\gamma))-2\cos((\alpha-\beta)x)\text{sign}(-(\alpha+\beta+\gamma))\\
&-2\cos((\alpha+\beta)x)\text{sign}(-(\alpha-\beta-\gamma))+
2\cos((\alpha+\beta)x)\text{sign}(-(\alpha-\beta+\gamma)]\\
-& [\cos((2\alpha-\gamma)x)\text{sign}(-(\alpha+\beta-\gamma) )-\cos((2\alpha+\gamma)x)\text{sign}(-(\alpha+\beta+\gamma))\\
&-\cos((2\alpha-\gamma)x)\text{sign}(-(\alpha-\beta-\gamma) )+
\cos((2\alpha+\gamma )x)\text{sign}(-(\alpha-\beta+\gamma) z]\\
+& [\cos((-2\beta+\gamma)x)\text{sign}((\alpha+\beta-\gamma) )-\cos(-2\beta-\gamma )x)\text{sign}((\alpha+\beta+\gamma) )\\
&-\cos((2\beta+\gamma )x)\text{sign}(\alpha-\beta-\gamma)+
\cos((2\beta-\gamma )x)\text{sign}(\alpha-\beta+\gamma) ]
\end{split}
\end{equation}
We assume $\alpha>\gamma>\beta$, so some of the signs are unambiguous
\begin{equation}
\begin{split}
& [-2\cos((\alpha-\beta)x)+2\cos((\alpha-\beta)x)\\
&-2\cos((\alpha+\beta)x)\text{sign}(-(\alpha-\beta-\gamma))-
2\cos((\alpha+\beta)x)]\\
-& [-\cos((2\alpha-\gamma)x) +\cos((2\alpha+\gamma)x)\\
&-\cos((2\alpha-\gamma)x)\text{sign}(-(\alpha-\beta-\gamma) )-
\cos((2\alpha+\gamma )x)]\\
+& [\cos((-2\beta+\gamma)x)-\cos((-2\beta-\gamma )x) \\
&-\cos((2\beta+\gamma )x)\text{sign}(\alpha-\beta-\gamma)+
\cos((2\beta-\gamma )x) ]
\end{split}
\end{equation}
If we further assume $\alpha-\beta<\gamma$ we get that the integral is proportianal to 
\begin{equation}
\frac{\cos((2\alpha-\gamma )x)+\cos((-2\beta+\gamma )x)-2\cos((\alpha +\beta)x)}{x^2}.
\label{eq:ker_sinc_out}
\end{equation}
In order to find the approximate width $\gamma$ we wish to find values of $\gamma$ for which \eqref{eq:ker_sinc_out} has a zero for $\pi/\gamma$.

\begin{figure}[htbp]
	\centering
	\includegraphics[width=0.5\textwidth]{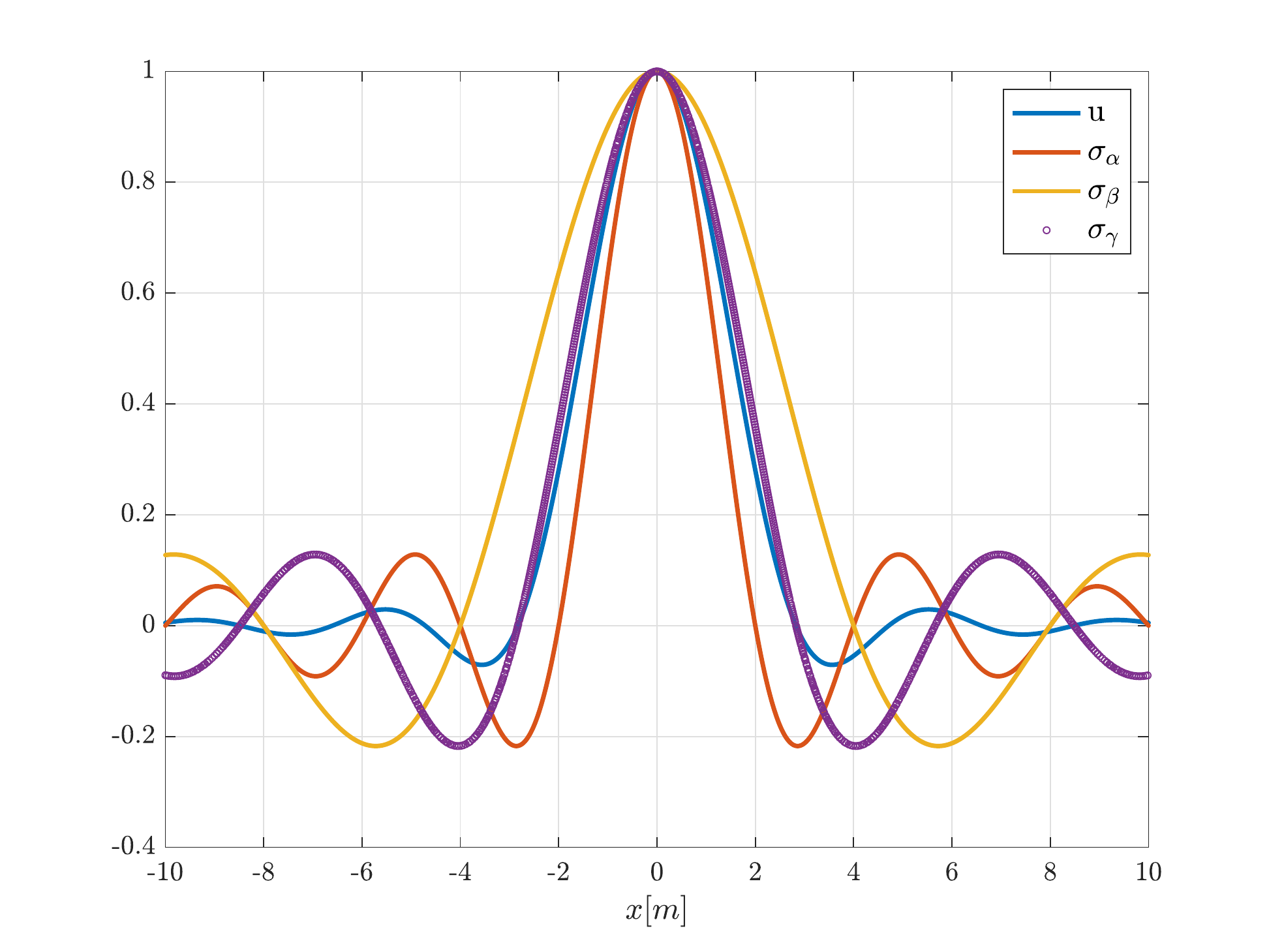}
	\caption{Comparison of the top singular value of the sinc kernel of form \eqref{eq:ker_sinc}. We can see that indeed the eigenfunction's main lobe is approximated by a sinc with the width $\sigma_\gamma$, though it decays much faster.}
	\label{fig:harm_mean}
\end{figure}

\begin{figure}[htbp]
	\centering
	\includegraphics[width=0.5\textwidth]{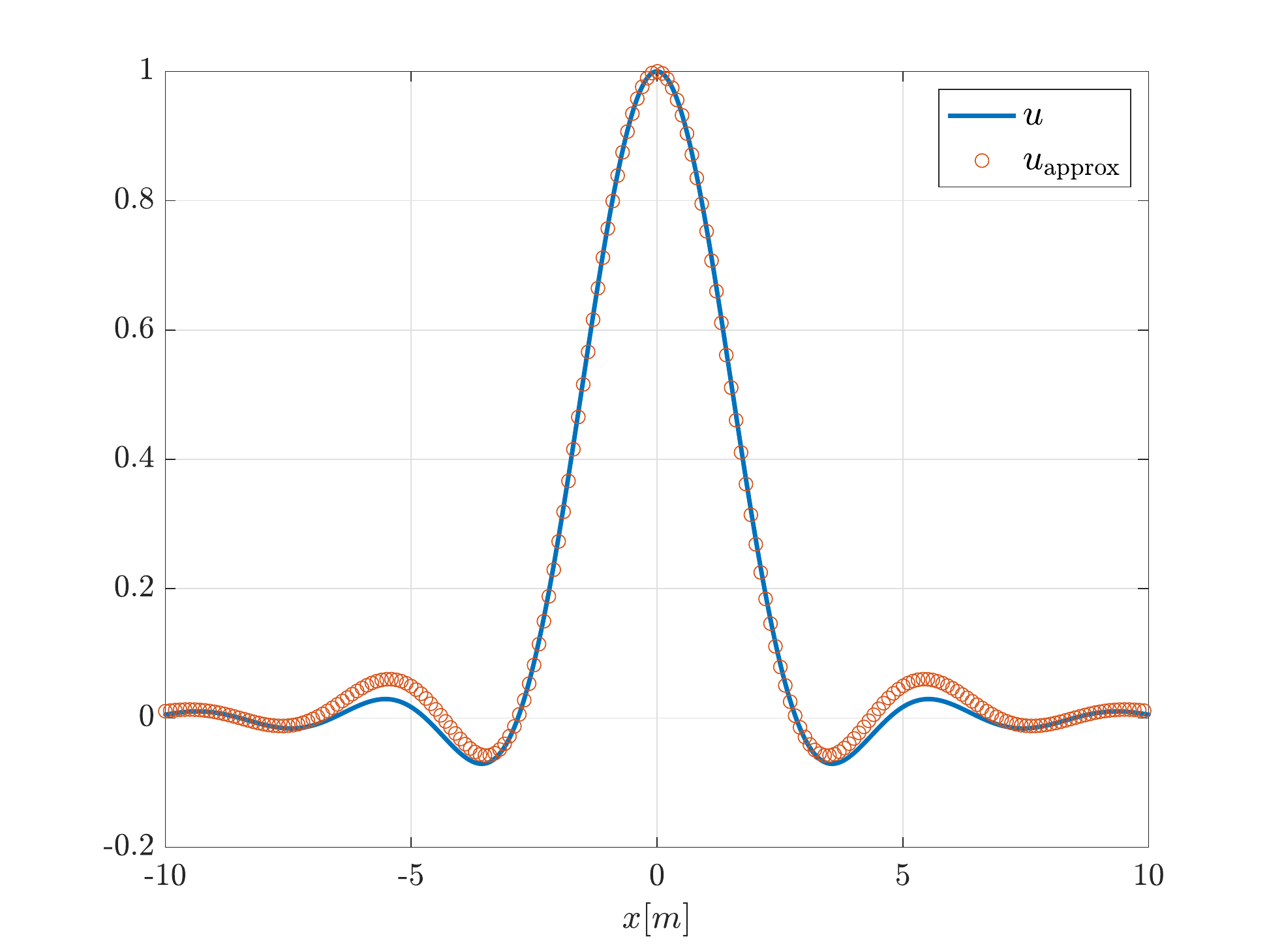}
	\caption{Comparison of the top singular value of the sinc kernel of form \eqref{eq:ker_sinc} with the approximate eigenfunction in \eqref{eq:ker_sinc_out}. We can see that \eqref{eq:ker_sinc_out} indeed approximates the top eigenfunction with a high degree of accuracy.}
	\label{fig:approx_example}
\end{figure}

We can rewrite the numerator as
\begin{equation}
2\cos((\alpha-\beta)x)\cos((\alpha+\beta-\gamma)x)-2\cos((\alpha+\beta)x)
\end{equation}
for $x=\pi/\gamma$ (the first zero of $\text{sinc}(\gamma x)$), using $\cos(a-\pi)=-\cos(a)$, the numerator becomes
\begin{equation}
-2\cos((\alpha+\beta)\pi/\gamma)(\cos((\alpha-\beta)\pi/\gamma)+1),
\end{equation}
which has zeros at $$\gamma=\frac{\alpha+\beta}{n+1/2},n\in\mathbb{Z},\quad \gamma=\frac{\alpha-\beta}{2n+1},n\in\mathbb{Z}$$
The most localized eigenvector would be for the largest value of $\gamma$, e.g. for $\alpha=8,\beta=2$, we get $\gamma=\frac{2}{3} (\alpha+\beta)$, which indeed matches the numerical estimation as can be seen in figures \ref{fig:harm_mean} and \ref{fig:approx_example}. The effective width of $\sin(c(x\pm y))$, defined by the first zero in the $\frac{1}{\sqrt{2}}(\hat{x} \mp \hat{y})$ direction is $\frac{\pi}{\sqrt{2}c}$. So if we denote $$\sigma_\alpha=\frac{\pi}{\sqrt{2}\alpha},\sigma_\beta=\frac{\pi}{\sqrt{2}\beta},\sigma_\gamma=\frac{\pi}{\gamma},$$ we get the equation
\begin{equation}
\sigma_{\mb u}\approx \sigma_\gamma=\frac{3}{\sqrt{2}}\frac{\sigma_\alpha \sigma_\beta}{\sigma_\alpha+\sigma_\beta},
\end{equation}
which is very close to the harmonic mean. 
It is easy to verify that with $\alpha-\beta>\gamma$ the integral is 0 in principal value.

We see that in the sinc case the top eigenfunction is even more favorable compared to the case of a Gaussian kernel in the previous subsection. Now $\sigma _\mb u$ is close to the harmonic mean which is biased toward the smaller width, that is, we gain in resolution compared to the Gaussian case, where we got the geometric mean. This makes sense as a Gaussian case could represents a random medium, as noted before, and we expect better resolution in media with weak fluctuations or no randomness. We also note that the top eigenfunction has a faster decay rate compared to the diagonal of the kernel, that is, the single point migration image. The estimation suggests $x^{-2}$ as a lower bound, compared to $x^{-1}$ decay of the Kernel itself.

\subsection{Eigenfunction of the multiple stationary points}
If now $\mathcal{K}(x,y)$ is a linear combination of translations of \eqref{eq:kernel_form_1D},
$$\mathcal{K}{}(x,y)=\sum\limits_{i,j} c_{ij}K(x-a_i,y-a_j),$$ then as long as the translations are far enough apart such that
$$
\int dyK(x-a_i,y-a_j)u(y-a_k)= c_{ij} u(x-a_i)\delta_{jk},$$
the top eigenfunction of $\mathcal{K}$ is 
$$
\mathcal{U}(x)=\sum\limits_i \alpha_i u(x-a_i)$$
with the restriction
$$
\sum\limits_{i,j} c_{ij} \alpha_j=\lambda\alpha_i$$
i.e, the eigenfunction's coefficient vector is an eigenvector of the coefficient matrix $c_{ij}$.

To summarize we have shown that the top eigenfunction of the superposition of anisotropic kernels is more localized than the maximal width of the kernel. Hence, on top of other considerations considered in Section~\ref{sec:generalized_migration}, the rank-1 image provides better resolution than the single point migration image. 
\bibliographystyle{plain} \bibliography{SBIR_SAT}

\end{document}